\documentclass[12pt]{article}

\usepackage[margin=0.9in]{geometry}

\usepackage{dsfont}
\usepackage{amsmath}       
\usepackage{natbib}
\usepackage{scrextend}  
\usepackage{amssymb}  

\usepackage{hyperref}
\hypersetup{
  colorlinks   = true,    
  urlcolor     = blue,    
  linkcolor    = blue,    
  citecolor    = blue     
}

\usepackage[utf8]{inputenc} 
\usepackage[T1]{fontenc}    
\usepackage{url}            
\usepackage{booktabs}       
\usepackage{amsfonts}       
\usepackage{nicefrac}       
\usepackage{microtype}      

\usepackage{graphicx}

\usepackage{algorithm}
\usepackage{algorithmic}

\usepackage{amsthm}  

\newtheorem{theorem}{Theorem}  
\newtheorem{lemma}{Lemma} 
\newtheorem{corollary}{Corollary} 
\newtheorem{pro}{Proposition} 

\newtheorem{remark}{Remark} 

\newcommand{\ind}[1]{\mathds{1}{\Big\{ {#1} \Big\} }}   
\newcommand{\indi}[1]{\mathds{1}{\big\{ {#1} \big\} }}   
\newcommand{\bw}{\boldsymbol{w}}    

\usepackage{authblk}
\title{Enhanced Nearest Neighbor Classification for Crowdsourcing}
\author[1]{Jiexin Duan\thanks{Senior Financial Modeler, Moody's Analytics, Inc., Newark, CA 94560. (Email: jiexin.duan@gmail.com).}}
\author[2]{Xingye Qiao\thanks{Associate Professor, Binghamton University, State University of New York, Binghamton, NY 13902. (Email: qiao@math.binghamton.edu).}}
\author[3]{Guang Cheng\thanks{Professor, Department of Statistics, University of California, Los Angeles, CA 90095. (Email: guangcheng@ucla.edu). }}
\affil[1]{Department of Statistics, Purdue University}
\affil[2]{Department of Mathematical Sciences, Binghamton University}
\affil[3]{Department of Statistics, UCLA}

\date{}

\begin{document}

\maketitle

\begin{abstract}
In machine learning, crowdsourcing is an economical way to label a large amount of data. However, the noise in the produced labels may deteriorate the accuracy of any classification method applied to the labelled data. We propose an enhanced nearest neighbor classifier (ENN) to overcome this issue. Two algorithms are developed to estimate the worker quality (which is often unknown in practice): one is to construct the estimate based on the denoised worker labels by applying the $k$NN classifier to the expert data; the other is an iterative algorithm that works even without access to the expert data. Other than strong numerical evidence, our proposed methods are proven to achieve the same regret as its oracle version based on high-quality expert data. As a technical by-product, a lower bound on the sample size assigned to each worker to reach the optimal convergence rate of regret is derived.
\end{abstract}


\newpage

\section{Introduction}\label{sec:intro_E}
In light of the needs of a large amount of labeled data as the training sets, machine learning researchers pay increasing attentions to crowdsourcing services such as the Amazon Mechanical Turk \footnote{\label{note:AMT}https://www.mturk.com/mturk/welcome}(AMT). In crowdsourcing, many independent and relatively inexpensive workers produce their labels that collectively determine a solution by aggregating these crowd opinions. Ideally, the ground truth labels are inferred from these noisy labels. In the literature, many methods \citep{dawid1979maximum,raykar2009supervised,whitehill2009whose} build probabilistic models for the crowdsourcing process and then derive the labels using Expectation Maximization-type algorithms \citep{dempster1977maximum}. Recently, classifiers that predict the labels for future observations directly from the crowdsourcing data has also been proposed \citep{dekel2009vox,wauthier2011bayesian,kajino2012convex}.

A commonly recognized challenge to classification using crowdsourcing data is the low quality of the workers \citep{sheng2008get,wauthier2011bayesian}. Previous proposals heavily depend on prior knowledge of the ground truth distribution \citep{raykar2010learning,yan2010modeling}. Additionally, many methods \citep{kajino2012learning,wang2015crowdsourcing} require the availability of the so-called expert data, whose labels are generated by ground truth distribution. To overcome the issue of low-quality workers, in this article, we propose a nonparametric classification method based on crowdsourcing data that requires neither the expert data nor prior knowledge of the ground truth distribution.

The nearest neighbor (NN) classifier \citep{FH51,CH67} is among the conceptually simplest and prevalent classification methods. Its statistical properties have been studied in \cite{DGKL94,S12,CD14,gottlieb2014near,gkm16,SQC16,doring2017rate,xue2017achieving}. See extensive surveys of $k$-NN classifiers in \citet{devroye2013probabilistic,biau2015lectures,chen2018explaining}. Applications of the NN classifier in crowdsourcing data have been studied in \citet{diab2012musical,hwang2012environmental, burrows2013paraphrase,li2019double}. To the best of our knowledge, there is no theoretical study on how NN classifiers work with crowdsourcing data.

We proposed a new NN classifier for crowdsourcing data that overcomes the noise in the low-quality worker labels. Our major contribution is the investigation of a type of crowdsourcing method where the worker labels data are first enhanced (hence dubbed as ``ENN'') and then a test data prediction is made  through a weighting scheme to aggregate the enhanced labels. This concise enhancement effort can substantially reduce the noise in worker labels. It has a potential to generalize to other methods than the NN classifier.

As the second contribution, we derive an asymptotic expansion form of the regret of the ENN classifier. This technical result is a nontrivial extension from \citet{S12}. Specifically, we enhance the noisy worker data with different quality and sizes, which leads to remainder terms bounded in a nontrivial way. With carefully chosen weights, the regret of ENN achieves the same optimal regret on the expert data as the ``oracle'' optimal weighted nearest neighbor (OWNN) \citep{S12}, in terms of both the rate of convergence and the multiplicative constant. Here, we define an ``oracle'' classifier as the classifier trained on an expert data set with the sample size. \citet{cannings2020classification} analyzed a special case with only one worker sample, and they assume that the Bayes classifier given noisy labels predicts as well as its ground truth version. This unrealistic assumption is not required in our analysis because of the use of the enhancing technique. 

Our proposed ENN requires quantifying the worker quality, which is often unknown in practice. Our third contribution is the development of two estimators for the worker quality. One method (ENN2) constructs the estimators based on the denoised worker labels through applying $k$NN classifier to the expert data. Unlike previous worker quality estimation methods, which had no statistical guarantee, ENN2 is proven to achieve the same regret as ENN with known worker quality. The other method (ENN3) uses ENN to estimate the worker quality in an iterative manner, and works well even without access to the expert data.

In summary, we have made the following contributions:\\
(1) A denoising enhancement to the worker data labels, which can be easily extended to other classifiers.\\
(2) A solid theoretical study of the statistical guarantee for the crowdsourcing data classification.\\
(3) Repetition of instance is not required, lowering the cost of label collection.\\
(4) Expert data is not required for ENN and ENN3, which is more practical for crowdsourcing data.

The rest of this article is organized as follows. Section~\ref{sec:pre_E} introduces the setting and notations. The asymptotic expansion form for the regret is presented in Section~\ref{sec:ENN}, followed by some comparisons between ENN and the oracle WNN. Section~\ref{sec:quality} focuses on the estimation of worker quality. Section~\ref{sec:exp_E} and Section~\ref{sec:conclusion_E} include numerical experiments and some concluding discussions.

\section{Preliminaries}\label{sec:pre_E}
Consider $s$ workers and $n$ instances in the crowdsourcing problem. Let $\mathcal{J}_j \subseteq \{1,\dots,n\}$ be an index set of instances that the $j$-th worker has labeled, and $n_j=|\mathcal{J}_j|$ be the number of instances the $j$-th worker has labeled. In total, the crowdsourcing data has $N={\textstyle\sum}_{j=1}^s n_j$ observations. $P^j$ defined on ${\mathbb R}^d\times\{0,1\}$ represents the joint distribution of the labeled data from the $j$-th worker. The ground truth distribution is denoted as $P^0$. We observe data from $s$ workers, ${\cal D}^{C}=\cup_{j=1}^s{\cal D}_j$ where ${\cal D}_j=\{(X_i^j, Y_i^j)\}_{i\in\mathcal{J}_j}\overset{iid}{\sim}P^j$. $Y_{i}^j$ is the label tagged by the $j$-th worker to the $i$-th instance. Denote the probability that an instance is labeled as class $r$ by worker $j$ as $\pi_r
^j:={\mathbb P}^j(Y=r)$. The conditional distribution of $X^j$ given $Y^j=r$ is denoted as $P_r^j$ for $r=0,1$. Hence, the marginal distribution of $X$ by worker $j$ is $$\bar{P}^j =\pi_1^j P_1^j + (1-\pi_1^j) P_0^j.$$ As the instances are randomly assigned to workers, we assume all worker data and the ground truth data share the same marginal distribution, i.e., $\bar{P}^j=\bar{P}$, for $j=0,\dots,s$.

Given $x$, the probability that worker $j$ would label it to be class 1 (i.e., the regression function) is defined as,
$$
\eta^j(x)={\mathbb P}^j(Y=1|X=x), \mbox{ for } j\in\{0,\dots,s\}.
$$
To model the labeling process, we assume the well known two-coin model \citep{raykar2010learning, kajino2012learning} below. The sensitivity and the specificity\footnote{In this paper, we assume that worker quality $a^j$ and $b^j$ are both constants, depending only on the unobserved ground truth label, but not on $x$, i.e., worker $j$ has the same quality on all instances.} for worker $j$ are defined as
\begin{align*}
    a^j={\mathbb P}^j(Y^j=1|Y^0 = 1),\makebox{ and }b^j = {\mathbb P}^j(Y^j = 0|Y^0 = 0),
\end{align*}respectively. 
Therefore, we have the following relationship between the $j$th worker's regression function and the ground truth regression function:
\begin{align}
\eta^j(x)=&a^j \eta^0(x) + (1-b^j)(1-\eta^0(x)). \label{eq:eta^j}
\end{align}
A worker who always gives the labels based on the $\eta^0(x)$ (i.e., $a^j=b^j=1$) is called an expert.

Our goal is to design classifiers $\phi$: ${\mathbb R}^d \rightarrow \lbrace 0, 1 \rbrace$, based on crowdsourced data, which minimizes the classification risk $
R(\phi)={\mathbb P}^0(\phi(X) \neq Y),$ under the ground truth distribution $P^0$. The theoretical minimizer of $R(\phi)$ is the so-called Bayes classifier $\phi^{\ast}(x)=\indi{\eta^0(x) \geq 1/2}$ with the corresponding Bayes risk $R(\phi^*)$. For any classifier $\widehat{\phi}_n := \Psi({\cal D})$ obtained by following a classification procedure $\Psi$ given the data $\cal D$, its regret is defined as: $$
{\rm Regret}(\Psi)={\mathbb E}_{\cal D} [ R(\widehat{\phi}_{n})] - R(\phi^{\ast}),
$$ where ${\mathbb E}_{\cal D}$ is with respect to the distribution
of the data $\cal D$.

We now introduce a general weighted nearest neighbor  (WNN) classifier. For a query point $x$, let $(X_{(1)},Y_{(1)})$, $(X_{(2)},Y_{(2)})$, $\ldots$ $(X_{(n)},Y_{(n)})$ be the sequence of observations with ascending distance to $x$, and denote $w_{ni}$ as the (non-negative) weight assigned to the $i$-th neighbor of $x$ with $\sum^n_{i=1} w_{ni}=1$. 
Define $\widehat{S}_{n,\bw_n}(x):=\sum^n_{i=1} w_{ni} Y_{i}$ as the WNN estimate of $\eta^0(x)$. The WNN prediction is thus
\begin{equation*}
\widehat{\phi}_{n, \bw_n}(x)=\ind{ \widehat{S}_{n,\bw_n}(x) \geq 1/2 },
\end{equation*} 
where $\bw_n$ denotes the weight vector. When $w_{ni}=k^{-1}$ for $1 \leq i \leq k$, or $0$ for $i>k$, WNN reduces to the standard $k$NN classifier, denoted as $\widehat{\phi}_{n,k}(x)$. Denote the WNN classifier on the expert data with size $N$ and on the crowdsourcing data ${\cal D}^{C}$ with the same size as $\widehat{\phi}_{N, \bw_N}^{0}(x)$ and $\widehat{\phi}_{N, \bw_N}^{C}(x)$, respectively. Proposition~\ref{thm:WNN_re_E} in \citet{S12} provides an asymptotic expansion of the WNN regret on the expert data.

\begin{pro}
\label{thm:WNN_re_E} (Asymptotic Regret for WNN) 
Assuming (A1)--(A4) stated in Appendix~\ref{sec:assumptions_E}, for each $\beta\in (0,1/2)$, we have, uniformly for $\bw_{N}\in W_{N,\beta}$,
\begin{align}
 {\rm Regret}(\widehat{\phi}_{N,\bw_{N}}^{0})&= \Big[B_1 \sum_{i=1}^{N} w_{Ni}^2 + B_2 \Big (\sum_{i=1}^{N} \frac{\alpha_i w_{Ni}}{N^{2/d}}\Big)^2 \Big]\{1+o(1)\},\label{eq:WNN_re_E}
\end{align}as $N \rightarrow \infty$,
where $\alpha_i=i^{1+\frac{2}{d}}-(i-1)^{1+\frac{2}{d}}$. Constants $B_1, B_2$ and $W_{N,\beta}$\footnote{In the case of $k$NN, it means $k$ satisfies $\max(n^{\beta}, (\log n)^2) \le k \le \min(n^{(1-\beta d /4)}, n^{1-\beta})$.} are defined in Appendix~\ref{sec:defwnb_E}.
\end{pro}

We remark that the first term in \eqref{eq:WNN_re_E} can be viewed as the variance component of regret, and the second term the squared bias. By minimizing the asymptotic regret \eqref{eq:WNN_re_E} over weights, \citet{S12} obtained the optimal weighted nearest neighbor (OWNN) classifier.

\section{Enhanced crowdsourcing classification}\label{sec:ENN}  
In this section, we propose an enhanced version of nearest neighbor classifier (ENN) and further prove that the ENN and its oracle counterpart share the same asymptotic regret, given that the weight in each worker is carefully chosen.

After a transformation of \eqref{eq:eta^j}, we have
\begin{equation}
    \eta^j(x) -\frac{a^j-b^j}{2}-1/2 =(a^j+b^j-1) (\eta^0(x)-1/2). \label{eq:bias_j}
\end{equation}
\eqref{eq:bias_j} suggests that the worker data and ground truth distribution can have different decision boundaries (set of $x$ with $\eta^0(x)$ or $\eta^j(x)=1/2$). This assumption is weaker than those in \citet{cai2019transfer} which requires the same decision boundaries. For example, for points on the ground truth decision boundary $\eta^0(x)=1/2$, we have $\eta^j(x) =\frac{a^j-b^j}{2}+1/2$. This means there exists a bias $\frac{a^j-b^j}{2}$ in the $j$-th worker data when her sensitivity and specificity are different. When $a^j>b^j$, with a higher probability, worker $j$ would label the instance to be class 1 than to class 0. In addition, the deviation from the decision boundary $\eta^0(x)
-1/2$, scaled by a multiplicative factor $a^j+b^j-1$, is always smaller than $1$ for non-expert data, suggesting that instance $x$ is more difficult to classify when $a^j=b^j$ since it is closer to the decision boundary. Therefore, it is necessary to enhance the labels for better performance. Illuminated by another transformation of \eqref{eq:eta^j}
\begin{align*}
    \tilde{\eta}^j(x):=\frac{\eta^j(x)+b^j-1}{a^j+b^j-1}=\eta^0(x),
\end{align*}  
we can derive the enhanced labels adjusted by the worker quality. We propose to enhance label according to (\ref{eq:enhencement}) in Algorithm~\ref{algo:ENN}, which removes the noise due to worker quality.

The main idea of ENN in Algorithm~\ref{algo:ENN} is straightforward: \\
(1) the enhanced labels are derived to take into account the noise in worker data;\\
(2) a local WNN regression estimator is obtained based on the data for each worker with enhanced labels;\\
(3) the final classifier is an outcome of the weighted voting over the $s$ local WNN predictions.

\begin{algorithm}
	\caption{Enhanced Nearest Neighbor with crowdsourced data (ENN)}
	\label{algo:ENN}
	\begin{algorithmic}[1]
	\renewcommand{\algorithmicrequire}{\textbf{Input:}}
	\renewcommand{\algorithmicensure}{\textbf{Output:}}
	\REQUIRE Crowdsourced data $\{{\cal D}_j\}_{j=1}^s$, weight vector $\bw_{j,i}$, worker sensitivity $a^j$ and specificity $b^j$, and query $x$.
	\ENSURE ENN.
	\FOR {$j = 1$ to $s$}
		\STATE Enhanced labels for the $j$-th worker data: \begin{equation}\label{eq:enhencement}
		    \tilde{Y}_{(i)}^{j}=\frac{Y_{(i)}^{j}+b^j-1}{a^j+b^j-1}.
		\end{equation}
		\STATE Local WNN estimator $\widehat{S}_{n_j,\bw_{n_{j}}}^{E}(x)={\textstyle\sum}_{i=1}^{n_{j}} w_{j,i}\tilde{Y}_{(i)}^{j}.$
	\ENDFOR
	\STATE Weighted voting of local WNN estimators
\begin{equation}
\label{eq:ENN}
\widehat{\phi}_{n_j,s,\bw_j}^{E}(x)=\indi{{\textstyle\sum}_{j=1}^s W_j\widehat{S}_{n_j,\bw_j}^{E}(x) \ge 1/2},
\end{equation}
where the worker weight $W_j=n_j/N$.
\STATE {\bfseries return:} $\widehat{\phi}_{n_j,s,\bw_{n_j}}^{E}(x)$.
\end{algorithmic} 
\end{algorithm}

\begin{remark}\label{rem:algo:ENN}
In Algorithm~\ref{algo:ENN}, if the worker quality is unknown, we can estimate it by Algorithm~\ref{algo:ENN2} or Algorithm~\ref{algo:ENN3} to be stated later. Note that $\widehat{S}_{n_j,\bw_{n_{j}}}^{E}(x)$ may be negative in a worker dataset with small size under an extreme marginal distribution of $X$. However, its negative value does not affect its contribution in \eqref{eq:ENN} for decision making.
\end{remark}

Our first main result, Theorem~\ref{thm:ENN_re},  gives an asymptotic expansion for the regret of ENN. Note that neither variance nor bias terms depends on worker quality $a^j$ and $b^j$.

\begin{theorem} 
\label{thm:ENN_re}
(Asymptotic Regret for ENN) Assume the same conditions as in Proposition~\ref{thm:WNN_re_E}. We have uniformly for $\bw_{n_j}\in W_{n_j,\beta}$, for $\beta\in (0,1/2)$, $a^j+b^j>1$, as $n_j \rightarrow \infty$,
\begin{align}
&{\rm Regret}(\widehat{\phi}_{n_j,s,\bw_{n_j}}^{E}) = \Big[B_1 \sum_{j=1}^s \Big(\frac{n_j}{N}\Big)^2 \sum_{i=1}^{n_j} w_{j,i}^2 + B_2\Big(\sum_{j=1}^s \frac{n_j}{N}\sum_{i=1}^{n_j} \frac{\alpha_i w_{j,i}}{n_j^{2/d}}\Big)^2 \Big]\{1+o(1)\}.\label{eq:ENN_re}
\end{align}

\end{theorem}

\begin{remark}\label{rem:thm:ENN_re}
$a^j+b^j>1$ means a worker gives label with more than $50\%$ correctness on average. Otherwise, we consider this worker as an adversary which should be dropped.
\end{remark}

In contrast with Proposition~\ref{thm:WNN_re_E}, the first term in the asymptotic regret of ENN in Theorem~\ref{thm:ENN_re} is reduced by a factor of $(n_j/N)^2$, while the squared bias term becomes the weighted average of bias from each worker data.

We know that the minimal asymptotic regret of the oracle $k$NN (`oracle' means the classifier is obtained from the expert data with size $N$; we use $K$ to denote the number of neighbors, emphasizing its global nature) is achieved when
\begin{equation*}
K=K^{*}:=\Big(\frac{dB_1}{4B_2} \Big)^{d/(d+4)}N^{4/(d+4)}
\end{equation*}\cite{S12}.
Consider a variant of ENN in which $k$NN is trained at each worker data, dubbed as ENN($k$). An intuitive choice for $k$, the number of local neighbors for each local $k$NN classifier, here is $\lceil (n_j/N)K^{*} \rceil$, so that globally about $K^{*}$ neighbors are used. Theorem \ref{thm:ENN_re} implies that the optimal local choice of $k_j$ in ENN($k$) (which gives rise to the same regret as the optimal oracle $k$NN) is indeed the above intuitive choice.

Given the weight vector, Theorem~\ref{thm:ENN_rr} affords an asymptotic regret comparison between the ENN and the oracle WNN, as implied by Proposition~\ref{thm:WNN_re_E} and Theorem~\ref{thm:ENN_re}. Theorem~\ref{thm:ENN_rr} says that given an oracle WNN which uses the expert data only, one can find an ENN with matching regret. It is encouraging that this can be done {\em without} incurring any regret loss, whether on the rate level or the multiplicative constant.

\begin{theorem}
\label{thm:ENN_rr}
(Asymptotic Regret Comparison between ENN and Oracle WNN) Assume the conditions in Theorem~\ref{thm:ENN_re}. Given an oracle WNN classifier with weights $\bw_N$ on an expert data with size N, denoted as $\widehat{\phi}_{N,\bw_N}^{0}(x)$, there exists an ENN classifier with weight $\bw_{n_j}$ on the crowdsourcing data, so that as $n_j \rightarrow \infty$,
\begin{eqnarray*}
\frac{{\rm Regret}(\widehat{\phi}_{n_j,s,\bw_{n_j}}^{E})}{{\rm Regret}(\widehat{\phi}_{N,\bw_N}^{0})} &\longrightarrow& 1,
\end{eqnarray*}
uniformly for $\bw_{n_j}\in W_{n_j,\beta}$ and $\bw_N\in W_{N,\beta}$ satisfying
\begin{eqnarray} 
\sum_{j=1}^s \Big(\frac{n_j}{N}\Big)^2 \sum_{i=1}^{n_j} w_{j,i}^2/\sum_{i=1}^N w_{Ni}^2 &\longrightarrow& 1\;\;  {\rm and}  \label{eq:ENN_rr_weight1}\\
 \sum_{j=1}^s \frac{n_j}{N}\sum_{i=1}^{n_j} \frac{\alpha_i w_{j,i}}{n_j^{2/d}}/\sum_{i=1}^N \frac{\alpha_i w_{Ni}}{N^{2/d}} &\longrightarrow& 1. \label{eq:ENN_rr_weight2}
\end{eqnarray}
\end{theorem} 

Theorem~\ref{thm:ENN_rr} says if the local weights for ENN are chosen to align with the oracle weights according to \eqref{eq:ENN_rr_weight1} and \eqref{eq:ENN_rr_weight2}, then ENN can achieve the same regret as the oracle WNN.

As an illustration, we show how to find the local weights by applying the results in Theorem~\ref{thm:ENN_rr} to the OWNN method, which is the best oracle WNN method due to \citet{S12}, whose global weights are defined as
\begin{equation}\label{ownn_weights_E}
w_{i}^*(N, m^*)=\left\{
                \begin{array}{ll}
                \frac{1}{m^*}\Big[1+\frac{d}{2}-\frac{d\alpha_i}{2(m^*)^{2/d}} \Big], \;{\rm if}\;\;i=1,\ldots,m^*,
                  \\
                  0, \;{\rm if}\;\;i=m^*+1,\ldots,N,
                \end{array}
              \right.
\end{equation}
where
\begin{eqnarray*}
m^* &=& \lceil \Big\{\frac{d(d+4)}{2(d+2)}\Big\}^{\frac{d}{d+4}}\Big(\frac{B_1}{B_2} \Big)^{\frac{d}{d+4}}N^{\frac{4}{d+4}} \rceil.
\end{eqnarray*} 
According to \eqref{eq:ENN_rr_weight1} and \eqref{eq:ENN_rr_weight2}, the local weights in the optimal ENN (that can achieve the same OWNN regret convergence rate $N^{-4/(d+4)}$) should be set as $w_{j,i}^*:=w_i^*(n_j, l_j^*)$, where 
\begin{equation}
l_j^*=\lceil(n_j/N)m^*\rceil.\label{eq:l_star_E}
\end{equation}
Interestingly, the above scaling factor is the same as that in the case of ENN($k$) discussed earlier. Corollary~\ref{thm:opt_ENN} summarizes the above findings, and further discovers, in (ii), the lower bound for the size of each worker data in ENN.

\begin{corollary}
\label{thm:opt_ENN}
(Optimal ENN) Suppose the conditions in Theorem~\ref{thm:ENN_re} hold.

\noindent(i) If $n_j/N^{d/(d+4)}\rightarrow\infty$, the asymptotic minimum regret of ENN is achieved by setting $w_{j,i}^*=w_{i}^*(n_j, l_j^*)$ with $l_j^*$ defined in \eqref{eq:l_star_E} and $w_i^*(\cdot,\cdot)$ as in \eqref{ownn_weights_E}. In addition, we have as $n_j\rightarrow\infty$,
$${{\rm Regret}(\widehat{\phi}_{n_j,s,\bw_{n_j}^*}^{E})}/{\rm Regret}(\widehat{\phi}_{N,\bw_N^*}^{0}) \rightarrow 1.$$

\noindent(ii) If $n_j=O(N^{d/(d+4)})$, then uniformly for $\bw_{n_j}\in W_{n_j,\beta}$,\\
$$\liminf_{n_j\rightarrow\infty} {{\rm Regret}(\widehat{\phi}_{n_j,s,\bw_{n_j}}^{E})}/{{\rm Regret}(\widehat{\phi}_{N,\bw_N^*}^{0})} \rightarrow \infty.$$
\end{corollary} 
The upper bound on $n_j$ in (ii) makes sense, as if the size of each worker data is too small, the bias and variance would be too large. In the special case that all $n_j$ are equal, we have a sharp bound that $n/N^{d/(d+4)}\rightarrow\infty$ (i.e., $s=o(N^{4/(d+4)})$). This result is the same as the one for W-DiNN in \citet{duan2020statistical}.

\section{Estimation of worker quality}\label{sec:quality}
We propose two methods to estimate worker quality, $a^j$ and $b^j$. One method requires access to a set of expert data, and is proven to achieve the same statistical guarantee as if $a^j$ and $b^j$ were known. The other method applies ENN to estimate the worker quality in an iterative manner, and it works well even without access to the expert data.

In Algorithm~\ref{algo:ENN2}, we estimate the worker quality by applying the $k$NN classifier on a set of expert data to relabel each worker data. The new labels are used as the substitutions for ground truth to estimate the worker quality.

\begin{algorithm}
	\caption{ENN2 with worker quality estimation (expert data required)}
	\label{algo:ENN2}
	\begin{algorithmic}[1]
	\renewcommand{\algorithmicrequire}{\textbf{Input:}}
	\renewcommand{\algorithmicensure}{\textbf{Output:}}
	\REQUIRE Crowdsourcing data $\{{\cal D}_j\}_{j=1}^s$ where ${\cal D}_s$ is an expert data, with $a^s=b^s=1$, local weight vector $\bw_{j,i}$.
	\ENSURE ENN2, estimated worker sensitivity $\widehat{a}^j$ and specificity $\widehat{b}^j$, for $j\in\{1,\dots,s\}$.
    \FOR {$j = 1$ to $s-1$}
		\STATE Derive predicted labels $\widehat{\phi}_{n_s,\tilde{k}}(X_i^j)$ for all $X_i^j$ in ${\cal D}_j$ using $k$NN ($\tilde{k}=n_s^{4/(d+4)}$) on the expert data ${\cal D}_s$. 
		\STATE Estimate the worker quality:
		\begin{eqnarray*}
		\widehat{a}^j &=&  \frac{\sum_{i=1}^{n_j}\indi{\widehat{\phi}_{n_s,\tilde{k}}(X_i^j)=1,Y_i^j=1}}{\sum_{i=1}^{n_j}\indi{\widehat{\phi}_{n_s,\tilde{k}}(X_i^j)=1}},  \mbox{ and }		\\ 
		\widehat{b}^j &=& \frac{\sum_{i=1}^{n_j}\indi{\widehat{\phi}_{n_s,\tilde{k}}(X_i^j)=0,Y_i^j=0}}{\sum_{i=1}^{n_j}\indi{\widehat{\phi}_{n_s,\tilde{k}}(X_i^j)=0}}.
		\end{eqnarray*}
	\ENDFOR
	\STATE Derive $\widehat{\phi}_{n_j,s,\bw_{n_j}}^{E2}(x)$ using Algorithm~\ref{algo:ENN} with $a^j=\widehat{a}^j$ and $b^j=\widehat{b}^j$ ($j\in\{1,\dots,s\}$) where $\widehat{a}^s=\widehat{b}^s=1$.
	\STATE {\bfseries return:} $\widehat{\phi}_{n_j,s,\bw_{n_j}}^{E2}(x)$; $\widehat{a}^j$ and $\widehat{b}^j$, for $j\in\{1,\dots,s\}$.
\end{algorithmic} 
\end{algorithm}

Plug the estimated worker quality $\widehat{a}^j$ and $\widehat{b}^j$ from Algorithm~\ref{algo:ENN2} to Algorithm~\ref{algo:ENN}, we obtain the ENN with estimated worker quality (ENN2) which has a similar statistical guarantee as ENN. Theorem~\ref{thm:ENN2_re} gives an asymptotic expansion formula for the regret of the ENN classifier given weight vector $\bw_{n_j}$ based on estimated  $\widehat{a}^j$ and $\widehat{b}^j$ from Algorithm~\ref{algo:ENN2}. Specifically, when the size of expert data has a higher order than each worker data, ENN2 can achieve the same asymptotical regret as ENN as in Theorem~\ref{thm:ENN_re} when the worker quality was known.

\begin{theorem} 
\label{thm:ENN2_re}
(Asymptotic Regret for ENN with estimated worker quality) Assuming the same conditions as in Theorem~\ref{thm:ENN_re}, $n_j/n_s=o(1)$ for $j\in\{1,\dots,s-1\}$, we have for each $\beta\in (0,1/2)$, $a^j+b^j>1$, as $n_j \rightarrow \infty$,
\begin{align*}
&{\rm Regret}(\widehat{\phi}_{n_j,s,\bw_{n_j}}^{E2}) = \Big[B_1 \sum_{j=1}^s \Big(\frac{n_j}{N}\Big)^2 \sum_{i=1}^{n_j} w_{j,i}^2 + B_2\Big(\sum_{j=1}^s \frac{n_j}{N}\sum_{i=1}^{n_j} \frac{\alpha_i w_{j,i}}{n_j^{2/d}}\Big)^2 \Big]\{1+o(1)\},\nonumber
\end{align*}
uniformly for $\bw_{n_j}\in W_{n_j,\beta}$.
\end{theorem}

\begin{remark}\label{rem:thm:ENN2_re}
In Theorem~\ref{thm:ENN2_re}, the assumption $n_j/n_s=o(1)$ for $j\in\{1,\dots,s-1\}$ is used to bound the order of remainder terms due to worker quality estimation.
\end{remark}

\begin{algorithm}[!htb]
	\caption{ENN3 with worker quality estimation (expert data not required)}
	\label{algo:ENN3}
	\begin{algorithmic}[1]
	\renewcommand{\algorithmicrequire}{\textbf{Input:}}
	\renewcommand{\algorithmicensure}{\textbf{Output:}}
	\REQUIRE Crowdsourcing data $\{{\cal D}_j\}_{j=1}^s$, local weight vector $\bw_{j,i}$, and stop criteria $c$.
	\ENSURE ENN3; estimated worker sensitivity $\widehat{a}^j$ and specificity $\widehat{b}^j$, for $j\in\{1,\dots,s\}$.
	\STATE Initialization: $a^j=b^j=1$, for $j\in\{1,\dots,s\}$.
	\FOR {$l = 1,2,\dots$}
    	\FOR {$j = 1$ to $s$}
    		\STATE Derive predicted labels $\widehat{\phi}_{n_j,s,\bw_{n_j}}^{E}(X_i^j)$ for all $X_i^j$ in ${\cal D}_j$ using ENN in Algorithm~\ref{algo:ENN}.
    		\STATE Estimate worker quality:
    		\begin{eqnarray*}
    		\widehat{a}^j &=&  \frac{\sum_{i=1}^{n_j}\indi{\widehat{\phi}_{n_j,s,\bw_{n_j}}^{E}(X_i^j)=1,Y_i^j=1}}{\sum_{i=1}^{n_j}\indi{\widehat{\phi}_{n_j,s,\bw_{n_j}}^{E}(X_i^j)=1}}, 	\\ 
    		\widehat{b}^j &=& \frac{\sum_{i=1}^{n_j}\indi{\widehat{\phi}_{n_j,s,\bw_{n_j}}^{E}(X_i^j)=0,Y_i^j=0}}{\sum_{i=1}^{n_j}\indi{\widehat{\phi}_{n_j,s,\bw_{n_j}}^{E}(X_i^j)=0}}.
    		\end{eqnarray*}
    	\ENDFOR
    	\STATE Compute $\Delta=\frac{1}{2s}{\textstyle\sum}_{j=1}^s (|\widehat{a}^j-a^j|+|\widehat{b}^j-b^j|).$
    	\STATE Update: $a^j=\widehat{a}^j$, $b^j=\widehat{b}^j$ for $j\in\{1,\dots,s\}$.
    	\IF {$\Delta \le c$} 
    	    \STATE {Break.}
    	\ENDIF
    \ENDFOR
    \STATE Derive $\widehat{\phi}_{n_j,s,\bw_{n_j}}^{E3}(x)$ using Algorithm~\ref{algo:ENN} with $a^j=\widehat{a}^j$ and $b^j=\widehat{b}^j$, for $j\in\{1,\dots,s\}$.
	\STATE {\bfseries return:} $\widehat{\phi}_{n_j,s,\bw_{n_j}}^{E3}(x)$, $\widehat{a}^j$ and $\widehat{b}^j$, for $j\in\{1,\dots,s\}$.
\end{algorithmic} 
\end{algorithm}

Algorithm~\ref{algo:ENN2} requires the size of expert data has a higher order than other worker data. However, this generally does not hold as expert data does not exist or has relatively smaller size in practice. Therefore, we propose a more practical algorithm that does not need expert data to estimate the worker quality. Specifically, we apply Algorithm~\ref{algo:ENN} to derive predicted labels for each worker to substitute the ground truth labels and update the worker quality iteratively. The main idea of this estimation procedure (summarized in Algorithm~\ref{algo:ENN3}) is straightforward:\\
(1) initialize all $a^j$ and $b^j$ with $1$;\\
(2) derive predicted labels for each worker data by applying ENN on the crowdsourcing data;\\
(3) update $a^j$ and $b^j$ by comparing observed labels and predicted labels for the $j$-th worker;\\
(4) iterate until convergence.

There are several advantages to this procedure. We do not need expert data to estimate worker quality, unlike most previous methods. It also converges quickly in practice if we choose a suitable stop criterion, such as $2\%$.

\section{Numerical studies}\label{sec:exp_E}
In this section, we check the accuracy of the ENN methods using simulations and real examples. All experiments are conducted in R environment on HPC clusters with two 12-core Intel Xeon Gold Skylake processors and two 10-core Xeon-E5 processors, with memory between 96 and 128 GB.

\subsection{Simulations}\label{sec:sim_E}
In the simulated studies, we compare ENN methods with naive $k$NN, oracle $k$NN, and oracle OWNN from different aspects. Here, naive kNN denotes kNN classifiers on the original crowdsourcing data directly, and oracle kNN denotes classifiers run on the expert data with size $N$.  In comparing ENN($k$) ($k$NN is trained at each worker data) with the oracle $k$NN, we aim to verify the main results in Theorem~\ref{thm:ENN_rr}, namely, the ENN can attain the same performance as the oracle method. In comparing the ENN methods with optimal local weights and the oracle OWNN method, we aim to verify the sharpness of upper bound on the number of workers in Corollary~\ref{thm:opt_ENN}. It is verified by showing that the difference in performance between the ENN methods and the oracle OWNN deviates when the theoretical upper bound is exceeded.

Three settings are considered for ground truth distribution. Simulation 1 allows a relatively easy classification task, Simulation 2 examines the bimodal effect, and Simulation 3 combines bimodality with dependence between variables. In Simulation 1, $N=\sum_{j=1}^5 n_j=20000$ and $d=4,6,8$. As ground truth distribution, the two classes are generated as $P_1^0\sim N(0_d,\mathbb{I}_d)$ and $P_0^0\sim N(\frac{2}{\sqrt{d}}1_d,\mathbb{I}_d)$ with the class probability $\pi_1^0={\mathbb P}(Y=1)=1/3$. The worker data are generated with \eqref{eq:eta^j} with different settings of quality and sizes in Table~\ref{tab:quality_setup}. Simulation 2 has the same setting as Simulation 1, except both classes are bimodal with $P_1^0\sim 0.5N(0_d,\mathbb{I}_d)+0.5 N(3_d,2\mathbb{I}_d)$ and $P_0^0\sim 0.5N(1.5_d,\mathbb{I}_d)+0.5 N(4.5_d,2\mathbb{I}_d)$. Simulation 3 has the same setting as Simulation 2, except $P_1^0\sim 0.5N(0_d,\Sigma)+0.5 N(3_d,2\Sigma)$ and $P_0^0\sim 0.5N(1.5_d,\Sigma)+0.5 N(4.5_d,2\Sigma)$ with $\pi_1^0=1/2$, and $\Sigma$ the Toeplitz matrix whose $(1,j)$th entry is $0.6^{j-1}$.

\begin{table*}[htb!] 
	\vspace{-0.5em}	
	\caption{Quality and size setups for worker data.}  \label{tab:quality_setup} 
	\vspace{+0.1em}	
	\setlength\tabcolsep{2.1pt} 
	\scriptsize
	\centering
	\begin{tabular}{c|ccccc|ccccc|ccccc|l}	
	\hline
	setup &$a_1$&$a_2$&$a_3$&$a_4$&$a_5$&$b_1$&$b_2$&$b_3$&$b_4$&$b_5$&$n_1$&$n_2$&$n_3$&$n_4$&$n_5$ & remarks on setups\\ 
	\hline
 1 & 0.90 & 0.90 & 0.95 & 0.90 & 1.00 & 0.80 & 0.80 & 0.85 & 0.85 & 0.90 & 2000 & 3000 & 4000 & 5000 & 6000 & no expert, $a>b$, high quality\\
 2 & 0.80 & 0.80 & 0.85 & 0.80 & 0.80 & 0.90 & 0.95 & 0.95 & 0.90 & 1.00 & 2000 & 3000 & 4000 & 5000 & 6000 & no expert, $a<b$, high quality\\
 3 & 0.60 & 0.65 & 0.85 & 0.80 & 0.80 & 0.75 & 0.75 & 0.95 & 0.90 & 0.95 & 2000 & 3000 & 4000 & 5000 & 6000 & no expert, some lower quality\\
 4 & 0.80 & 0.85 & 0.85 & 0.90 & 0.80 & 0.90 & 0.80 & 0.95 & 0.80 & 0.95 & 2000 & 3000 & 4000 & 5000 & 6000 & no expert, $a\neq b$\\
 5 & 0.80 & 0.85 & 0.95 & 0.85 & 0.90 & 0.80 & 0.85 & 0.95 & 0.85 & 0.90 & 2000 & 3000 & 4000 & 5000 & 6000 & no expert, $a=b$ \\
 \hline
   6 & 0.90 & 0.90 & 0.95 & 0.90 & 1.00 & 0.80 & 0.80 & 0.85 & 0.85 & 1.00 & 1000 & 2000 & 3000 & 4000 & 10000 & expert$(j=5)$, $a\ge b$, high quality\\
  7 & 0.80 & 0.80 & 0.85 & 0.80 & 1.00 & 0.90 & 0.95 & 0.95 & 0.90 & 1.00 & 1000 & 2000 & 3000 & 4000 & 10000 & expert$(j=5)$, $a\le b$, high quality\\
  8 & 0.60 & 0.65 & 0.85 & 0.80 & 1.00 & 0.75 & 0.75 & 0.95 & 0.90 & 1.00 & 1000 & 2000 & 3000 & 4000 & 10000 &expert$(j=5)$, some lower quality\\
  9 & 0.80 & 0.85 & 0.85 & 0.90 & 1.00 & 0.90 & 0.80 & 0.95 & 0.80 & 1.00 & 1000 & 2000 & 3000 & 4000 & 10000 &expert$(j=5)$, $a\neq b$\\ 
 10 & 0.80 & 0.85 & 0.95 & 0.85 & 1.00 & 0.80 & 0.85 & 0.95 & 0.85 & 1.00 & 1000 & 2000 & 3000 & 4000 & 10000 & expert$(j=5)$, $a=b$ \\
 	\hline
	\end{tabular}
	\vspace{-0.6em}
\end{table*}

When comparing the $k$NN methods, the number of neighbors $K$ in the oracle $k$NN is chosen as $K=N^{0.7}$. The number of local neighbors in ENN($k$) is chosen as $k_j=\lceil(n_j/N)K\rceil$ as suggested by Theorem~\ref{thm:ENN_rr}. These $k$ values are truncated at 1, since we cannot have a fraction of an observation. In comparing with the oracle OWNN method, the $m^{*}$ parameter in OWNN is tuned using cross-validation. The parameter $l_j$ in ENN for each worker data is chosen as $l_j^*= \lceil (n_j/N)m^*\rceil$ as stated in Corollary~\ref{thm:opt_ENN}. The test set is independently generated with $1000$ observations under the ground truth distribution for both comparisons. We repeat the simulation for $1000$ times for each quality setup and $d$.

We compare our proposed ENN methods with two benchmark NN classifiers (naive $k$NN and oracle $k$NN) in the first part of simulations.  We consider ten groups of worker quality setups. Table~\ref{tab:quality_setup} illustrates the setup of worker data with the remarks commenting on the purpose of each setup. There is no expert data in setups 1-5, while the worker $5$ is an expert in setups 6-10. All methods will run on each setup except that ENN2 is not applicable on setups 1-5.

\begin{figure*}[bt!]
	\centering\vspace{-0.5em}
	\includegraphics[width=0.28\textwidth,height=0.25\textwidth]{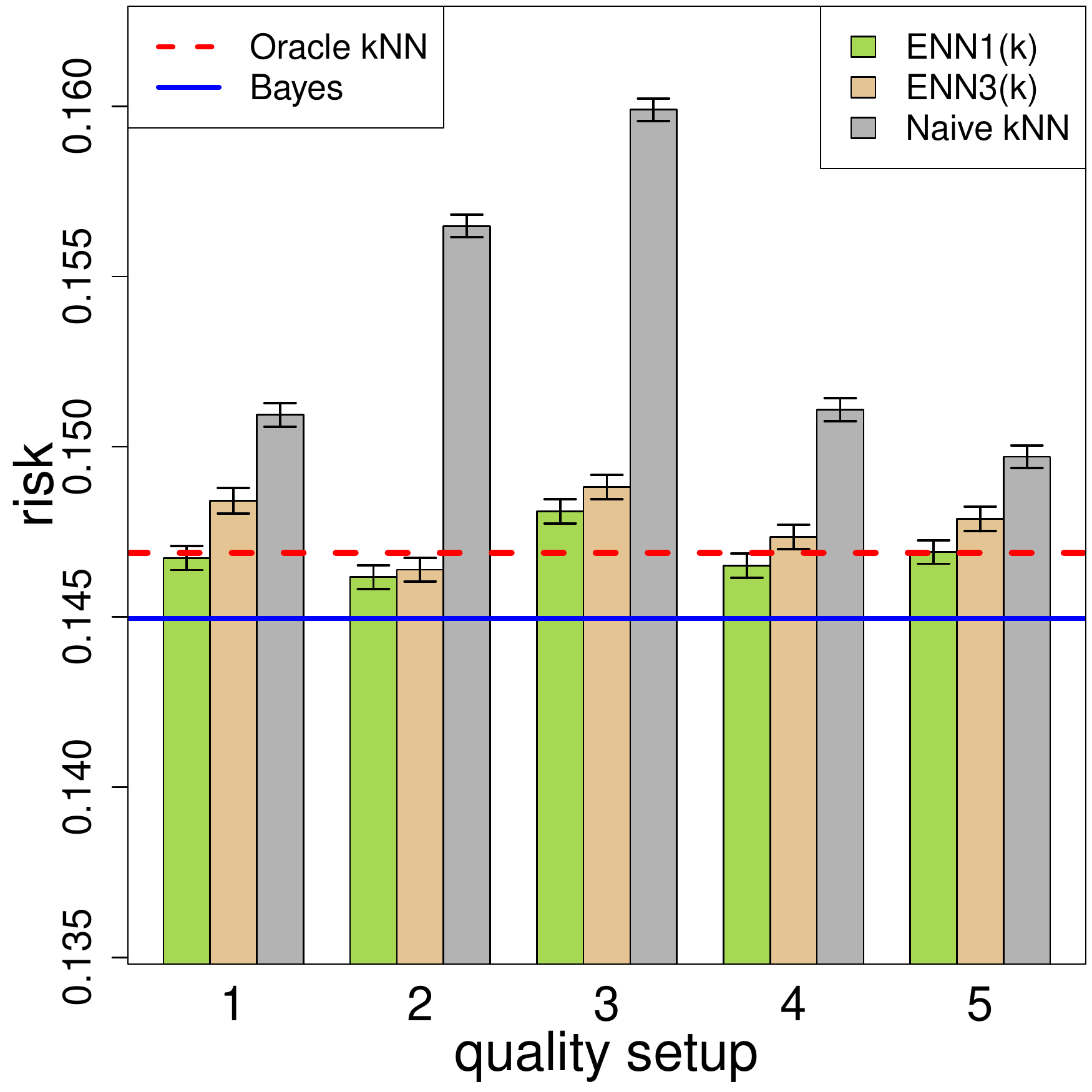}
    \includegraphics[width=0.28\textwidth,height=0.25\textwidth]{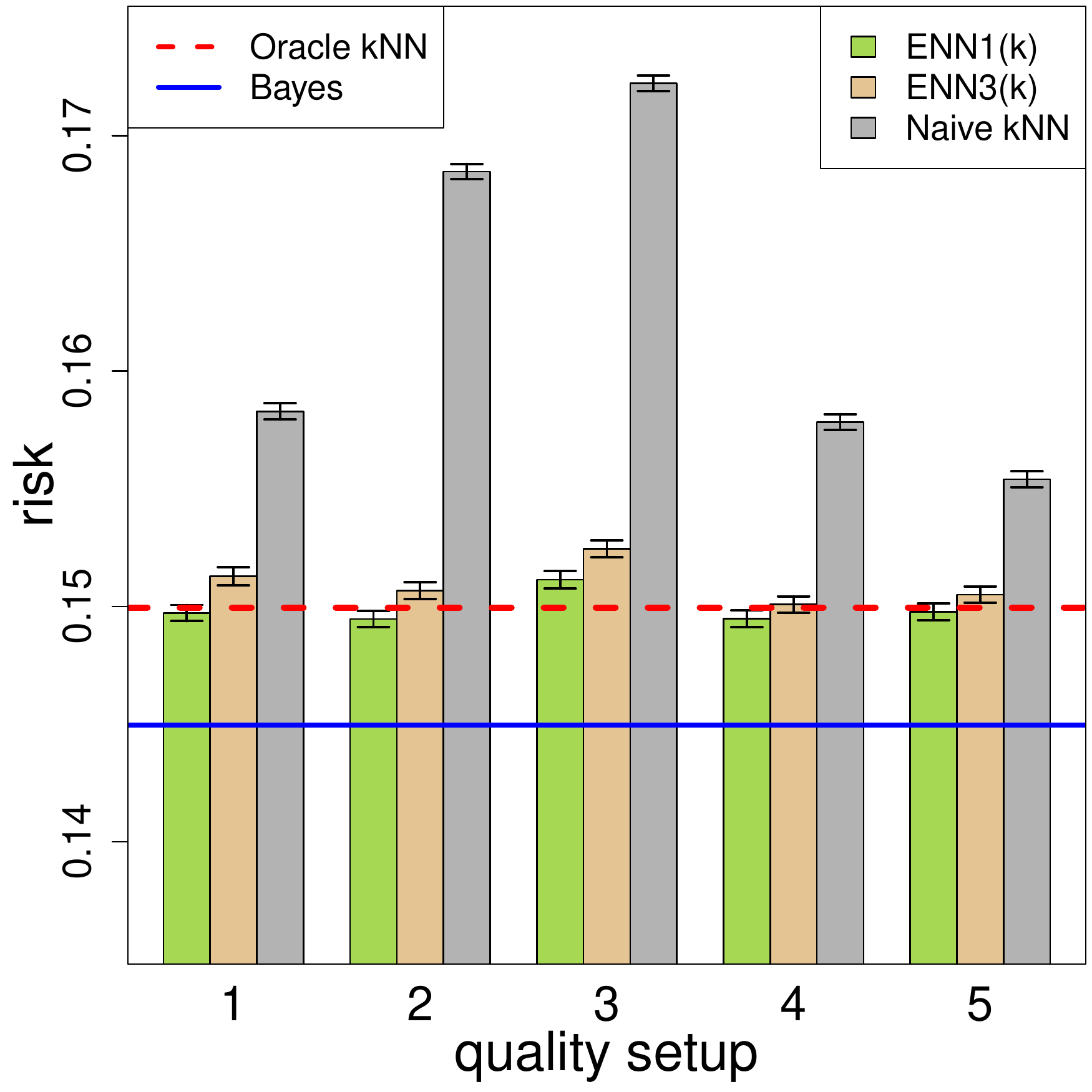}
    \includegraphics[width=0.28\textwidth,height=0.25\textwidth]{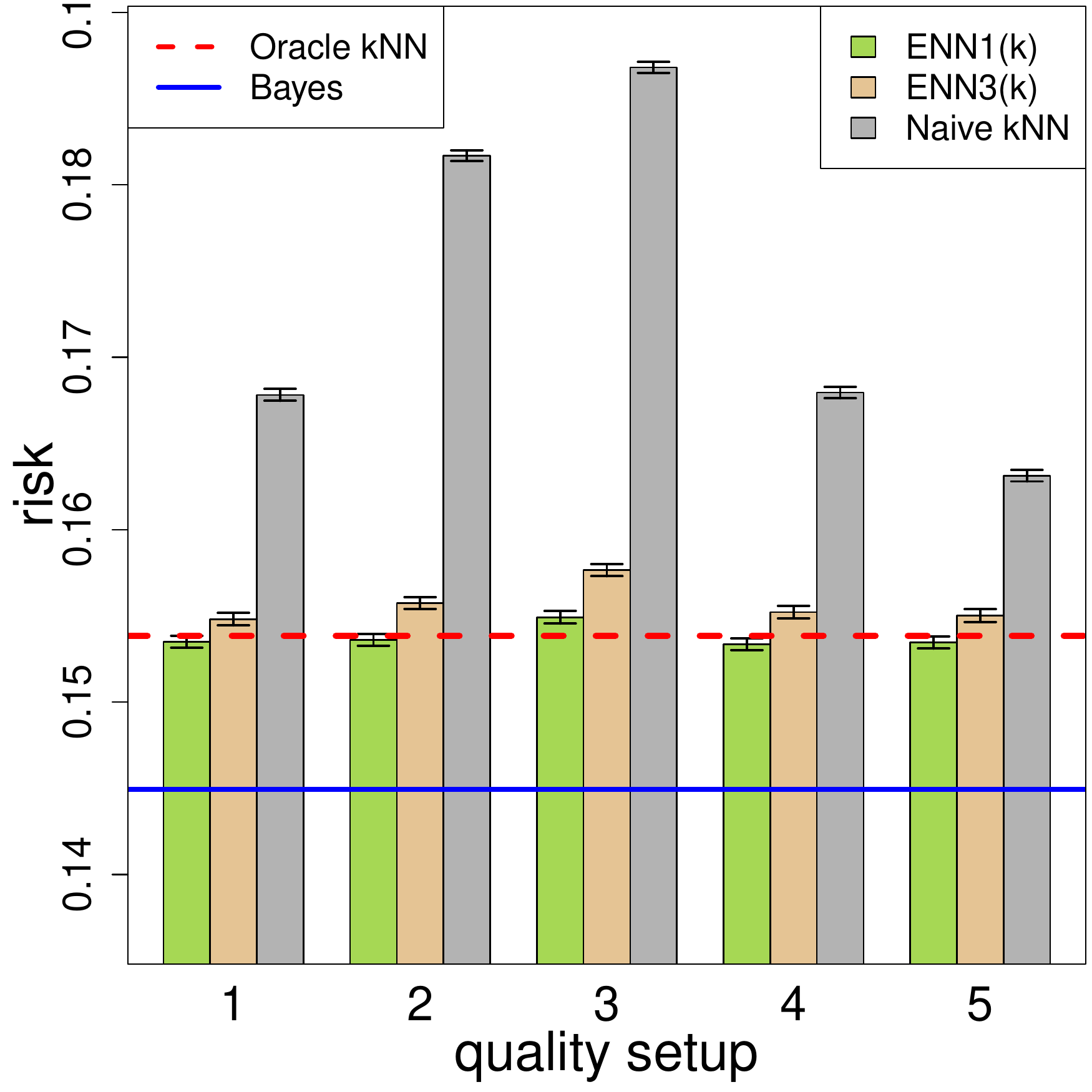}
    \includegraphics[width=0.28\textwidth,height=0.25\textwidth]{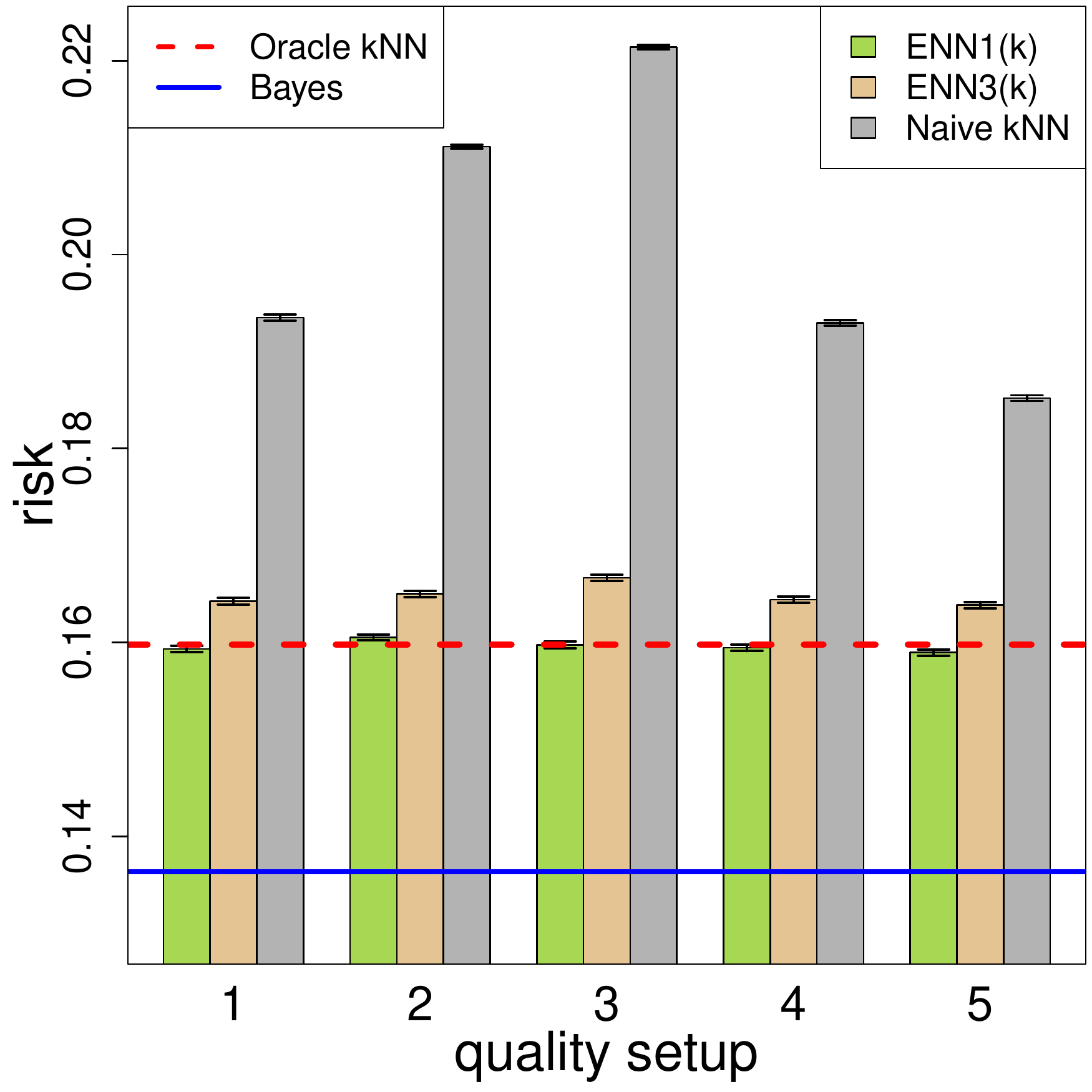}
    \includegraphics[width=0.28\textwidth,height=0.25\textwidth]{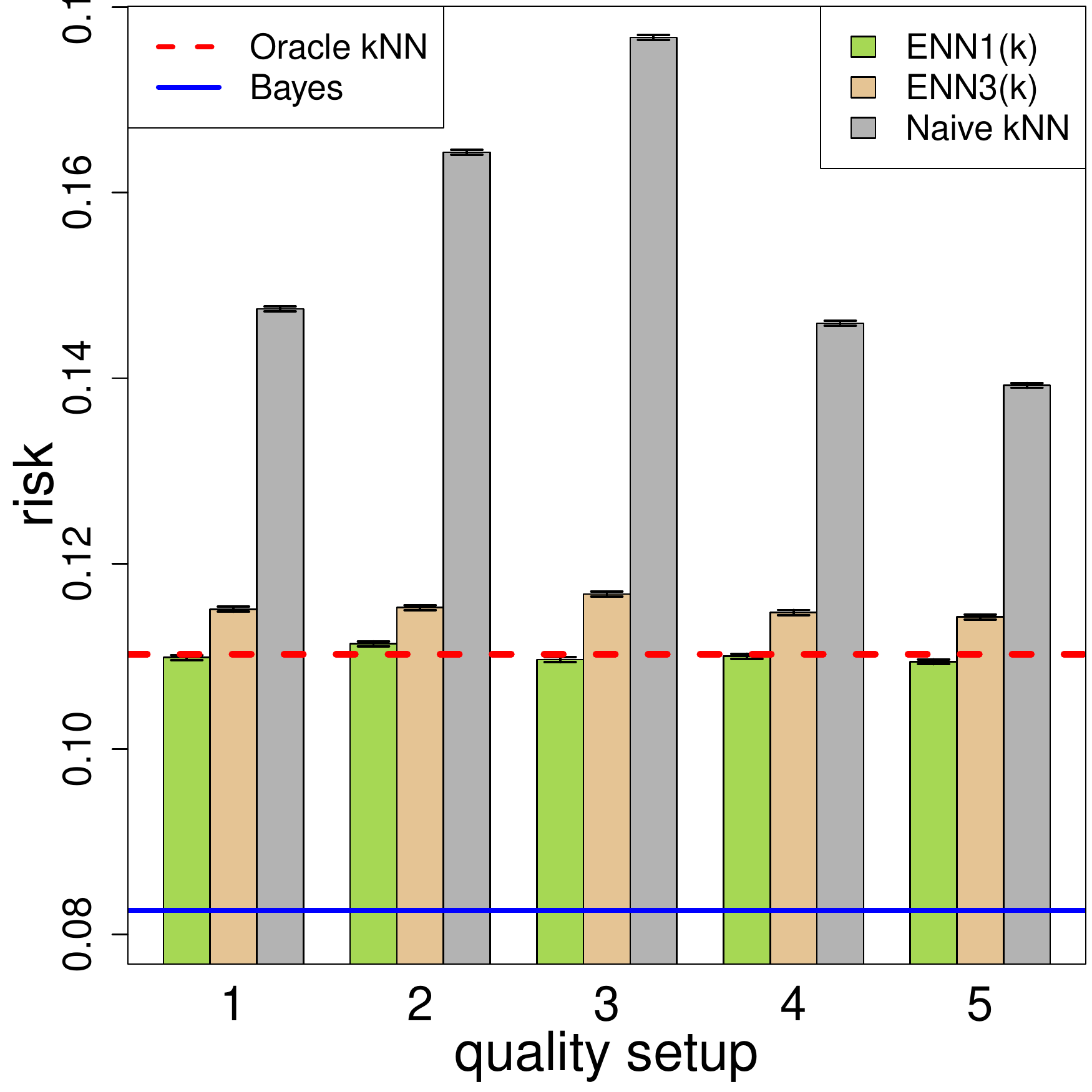}
    \includegraphics[width=0.28\textwidth,height=0.25\textwidth]{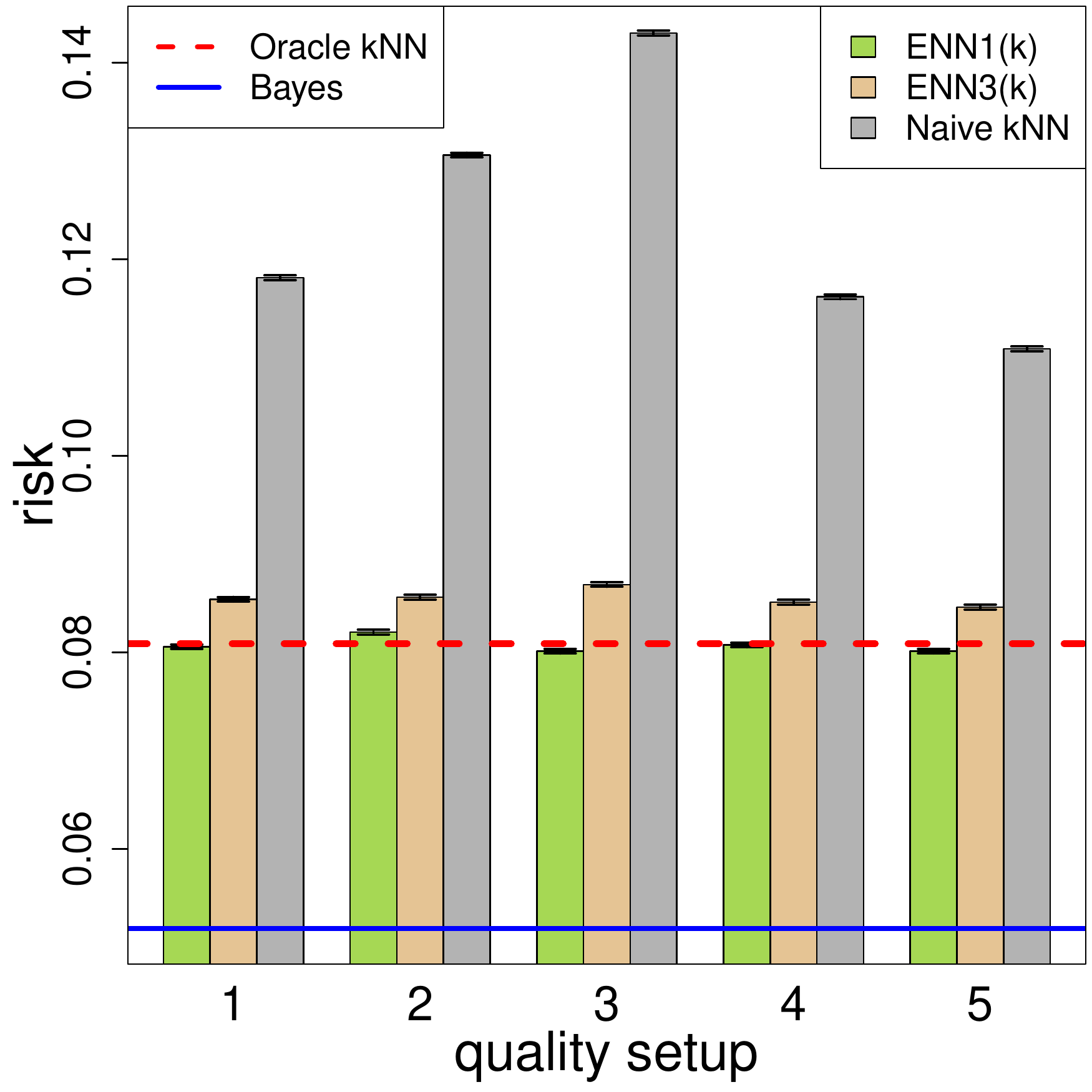}
    \includegraphics[width=0.28\textwidth,height=0.25\textwidth]{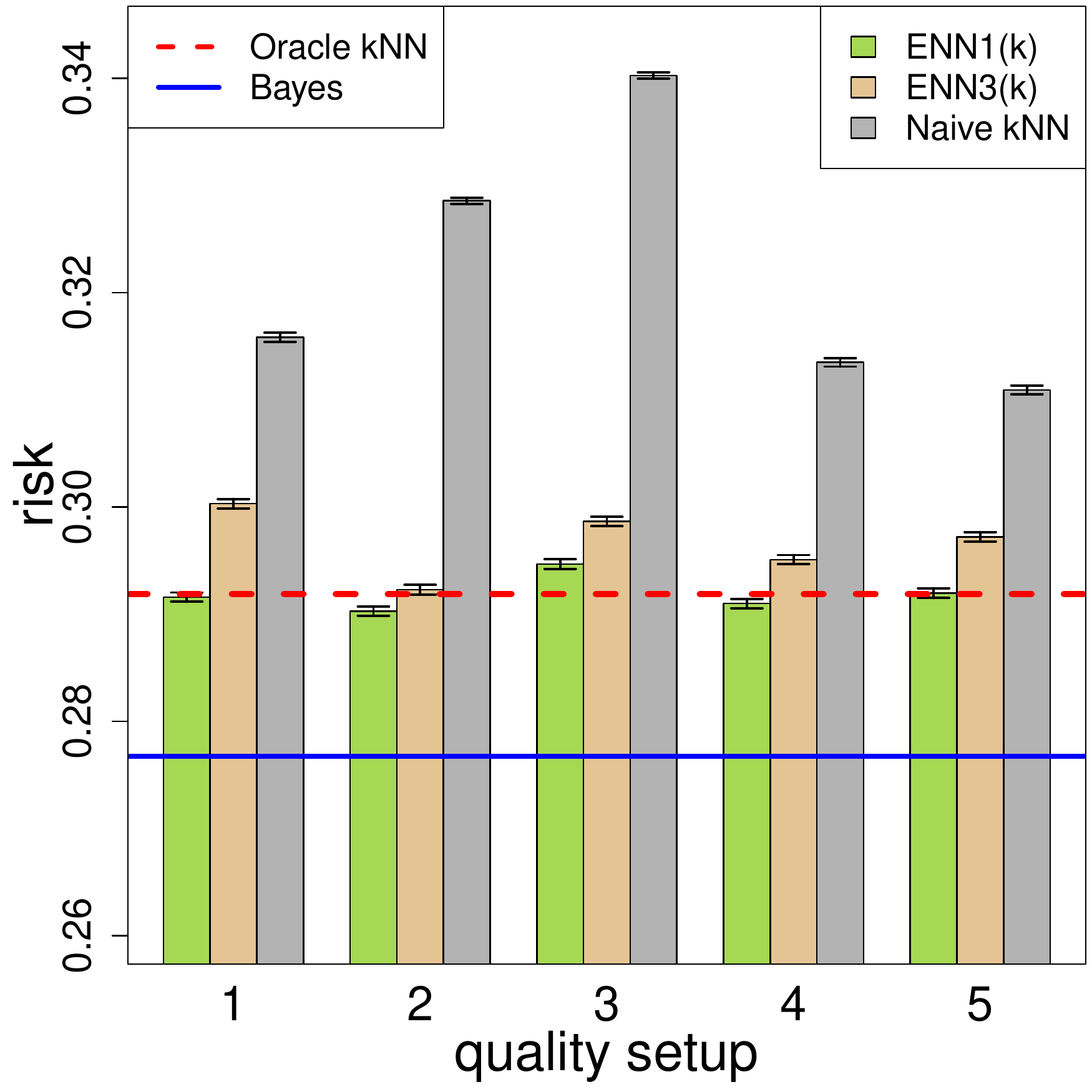}
    \includegraphics[width=0.28\textwidth,height=0.25\textwidth]{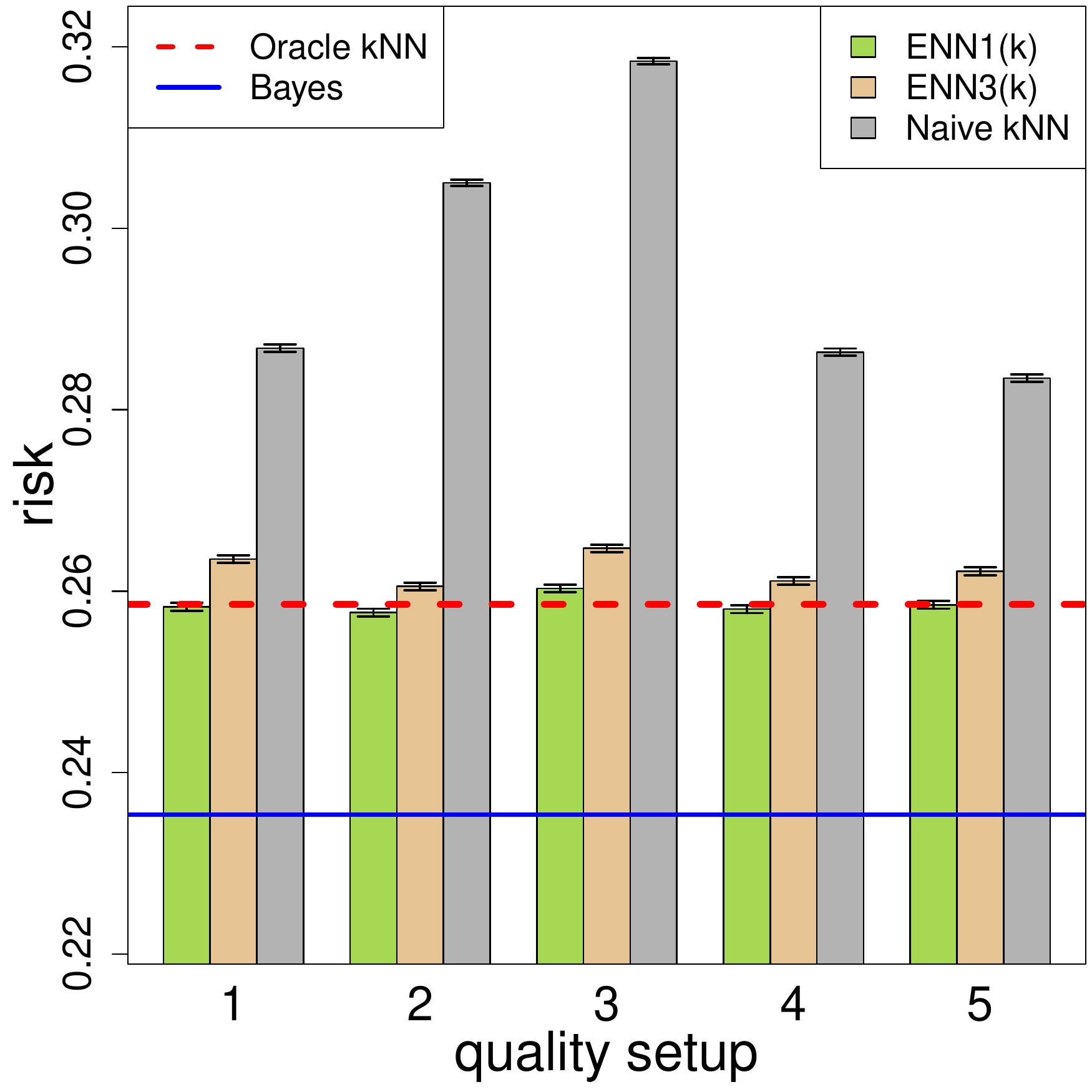}
    \includegraphics[width=0.28\textwidth,height=0.25\textwidth]{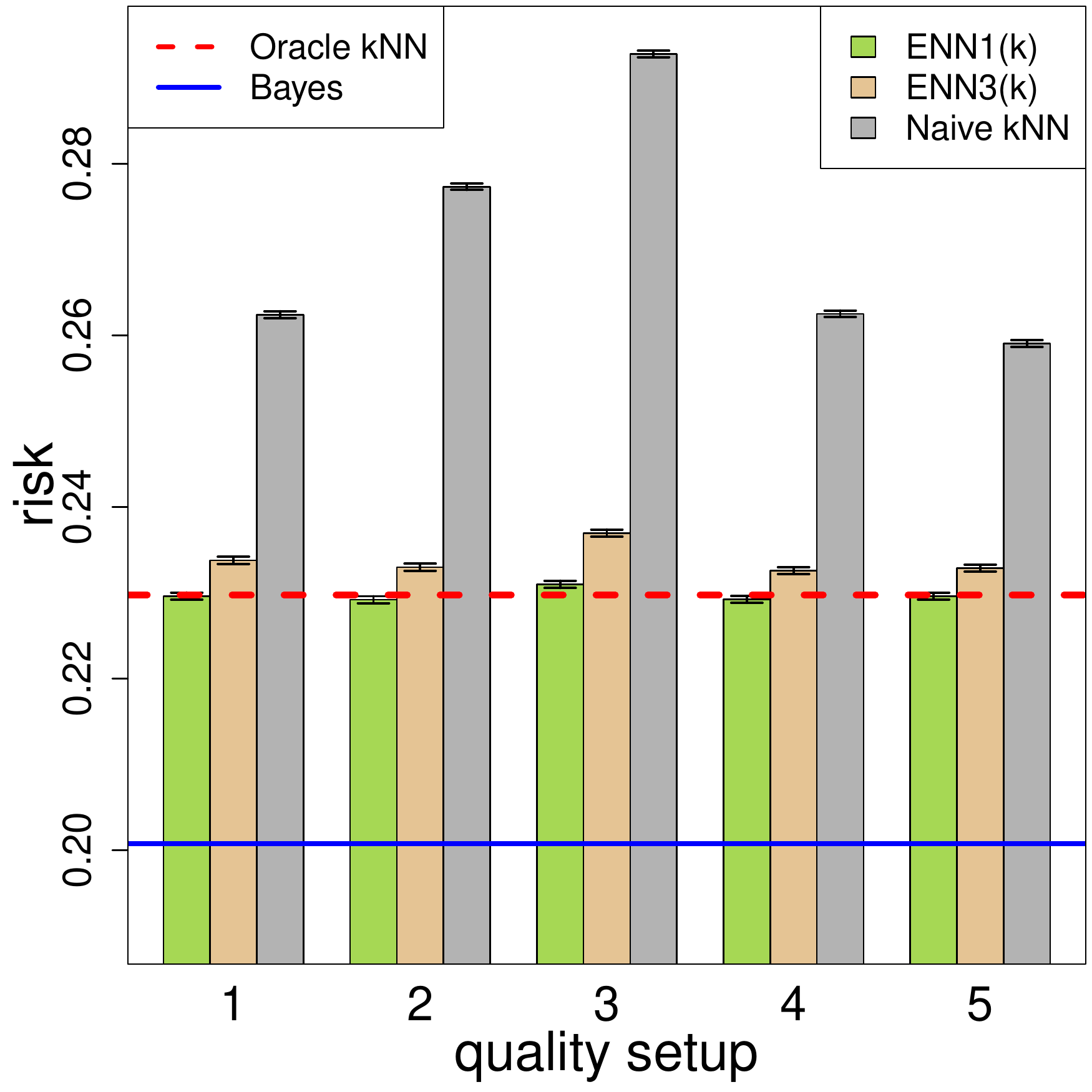}
\vspace{-1em}	
\caption{Risk (with standard error bar marked) of all methods (except ENN2)  and the Bayes rule without expert data. The x-axis indicates different settings with worker quality.  Top/middle/bottom: Simulation $1/2/3$; left/middle/right: $d=4/6/8$.} \label{fig:sim_risk_quality_no_expert}
\vspace{-0.5em}
\end{figure*}

The comparison between the risks of the four methods (two $k$NN and two ENN) on crowdsourcing data with no expert (setups 1-5) is reported in Figure~\ref{fig:sim_risk_quality_no_expert}. For all quality setups, the risk is similar between ENN1($k$) and the oracle kNN, while ENN3($k$) has a small gap with both. Naive kNN has a significantly larger risk as the original worker data contains some noise due to low worker quality. These verify the main results in Section~\ref{sec:ENN} and Section~\ref{sec:quality}. Therefore, ENN1 and ENN3 can achieve a similar performance as if the entire training data are labelled by an expert. Similar conclusions can be made for setups 6-10, shown in 
Figure~\ref{fig:sim_risk_quality_expert}. Moreover, if there exists expert data with a large size, ENN2($k$) performs well even with the worker quality unobserved.

\begin{figure*}[hbt!]
	\centering\vspace{-0.5em}
    \includegraphics[width=0.28\textwidth,height=0.25\textwidth]{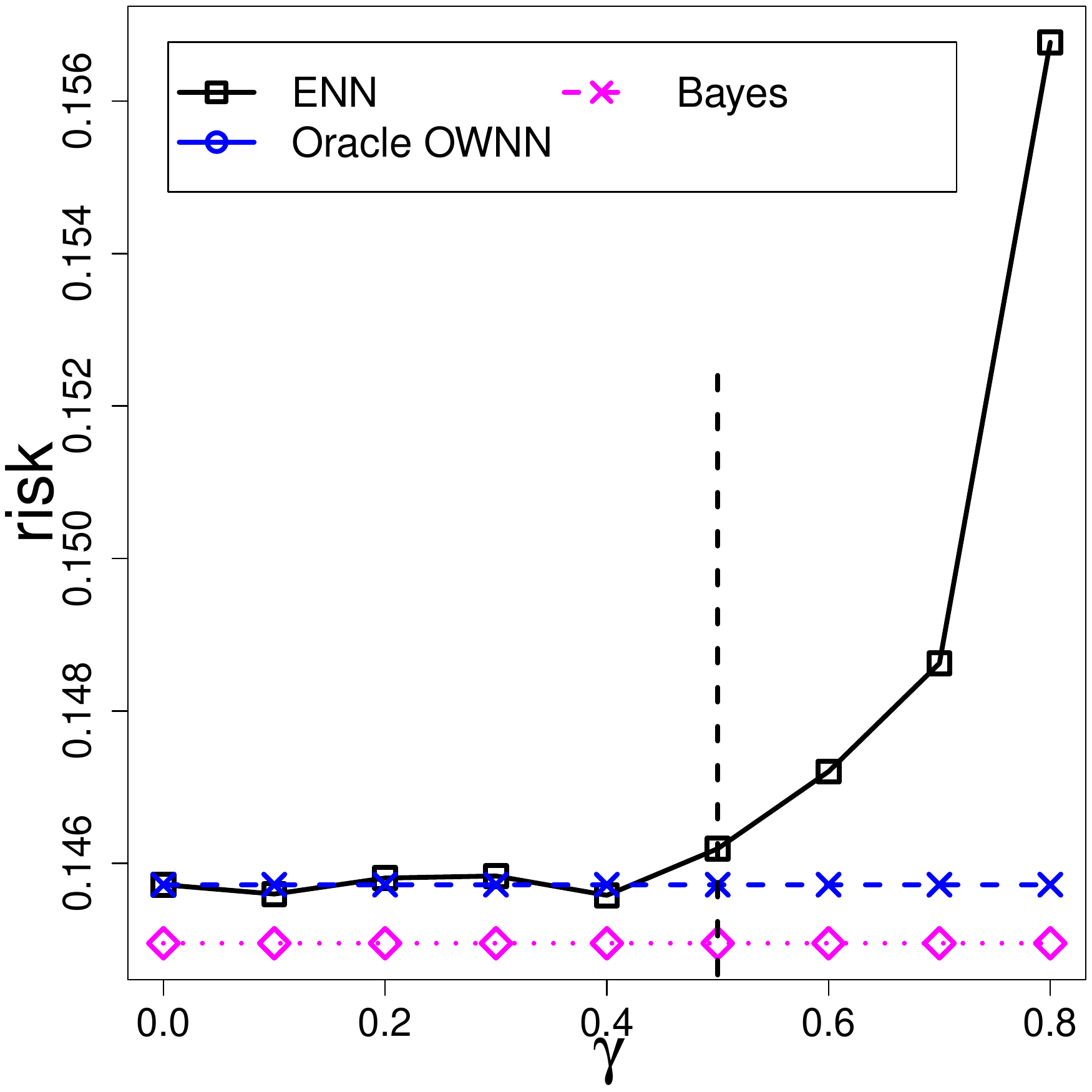}
    \includegraphics[width=0.28\textwidth,height=0.25\textwidth]{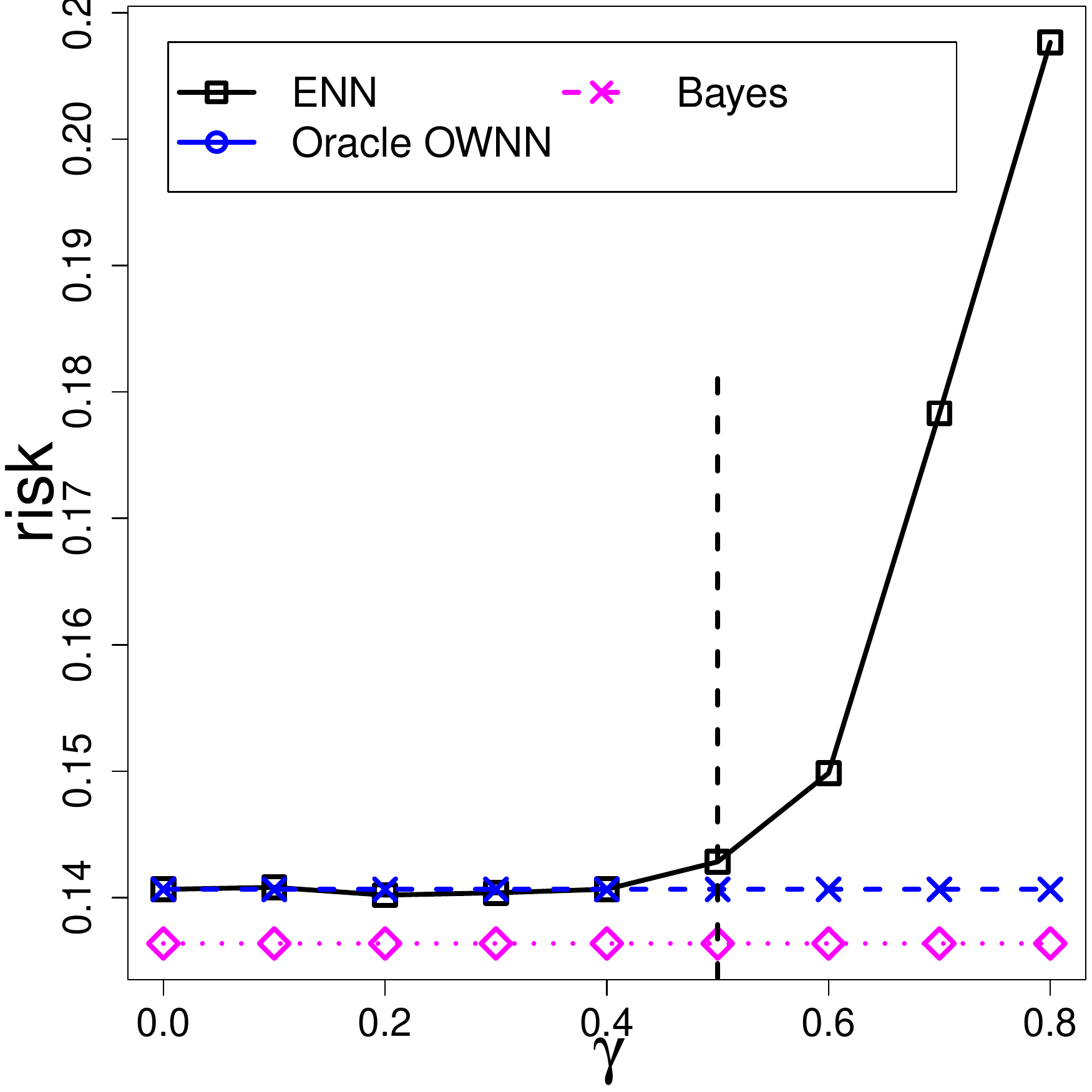}
    \includegraphics[width=0.28\textwidth,height=0.25\textwidth]{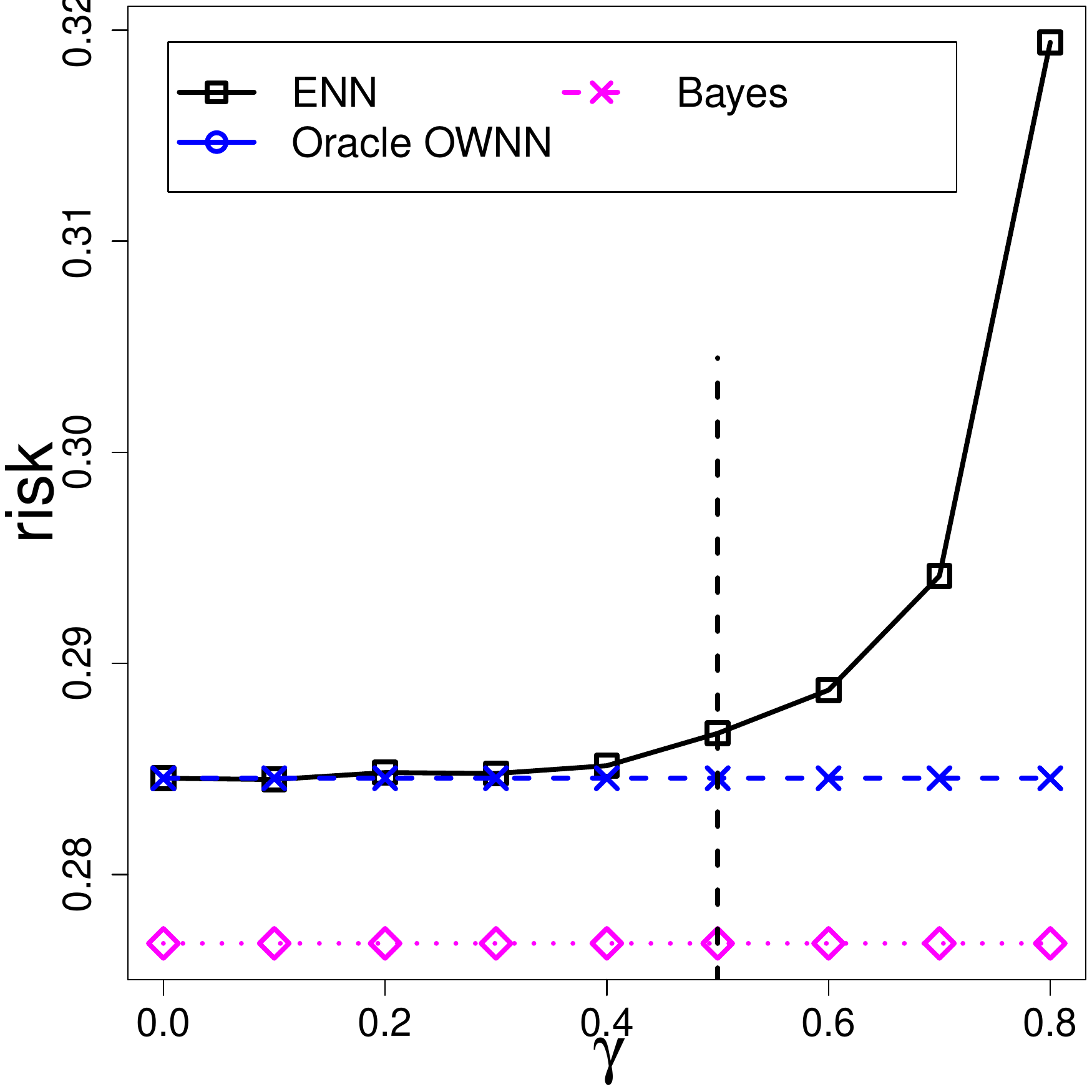}
\vspace{-1em}	
\caption{Risk of optimal ENN, oracle OWNN and the Bayes rule for different $\gamma$. Left/middle/right: Simulation $1/2/3$, $d=4$. Upper bound for number of worker data in optimal ENN ($\gamma=4/(d+4)=1/2$) is shown as a vertical line.} \label{fig:sim_opt_risk_gamma_E}
\vspace{-0.5em}
\end{figure*}

Table~\ref{tab:quality_results_ENN2} and Table~\ref{tab:quality_results_ENN3} show good estimation accuracy of the worker quality based on ENN2 and ENN3, respectively.

On the other hand, we apply a special worker data setup (five expert data with equal size $4000$) to verify the sharp upper bound for the number of worker data. Under this setup, the upper bound simplifies to $\gamma=4/(d+4)$ ($s=N^{\gamma})$. Since the comparison with the oracle OWNN is meant to verify the sharp upper bound for $\gamma$ in the optimal weight setting (Corollary~\ref{thm:opt_ENN}), we carefully tune the weights in the oracle OWNN method in order to reach the optimality. Figure~\ref{fig:sim_opt_risk_gamma_E} shows the comparison of risks for ENN and oracle OWNN methods. Our focus here is when the ENN method starts to have significantly worse performance than the oracle OWNN, and the answers lie in the upper bounds in Corollary~\ref{thm:opt_ENN}. For simplicity, we set $d=4$, which leads to an upper bound of $4/(d+4)=0.5$ for the ENN method. This upper bound is shown as vertical lines in Figure~\ref{fig:sim_opt_risk_gamma_E}. Specifically, the ENN has almost the same performance as the OWNN method for $\gamma\le 0.4$. However, ENN does not perform well enough for $\gamma\ge 4/(d+4)=0.5$ when compared to OWNN. These verify the results in Corollary~\ref{thm:opt_ENN}.

\begin{figure*}[hbt!] 
	\centering\vspace{-0.5em}
	\includegraphics[width=0.28\textwidth,height=0.25\textwidth]{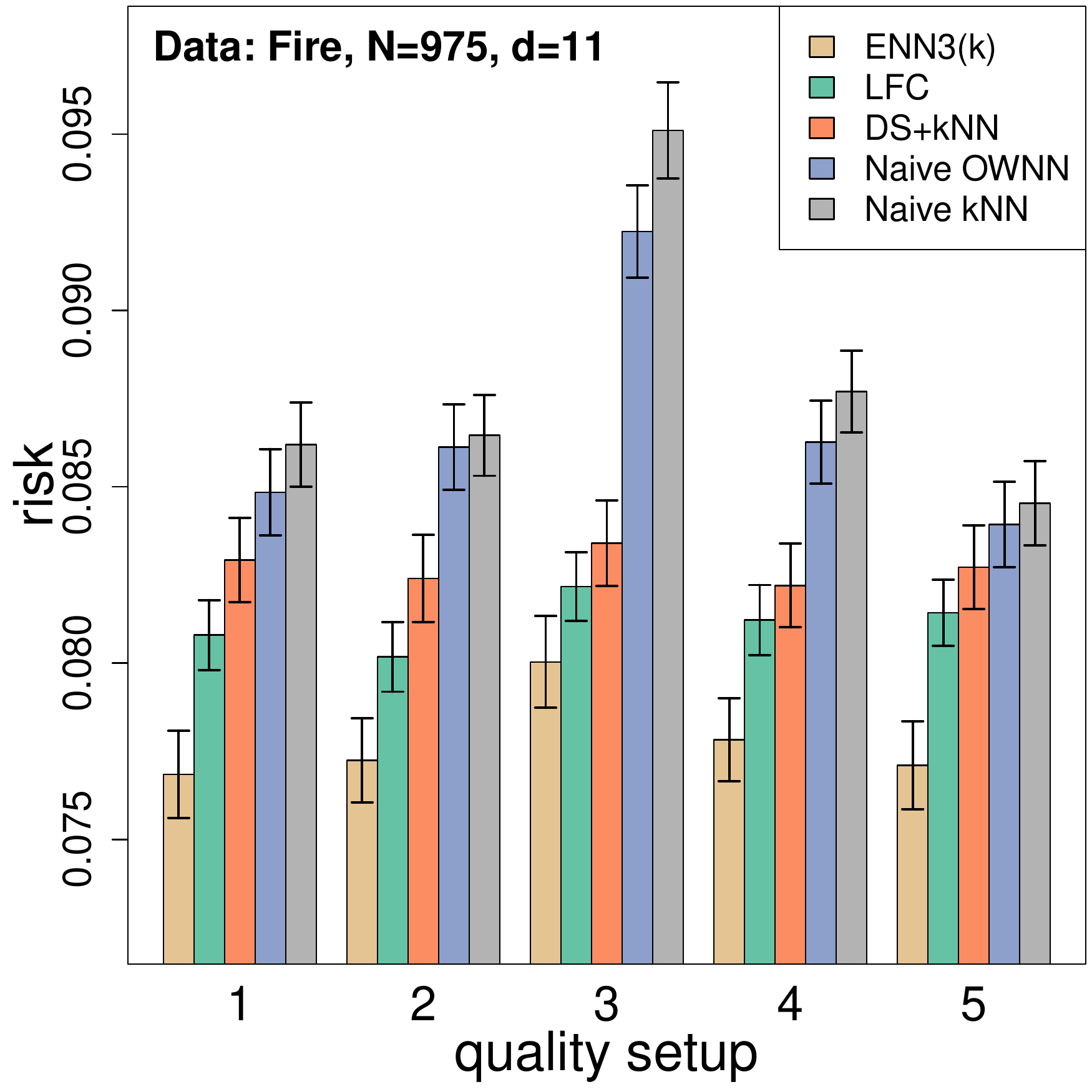}
    \includegraphics[width=0.28\textwidth,height=0.25\textwidth]{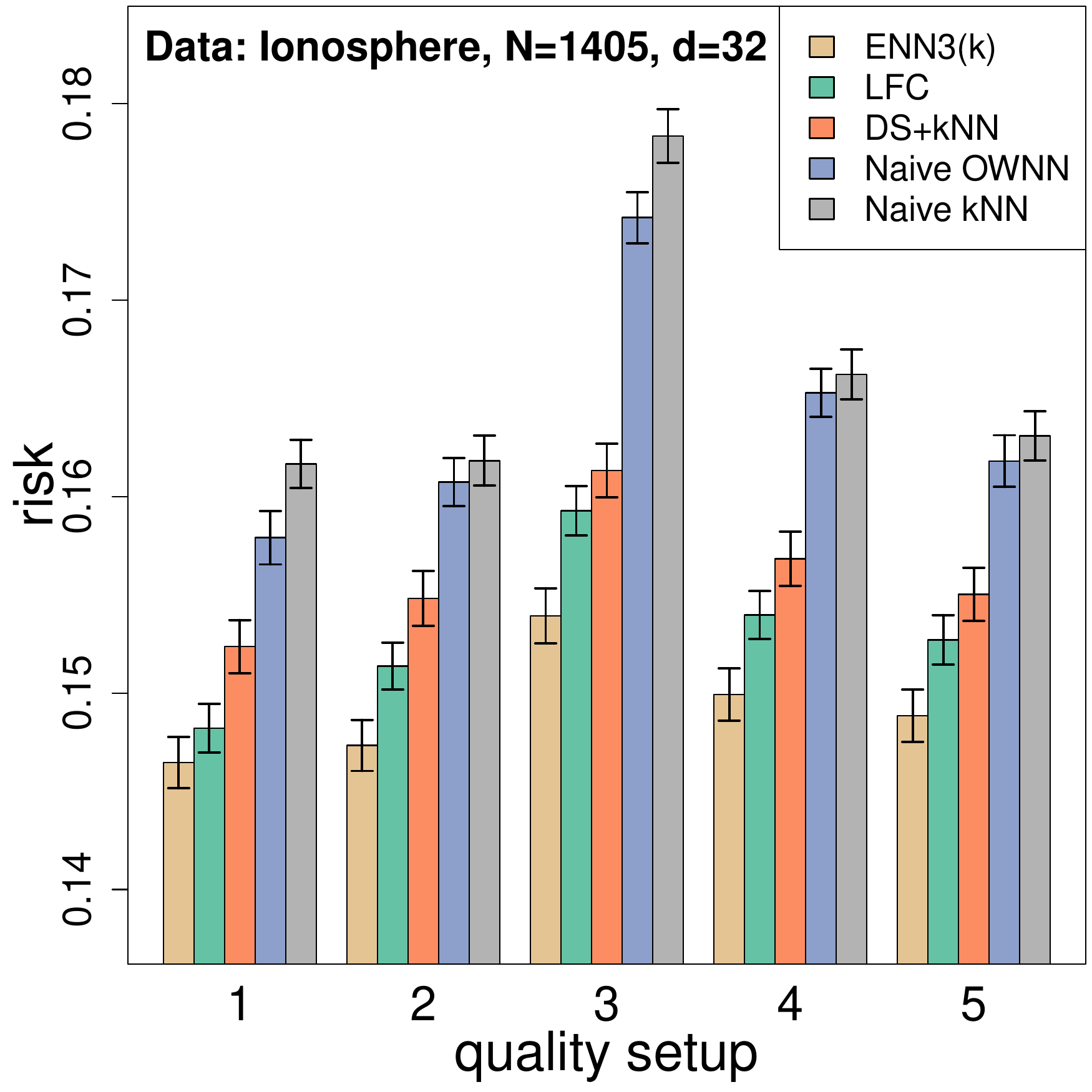}
    \includegraphics[width=0.28\textwidth,height=0.25\textwidth]{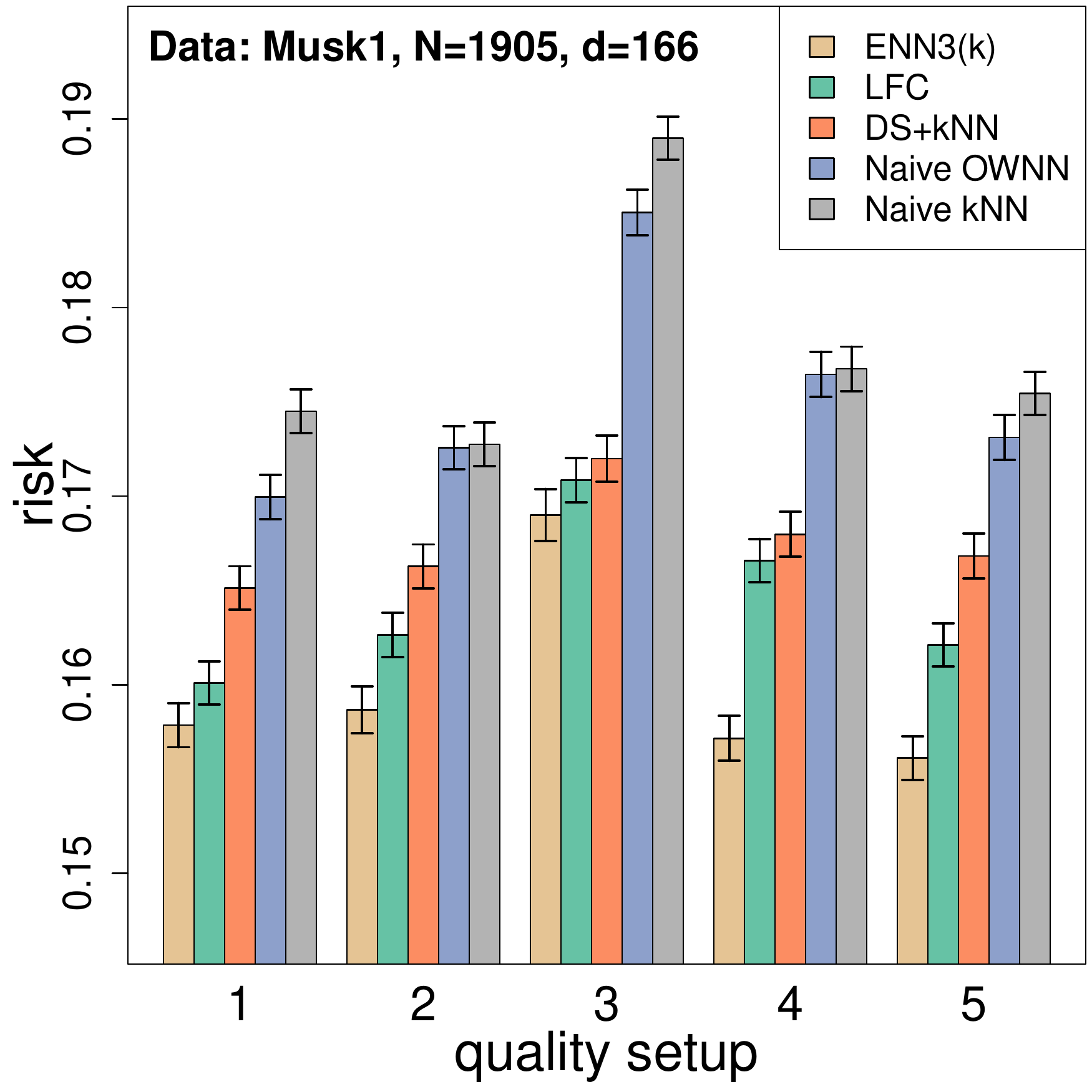}
    \includegraphics[width=0.28\textwidth,height=0.25\textwidth]{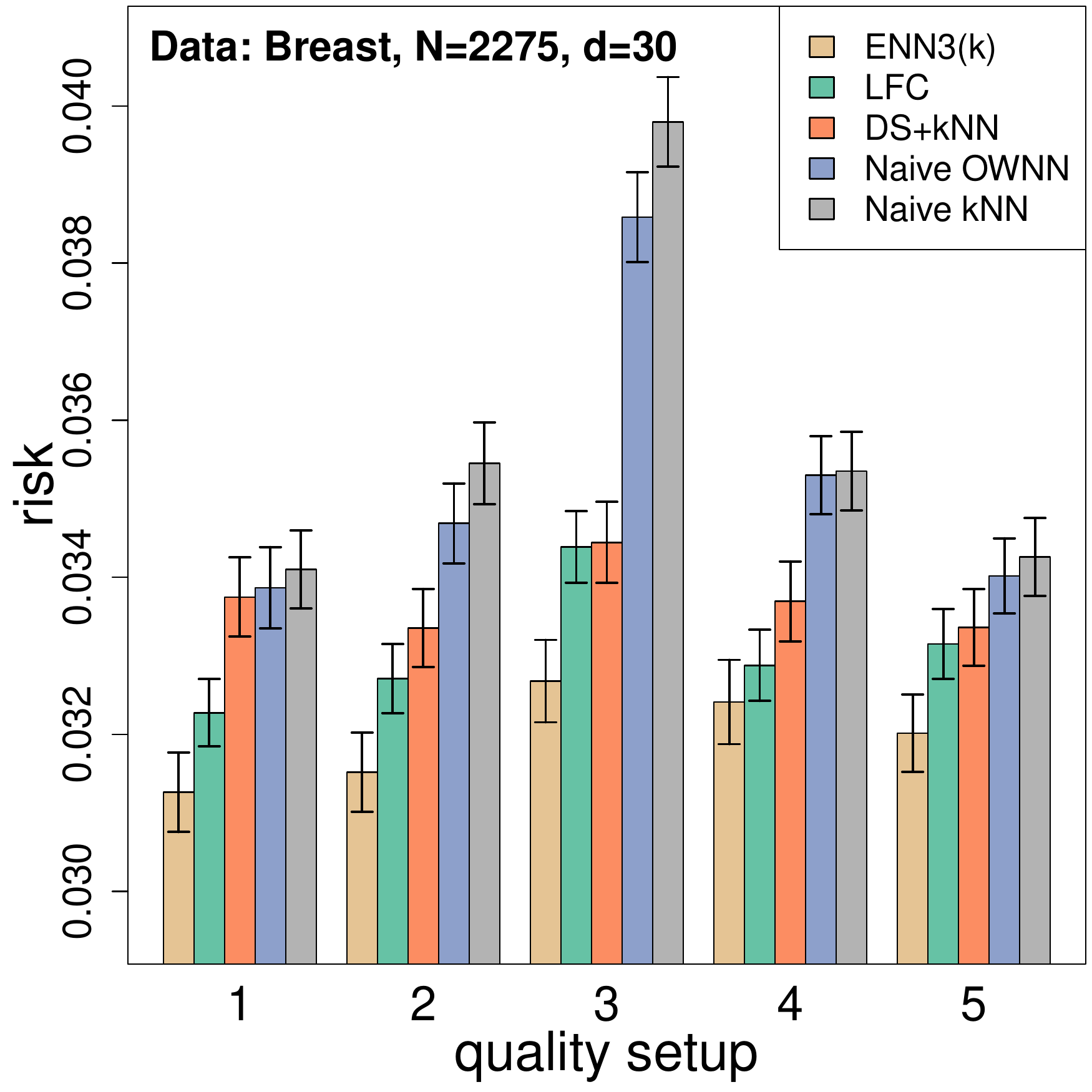}
    \includegraphics[width=0.28\textwidth,height=0.25\textwidth]{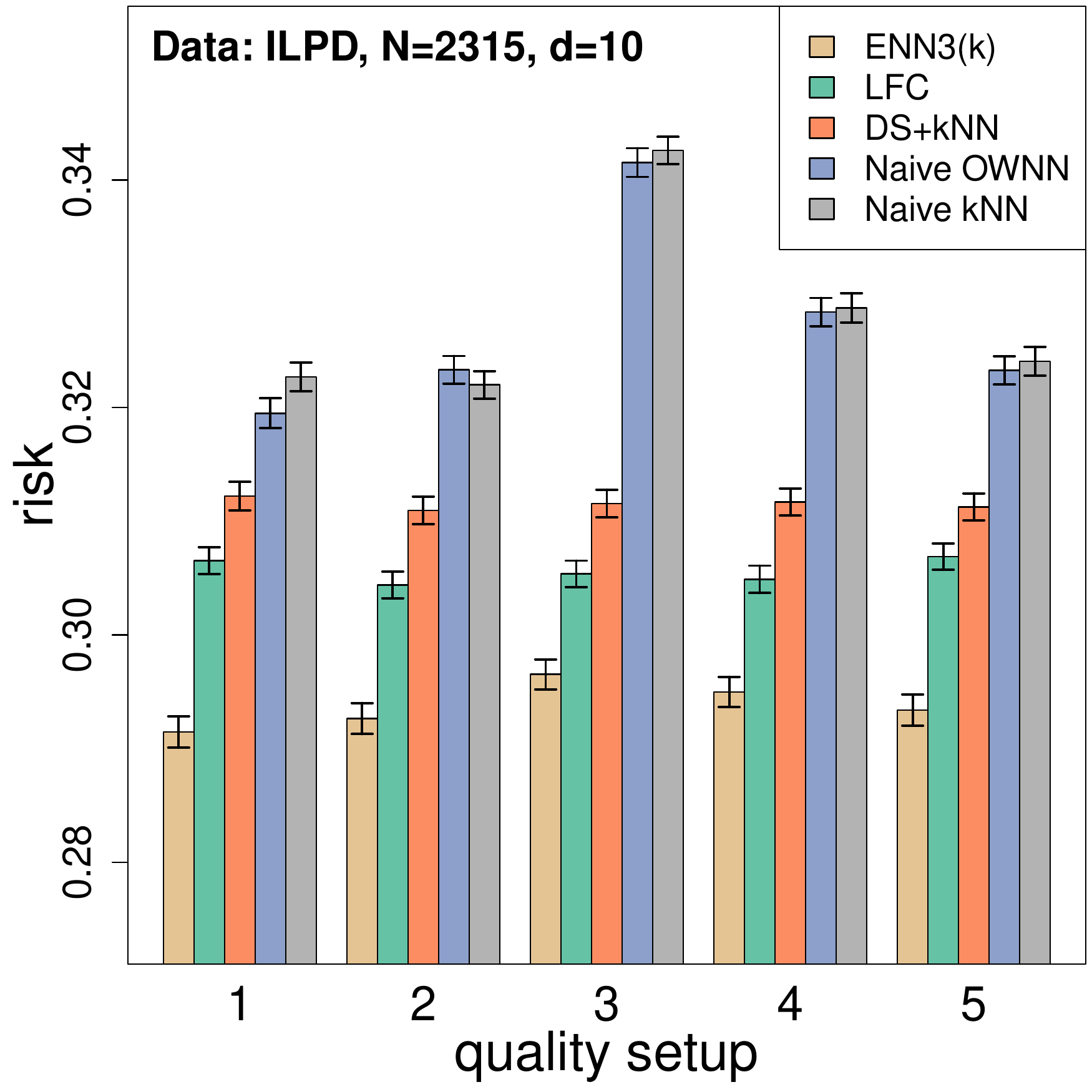}
    \includegraphics[width=0.28\textwidth,height=0.25\textwidth]{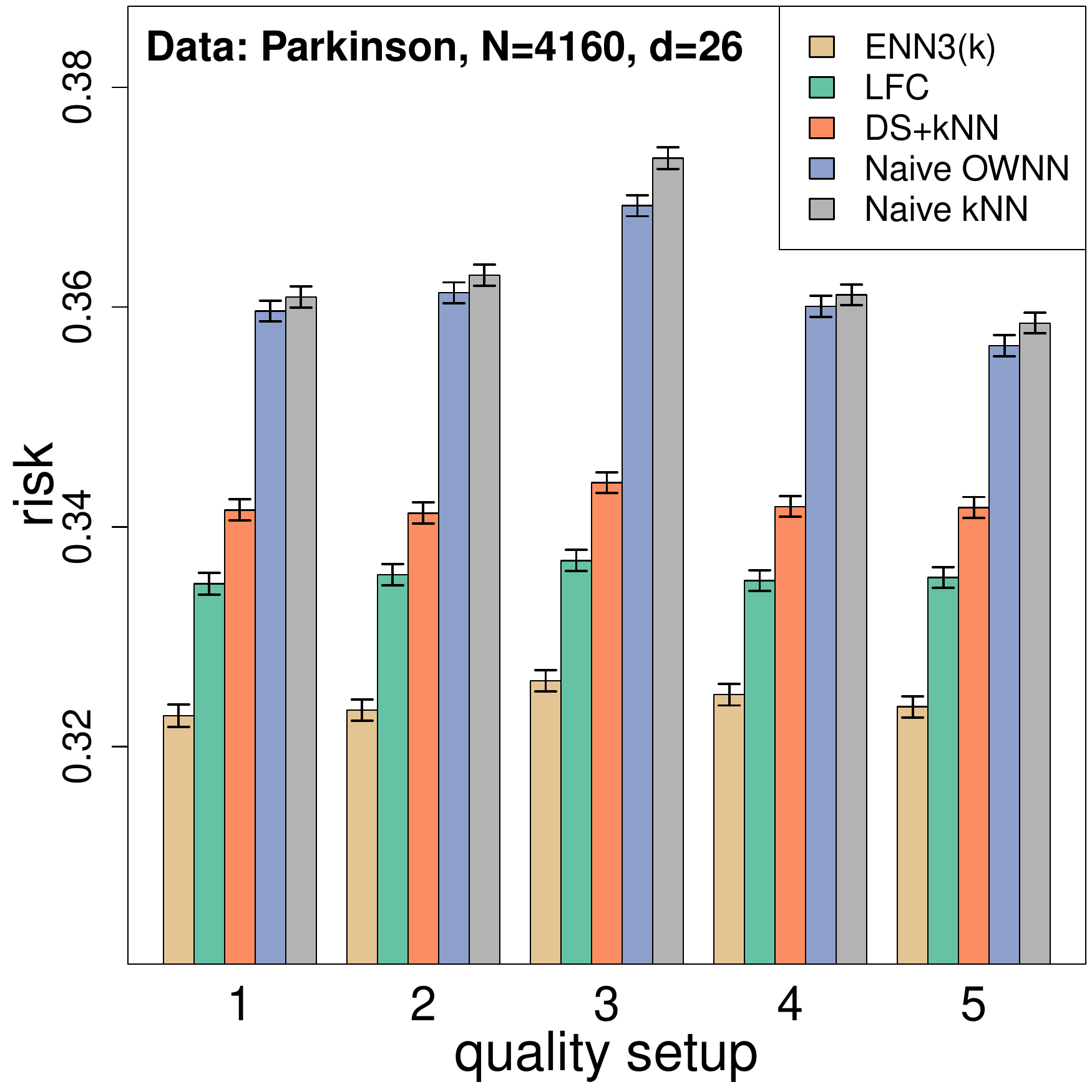}
    \includegraphics[width=0.28\textwidth,height=0.25\textwidth]{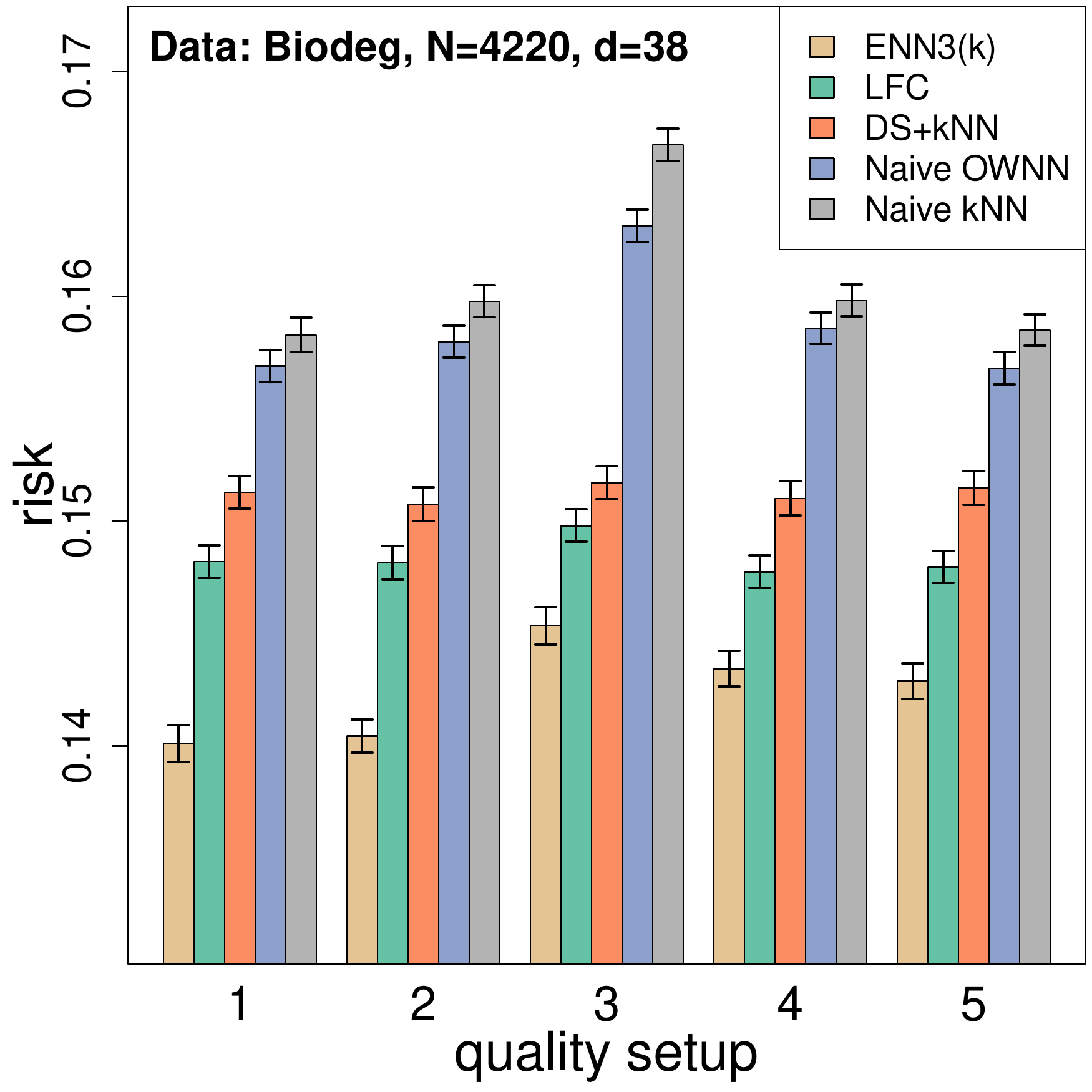}
    \includegraphics[width=0.28\textwidth,height=0.25\textwidth]{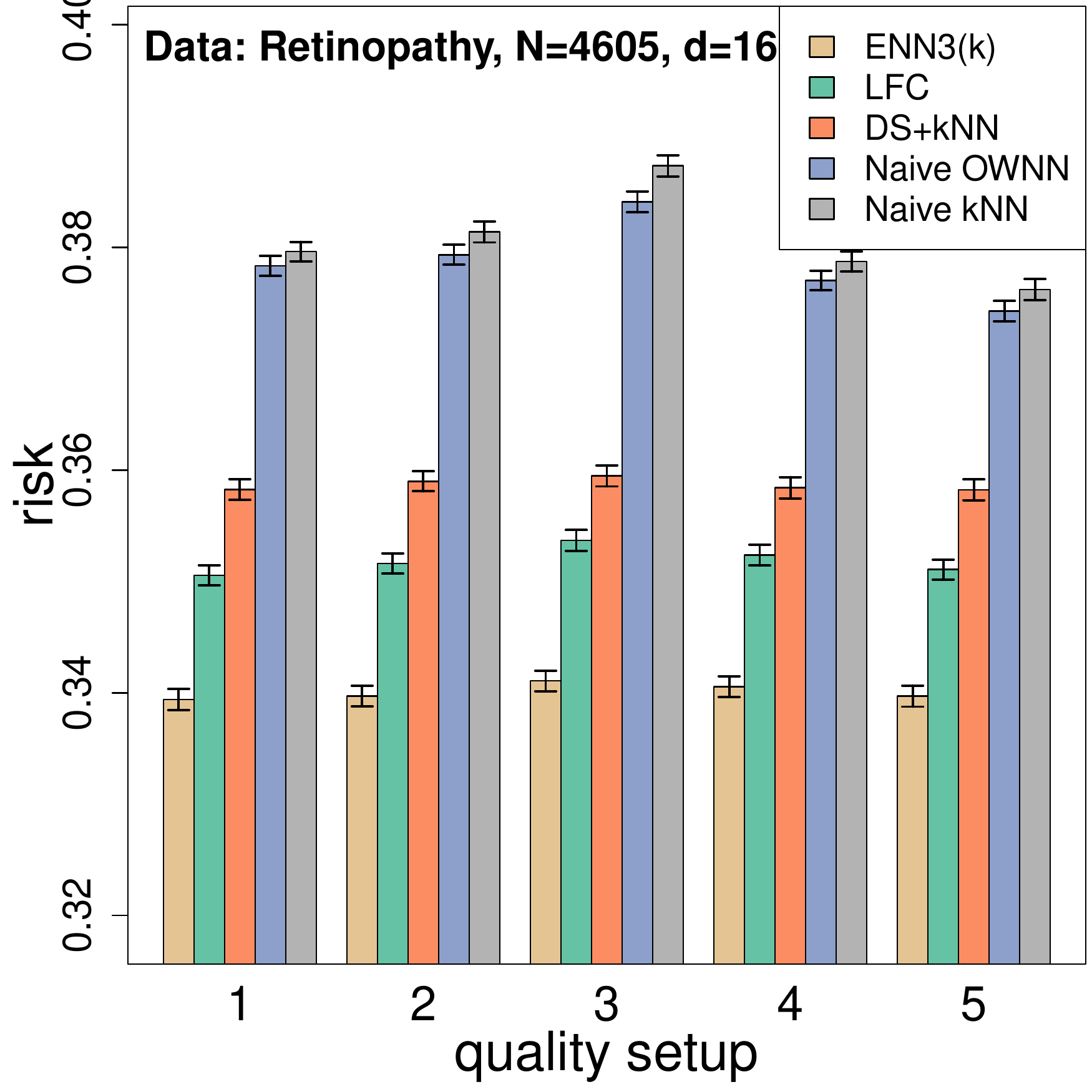}
    \includegraphics[width=0.28\textwidth,height=0.25\textwidth]{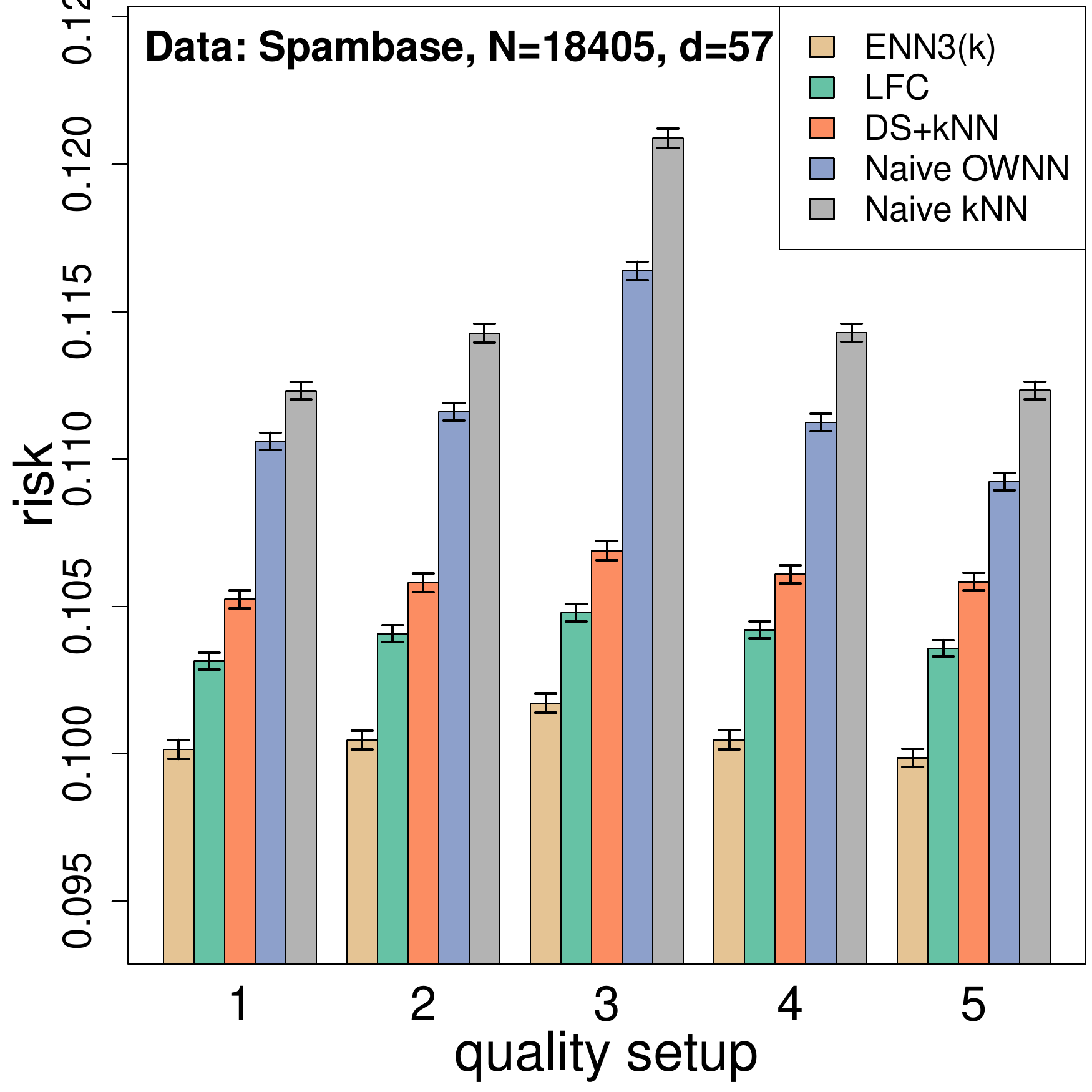}
\vspace{-0.5em}	
\caption{Risk (with standard error bar marked) of ENN3(k), LFC, DS, naive OWNN, and naive $k$NN on real data. The x-axis indicates different settings with worker quality. Dataset name, size and dimension are illustrated on top left.} \label{fig:real_risk_quality_no_expert}
\vspace{-0.5em}
\end{figure*} 
\subsection{Real examples}\label{sec:real_E}
In this section, we empirically check the accuracy of ENN compared with four benchmark methods: the naive $k$NN, naive OWNN methods, and two existing crowdsourcing methods (we denoted as DS and LFC). The DS method \citep{dawid1979maximum} applies confusion matrix and EM algorithm on the labels of the training set to estimate the truth labels. Based on updated labels from DS, we apply $k$NN on the testing set for prediction. The LFC method \citep{raykar2010learning} is a combination of a two-coin logistic model and EM algorithm.

We have retained benchmark data sets Fire \citep{abid2019predicting}, Ionosphere \citep{sigillito1989classification}, Musk1 \citep{dietterich1997solving}, Breast \citep{street1993nuclear}, ILPD \citep{ramana2012critical}, Parkinson \citep{sakar2013collection}, Biodeg \citep{mansouri2013quantitative}, Retinopathy \citep{antal2014ensemble}, and Spambase \citep{cranor1998spam}, from the UCI machine learning repository \citep{Dua:2019}. Following \citet{yan2010modeling} and \citet{raykar2010learning}, we simulate
five workers according to the two-coin model described in Section~\ref{sec:pre_E} with the quality setups 1-5 defined in Table~\ref{tab:quality_setup}. The test sample sizes are set as $(\mbox{total sample size)}/5$. Parameters in the naive $k$NN and OWNN are tuned using cross-validation, and the parameters $k_j$ in ENN(k) for each worker data are set using bridging formula stated in our theorems. The empirical risk is calculated over 1000 replications.

In Figure~\ref{fig:real_risk_quality_no_expert}, we compare the empirical risk (test error) of ENN3($k$) relative to LFC, DS, naive OWNN and $k$NN. From Figure~\ref{fig:real_risk_quality_no_expert}, we can see that the ENN3($k$) outperforms the other four benchmark methods in all cases. Both naive $k$NN and OWNN methods have significantly poor performance under different quality setups, especially on the setup $3$, which has a lower level of worker quality. The ENN method significantly enhances the case of poor quality. Lastly, we note that a larger sample size $N$ generates a more stable ENN method performance among different quality setups. As a larger sample size may increase the estimation accuracy for worker quality, the enhanced effect of ENN on the noisy worker data will thereby improve the performance.

\section{Discussions}\label{sec:conclusion_E}
There are a couple of interesting directions to be pursued in the future. The first two are extensions to the multicategory classification problem and high-dimensional data. The third direction is related to a realistic attack paradigm named adversarial examples that received a lot of recent attention \citep{szegedy2013intriguing,papernot2016limitations}. Some worker data may contain adversarial examples in practice, which might violate our quality assumption $a_j+b_j>1$. It leaves us to wonder how to take advantage of the quality-related nature of ENN to detect and deal with adversarial samples. In addition, it is also an interesting direction to explore strategies to relax the assumption that worker quality does not depend on the feature vector.

 
\bibliographystyle{asa}
\bibliography{ENN}

%
\newcommand\invisiblesection[1]{%
  \refstepcounter{section}%
  \addcontentsline{toc}{section}{\protect\numberline{\thesection}#1}%
  \sectionmark{#1}}
\invisiblesection{Supplementary}

\renewcommand{\theequation}{S.\arabic{equation}}
\renewcommand{\thetable}{S\arabic{table}}
\renewcommand{\thefigure}{S\arabic{figure}}
\renewcommand{\thesubsection}{S.\Roman{subsection}}
\setcounter{equation}{0}
\setcounter{table}{0}
\setcounter{figure}{0}
\setcounter{subsection}{0}

 
\newpage 
 
\begin{center}
\Large\bf Supplementary Materials
\end{center}

\subsection{Appendix 1: Assumptions (A1) - (A4)}\label{sec:assumptions_E}
For a smooth function $g$, we write $\dot{g}(x)$ for its gradient vector at $x$. The following conditions are assumed throughout this paper.

(A1) The set ${\cal R}\subset \mathbb R^d$ is a compact $d$-dimensional manifold with boundary $\partial{\cal R}$.

(A2) The set ${\cal S}=\{x\in {\cal R}: \eta^0(x)=1/2\}$ is nonempty. There exists an open subset $U_0$ of ${\mathbb R}^d$ which contains ${\cal S}$ such that: (1) $\eta^0$ is continuous on $U\backslash U_0$ with $U$ an open set containing ${\cal R}$; (2) the restriction of the conditional distributions of $X$, $P_1^0$ and $P_0^0$, to $U_0$ are absolutely continuous with respect to Lebesgue measure, with twice continuously differentiable Randon-Nikodym derivatives $f_1^0$ and $f_0^0$.

(A3) There exists $\rho>0$ such that $\int_{{\mathbb R}^d}\|x\|^{\rho} d\bar{P}(x) < \infty$. In addition, for sufficiently small $\delta>0$, $\inf_{x\in {\cal R}}\bar{P}(B_{\delta}(x))/(a_d\delta^d) \ge C_0 >0$, where $a_d=\pi^{d/2}/\Gamma(1+d/2)$, $\Gamma(\cdot)$ is gamma function, and $C_0$ is a constant independent of $\delta$.

(A4) For all $x\in {\cal S}$, we have $\dot{\eta}^0(x)\ne 0$, and for all $x\in {\cal S}\cap \partial{\cal R}$, we have $\dot{\partial \eta^0}(x)\ne 0$, where $\partial \eta^0$ is the restriction of $\eta^0$ to $\partial {\cal R}$. \hfill $\blacksquare$

\subsection{Appendix 2: Definitions of \texorpdfstring{$a^0(x)$, $B_1$, $B_2$, $W_{n_j,\beta}$ and $W_{N,\beta}$}{Lg}}\label{sec:defwnb_E}
For a smooth function $g$: $\mathbb{R}^d\rightarrow \mathbb{R}$, denote $g_m(x)$ as its $m$-th partial derivative at $x$ and $g_{mk}(x)$ the $(m,k)$-th element of its Hessian matrix at $x$. Let $c_{m,d}=\int_{v:\|v\|\le 1} v_m^2 dv$, $\bar{f}=\pi_1^0 f_1^0 + (1-\pi_1^0)f_0^0$. Define
\begin{align*}
a^0(x)&=\sum_{m=1}^d  \frac{c_{m,d}\{\eta_{m}^0(x)\bar{f}_m(x) + 1/2 \eta_{mm}^0(x)\bar{f}(x)\}}{a_d^{1+2/d} \bar{f}(x)^{1+2/d}}.\end{align*}
Moreover, define two distribution-related constants
\begin{eqnarray*}
B_1 = \int_{\cal S} \frac{\bar{f}(x)}{4\|\dot{\eta}^0(x)\|} d \textrm{Vol}^{d-1}(x),\quad B_2 = \int_{\cal S} \frac{\bar{f}(x)}{\|\dot{\eta}^0(x)\|} [a^0(x)]^2 d \textrm{Vol}^{d-1}(x),
\end{eqnarray*}
where $\textrm{Vol}^{d-1}$ is the natural $(d-1)$-dimensional volume measure that ${\cal S}$ inherits as a subset of $\mathbb{R}^d$. According to Assumptions (A1)-(A4) in Appendix~\ref{sec:assumptions_E}, $B_1$ and $B_2$ are finite with $B_1>0$ and $B_2\ge 0$, with equality only when $a^0(x)=0$ on ${\cal S}$.

In addition, for $\beta>0$, we define $W_{n_j,\beta}$ as the set of $\bw_j$ satisfying:
\begin{itemize}
    \item[(w.1)] $\sum_{i=1}^{n_j} w_{j,i}^2 \le n_j^{-\beta}$;
    \item[(w.2)] $n_j^{-4/d}(\sum_{i=1}^{n_j}\alpha_iw_{j,i})^2\le n_j^{-\beta}$, where $\alpha_i=i^{1+\frac{2}{d}}-(i-1)^{1+\frac{2}{d}}$;
     \item[(w.3)] $n_j^{2/d}\sum_{i=k_2^j+1}^{n_j} w_{j,i}/\sum_{i=1}^{n_j} \alpha_iw_{j,i}\le 1/\log n_j$ with $k_2^j=\lceil n_j^{1-\beta} \rceil$;
      \item[(w.4)] $\sum_{i=k_2^j+1}^{n_j} w_{j,i}^2/\sum_{i=1}^{n_j} w_{j,i}^2 \le 1/\log n_j$;
      \item[(w.5)] $\sum_{i=1}^{n_j} w_{j,i}^3/(\sum_{i=1}^{n_j} w_{j,i}^2)^{3/2} \le 1/\log n_j$.
\end{itemize}
When $n_j$ in (w.1)--(w.5) is replaced by $N$, we can define the set $W_{N,\beta}$. \hfill $\blacksquare$

\subsection{Appendix 3: Additional numerical results}\label{sec:quality_estimation_table}
\begin{figure*}[htb!] 
	\centering\vspace{-0.5em}
	\includegraphics[width=0.28\textwidth,height=0.25\textwidth]{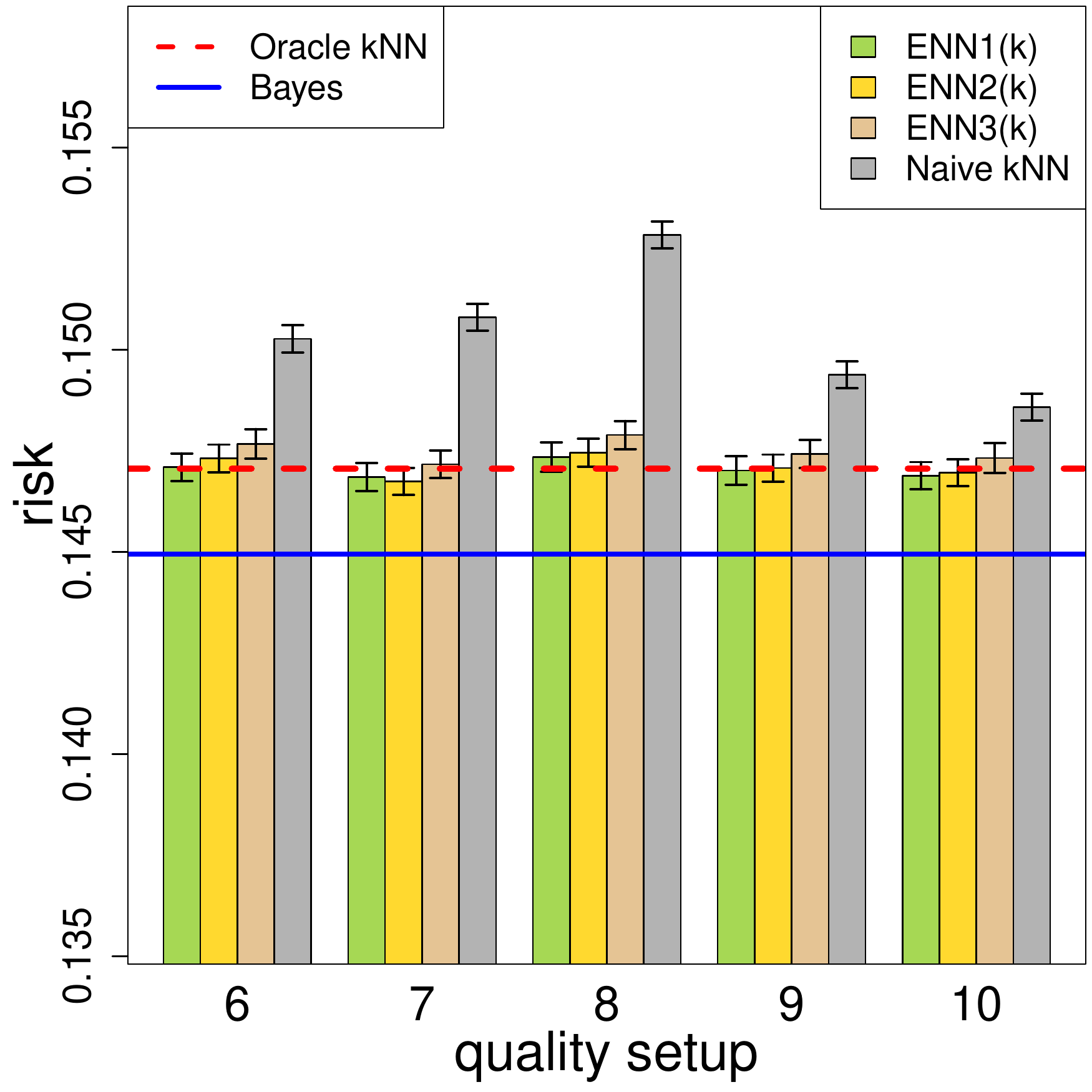}
    \includegraphics[width=0.28\textwidth,height=0.25\textwidth]{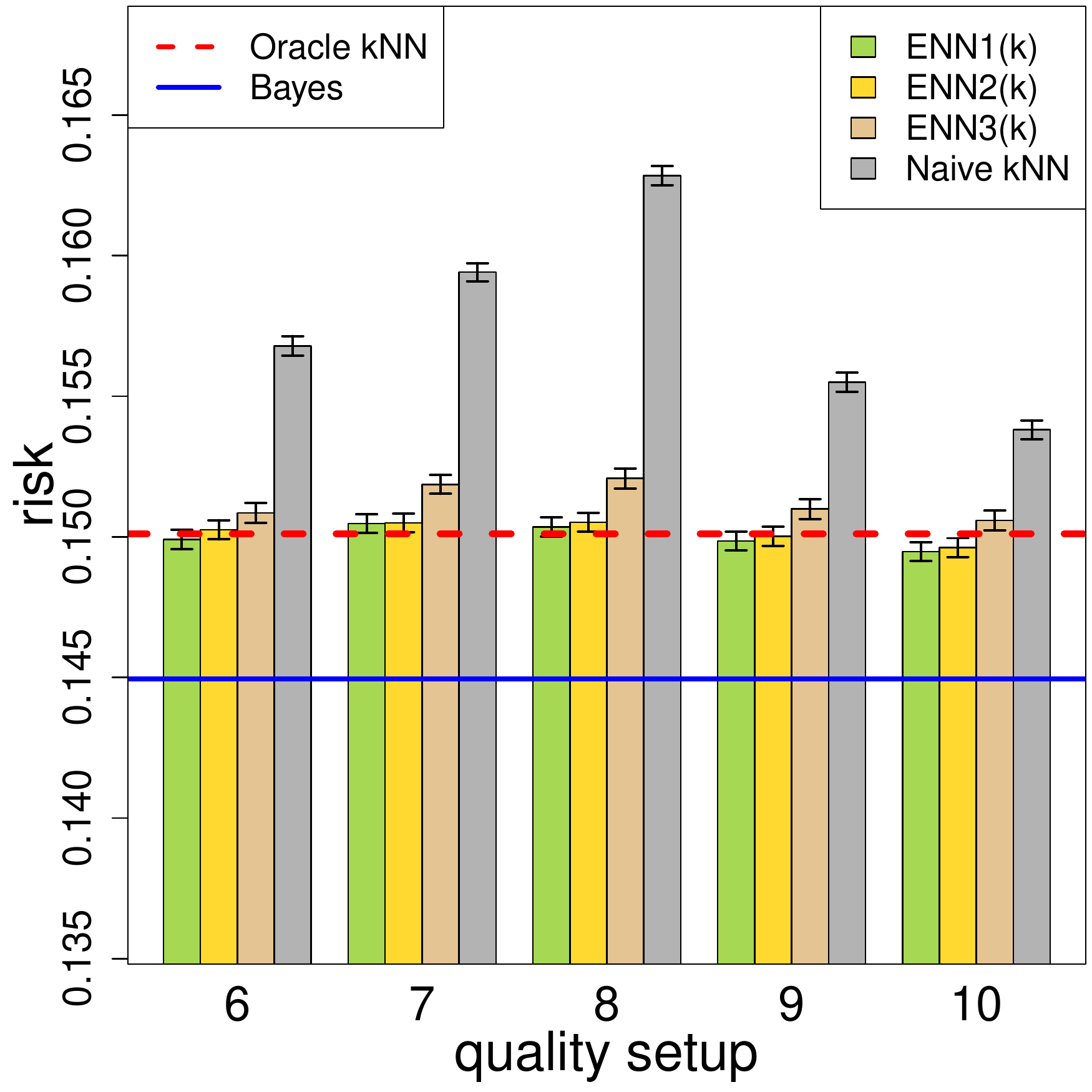}
    \includegraphics[width=0.28\textwidth,height=0.25\textwidth]{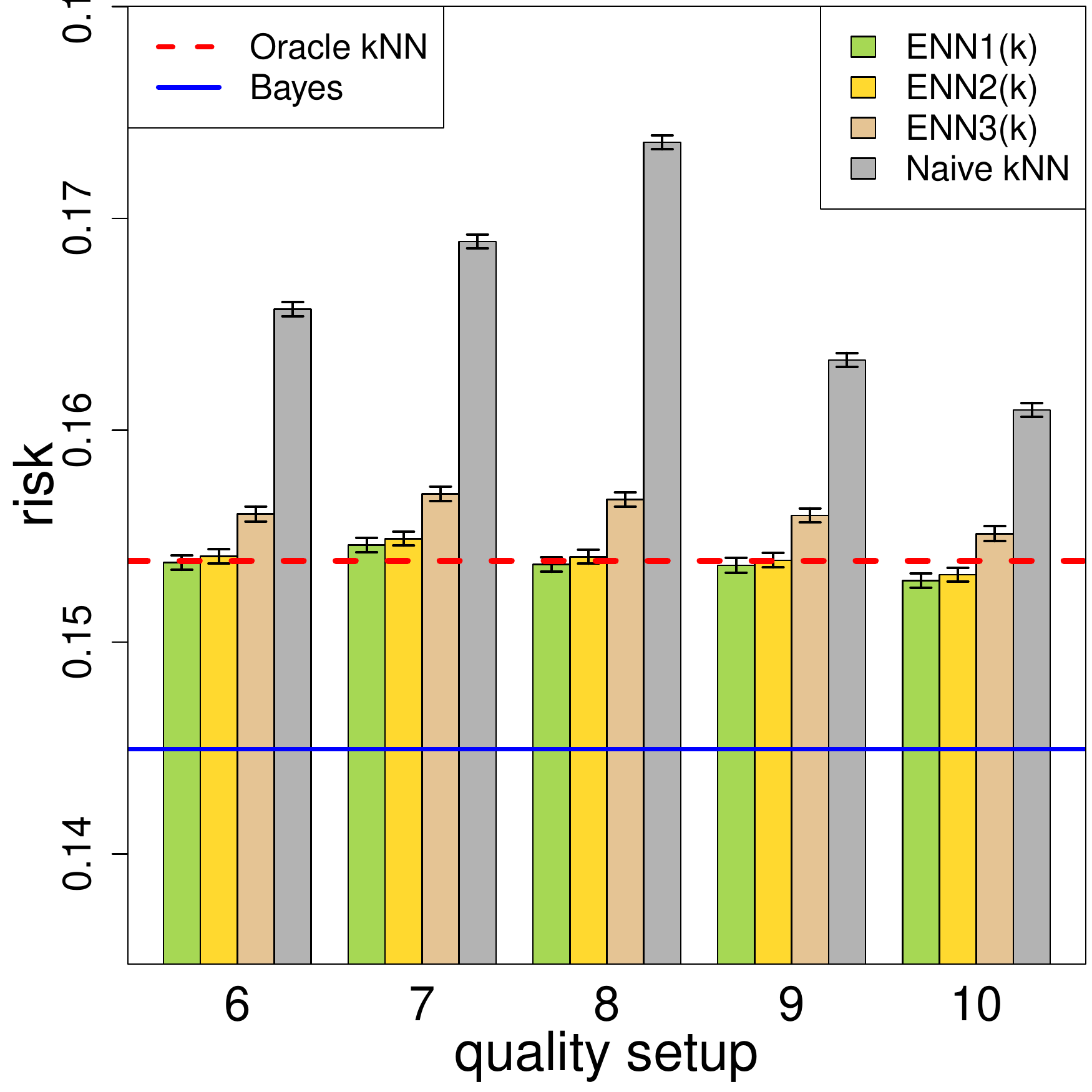}
    \includegraphics[width=0.28\textwidth,height=0.25\textwidth]{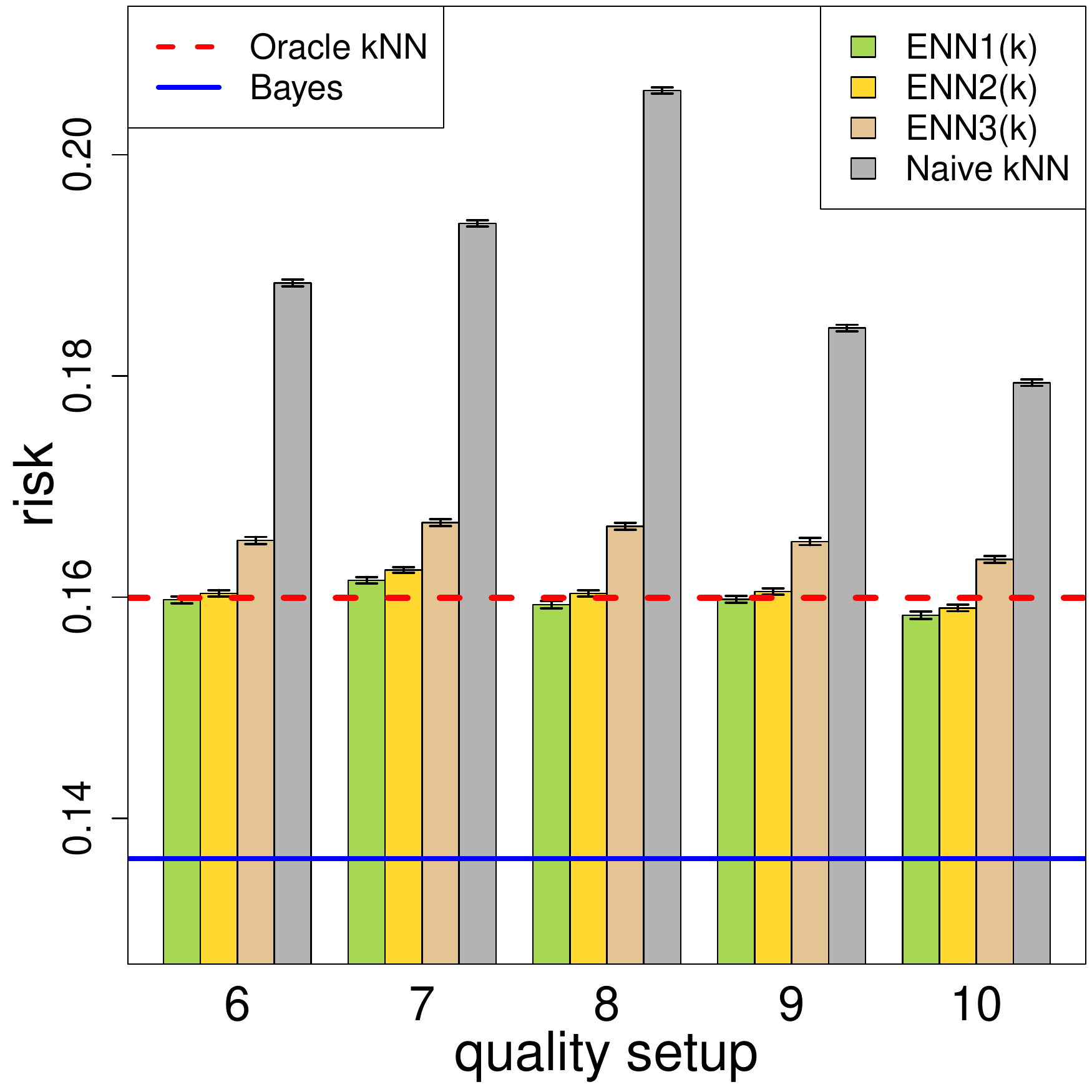}
    \includegraphics[width=0.28\textwidth,height=0.25\textwidth]{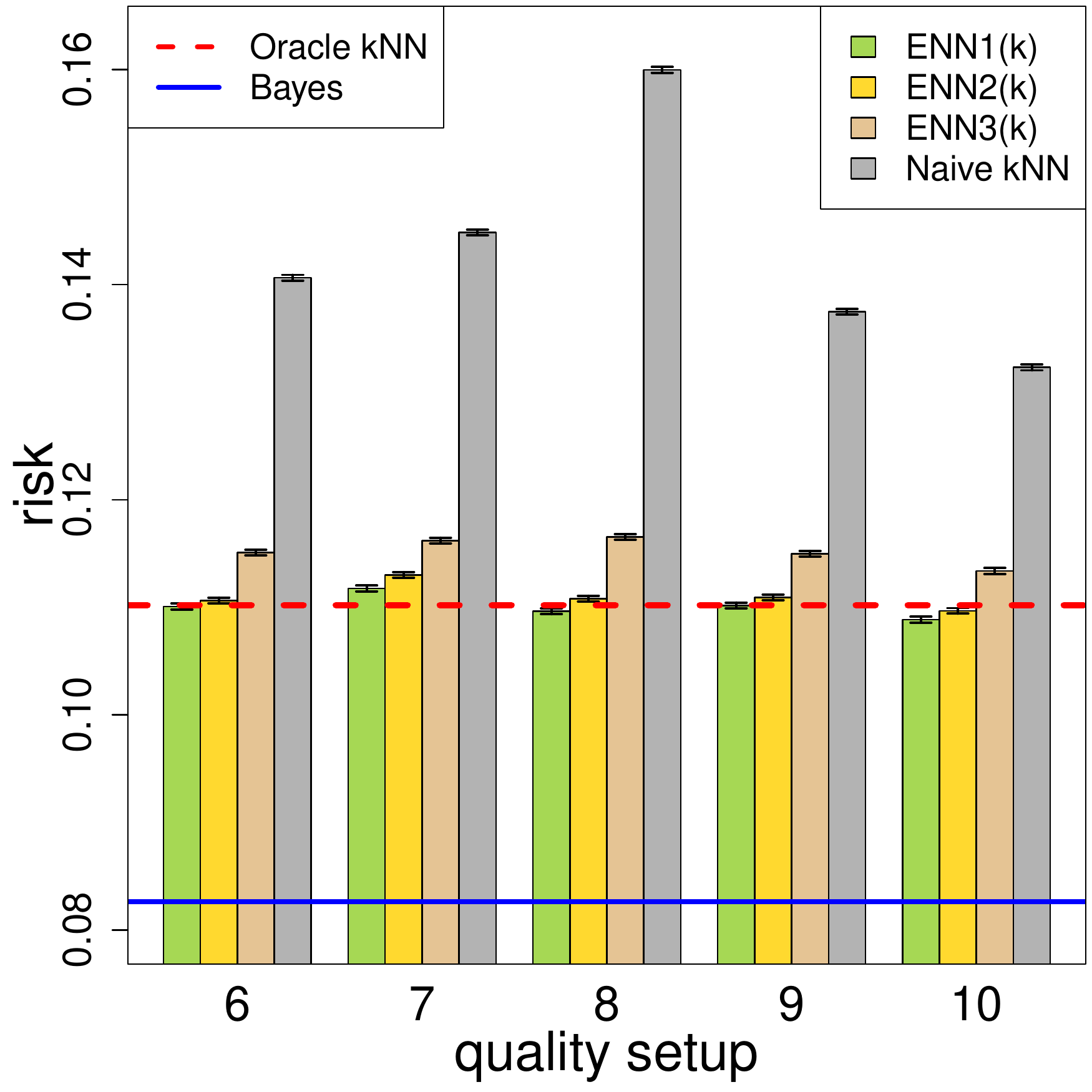}
    \includegraphics[width=0.28\textwidth,height=0.25\textwidth]{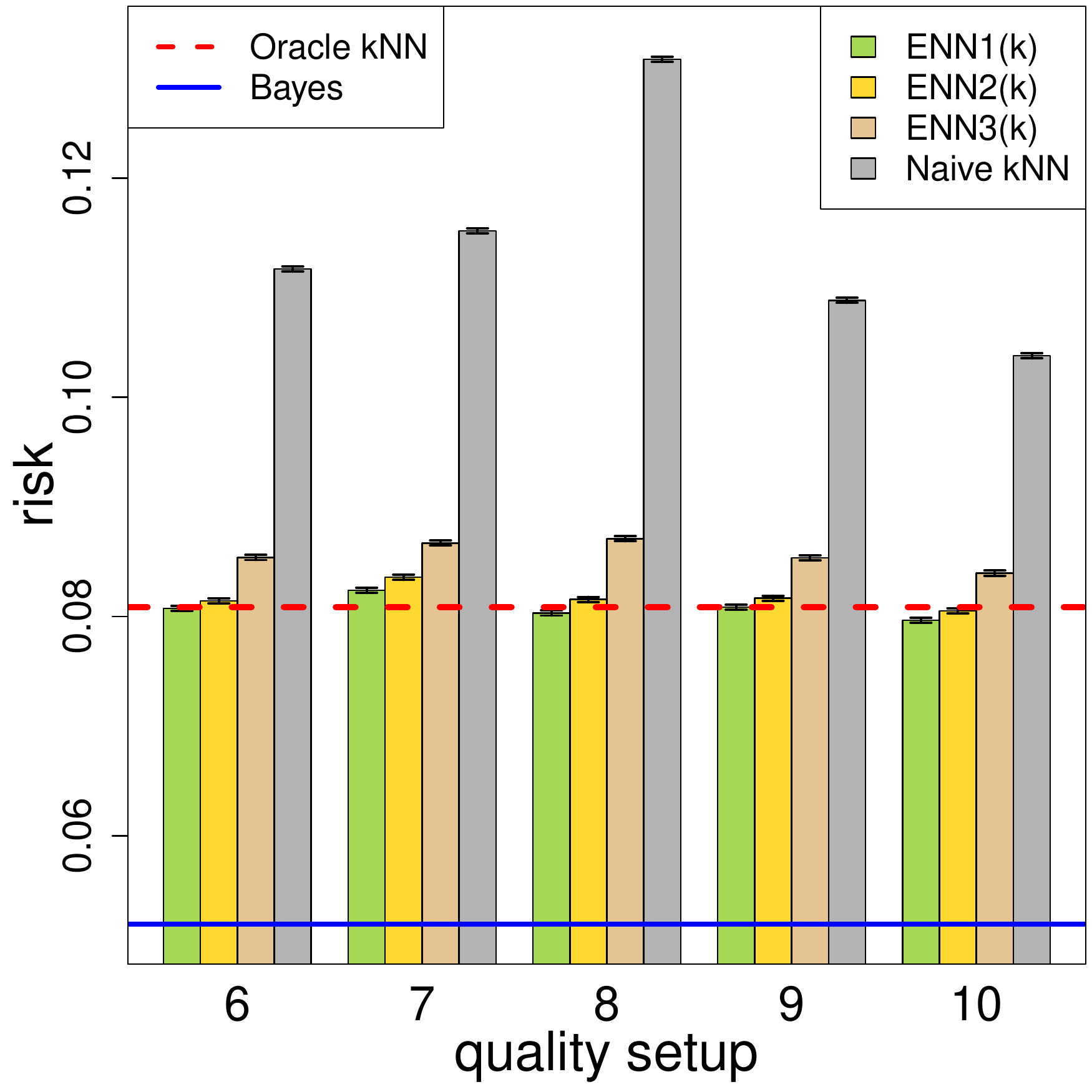}
    \includegraphics[width=0.28\textwidth,height=0.25\textwidth]{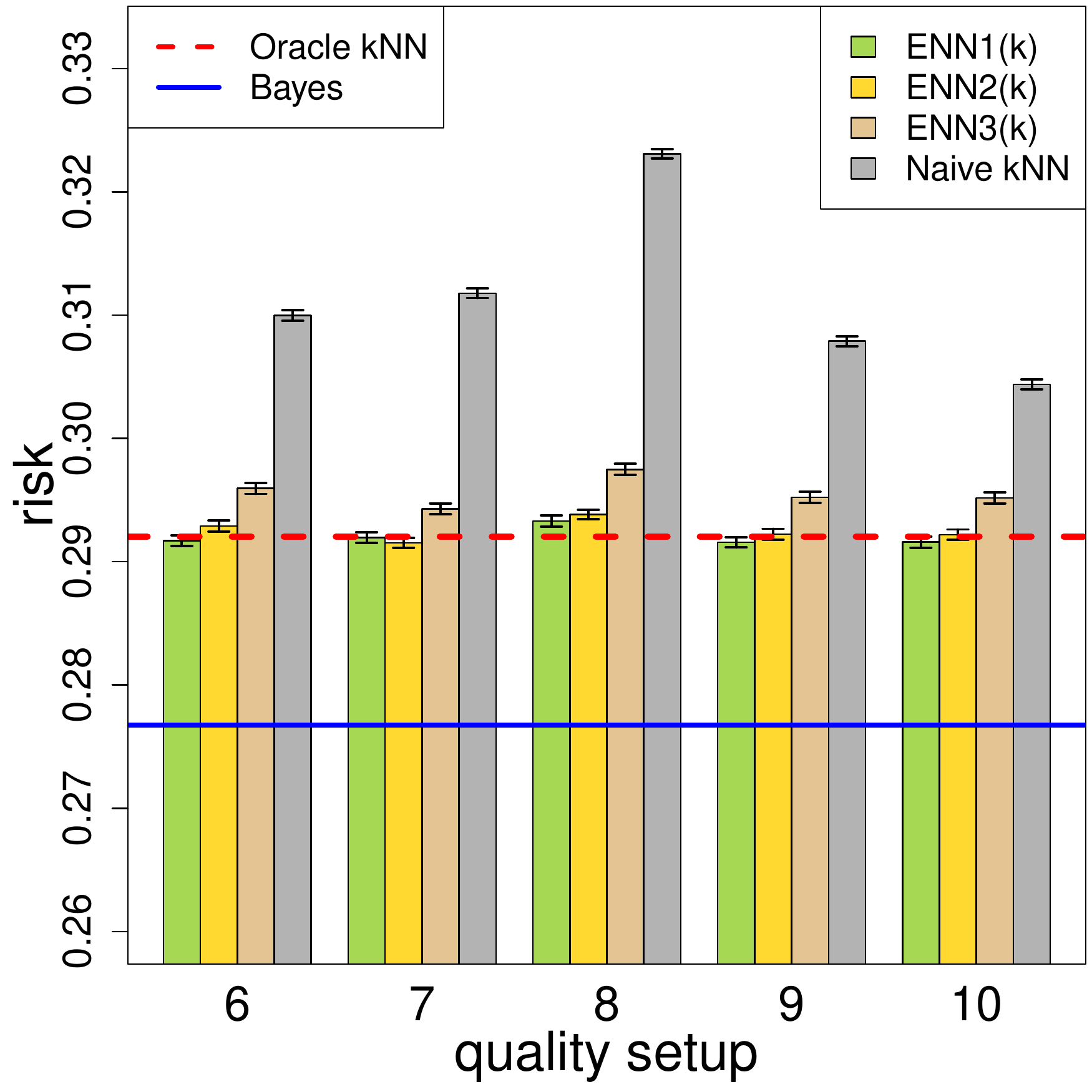}
    \includegraphics[width=0.28\textwidth,height=0.25\textwidth]{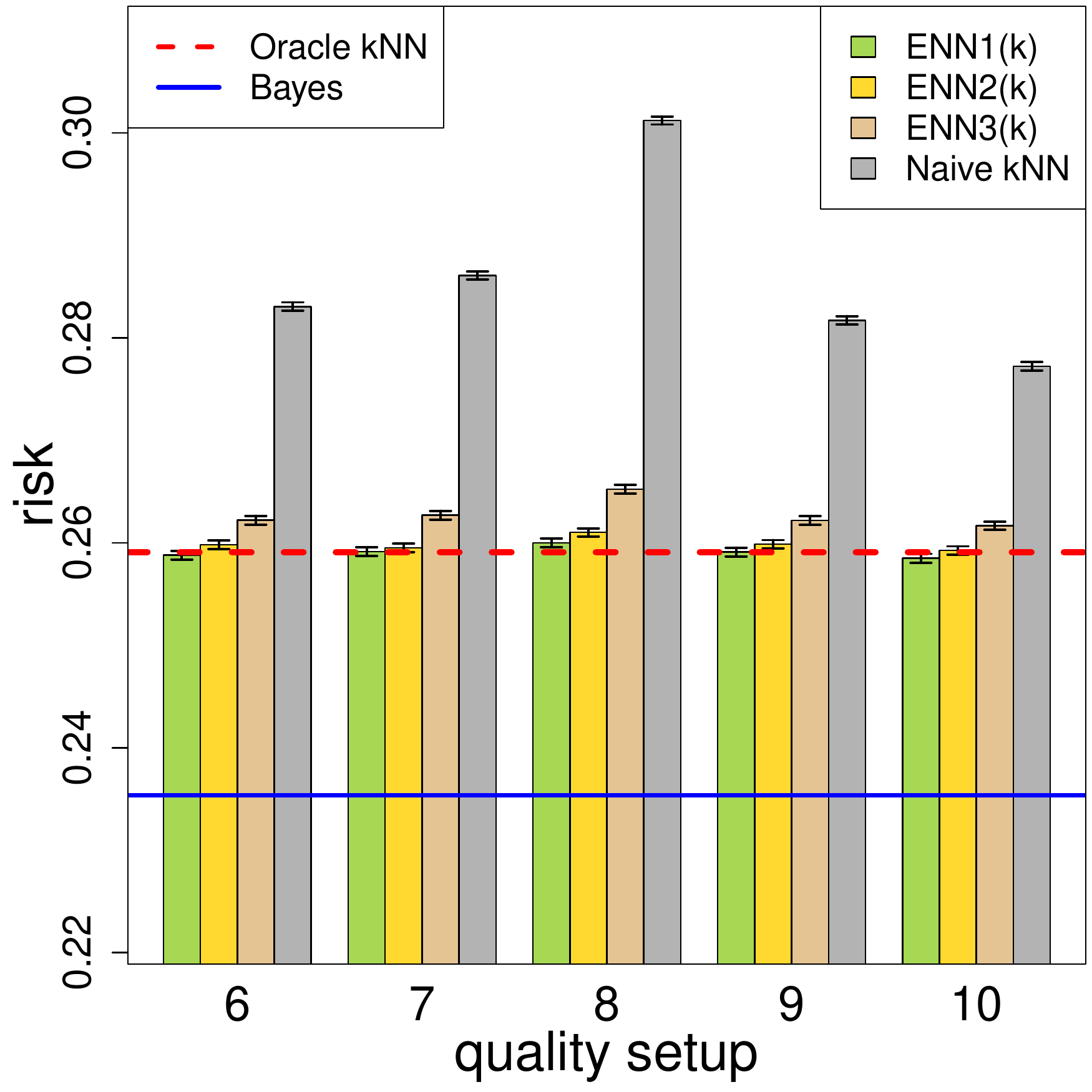}
    \includegraphics[width=0.28\textwidth,height=0.25\textwidth]{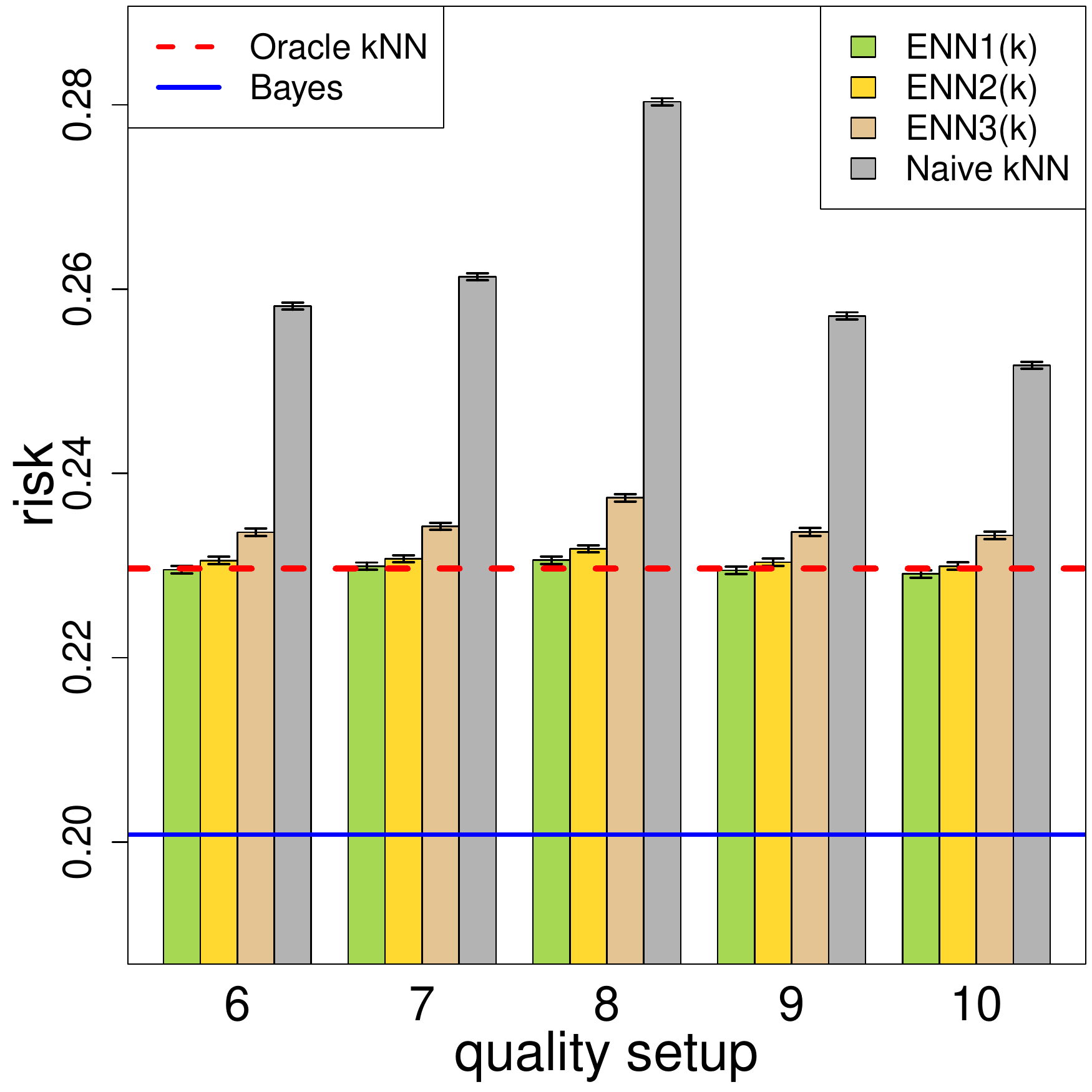}
\vspace{-1em}	
\caption{Risk (with standard error bar marked) of all methods and the Bayes rule, with expert data. The x-axis indicates different settings with worker quality. Top/middle/bottom: Simulation $1/2/3$; left/middle/right: $d=4/6/8$.} \label{fig:sim_risk_quality_expert}
\vspace{-0.5em}
\end{figure*}

Table \ref{tab:quality_results_ENN2} and Table \ref{tab:quality_results_ENN3} illustrate the comparison of true and estimated worker quality based on Algorithm \ref{algo:ENN2} (ENN2) and Algorithm \ref{algo:ENN3} (ENN3), respectively. Figure \ref{fig:sim_risk_quality_expert} shows the comparison of risks for setups 6-10.

\begin{table}[hbt!]
	\vspace{-0.5em}	
	\caption{Comparison of true and estimated worker quality based on Algorithm~\ref{algo:ENN2}.} \label{tab:quality_results_ENN2}
	\vspace{+0.1em}	
	\setlength\tabcolsep{2.1pt}
	\scriptsize
	\centering
	\begin{tabular}{lc|cc|cc|cc|cc|cc|cc|cc|cc|cc|cc}
	\hline
	Sim & d  &$a_1$&$\widehat{a}_1$&$a_2$&$\widehat{a}_2$&$a_3$&$\widehat{a}_3$&$a_4$&$\widehat{a}_4$&$a_5$&$\widehat{a}_5$&$b_1$&$\widehat{b}_1$&$b_2$&$\widehat{b}_2$&$b_3$&$\widehat{b}_3$&$b_4$&$\widehat{b}_4$&$b_5$&$\widehat{b}_5$ \\
	\hline
 1 & 4 & 0.90 & 0.901 & 0.90 & 0.901 & 0.95 & 0.945 & 0.90 & 0.897 & 1.00 & 1.000 & 0.80 & 0.807 & 0.80 & 0.807 & 0.85 & 0.852 & 0.85 & 0.854 & 1.00 & 1.000 \\
 1 & 4 & 0.80 & 0.798 & 0.80 & 0.794 & 0.85 & 0.842 & 0.80 & 0.798 & 1.00 & 1.000 & 0.90 & 0.904 & 0.95 & 0.951 & 0.95 & 0.949 & 0.90 & 0.904 & 1.00 & 1.000 \\
 1 & 4 & 0.60 & 0.616 & 0.65 & 0.664 & 0.85 & 0.842 & 0.80 & 0.798 & 1.00 & 1.000 & 0.75 & 0.769 & 0.75 & 0.768 & 0.95 & 0.949 & 0.90 & 0.904 & 1.00 & 1.000 \\
 1 & 4 & 0.80 & 0.798 & 0.85 & 0.853 & 0.85 & 0.842 & 0.90 & 0.901 & 1.00 & 1.000 & 0.90 & 0.904 & 0.80 & 0.808 & 0.95 & 0.949 & 0.80 & 0.807 & 1.00 & 1.000 \\
 1 & 4 & 0.80 & 0.805 & 0.85 & 0.849 & 0.95 & 0.938 & 0.85 & 0.849 & 1.00 & 1.000 & 0.80 & 0.810 & 0.85 & 0.855 & 0.95 & 0.946 & 0.85 & 0.855 & 1.00 & 1.000 \\
 \hline
 2 & 6 & 0.90 & 0.908 & 0.90 & 0.908 & 0.95 & 0.954 & 0.90 & 0.905 & 1.00 & 1.000 & 0.80 & 0.808 & 0.80 & 0.808 & 0.85 & 0.854 & 0.85 & 0.855 & 1.00 & 1.000 \\
 2 & 6 & 0.80 & 0.805 & 0.80 & 0.802 & 0.85 & 0.850 & 0.80 & 0.805 & 1.00 & 1.000 & 0.90 & 0.905 & 0.95 & 0.952 & 0.95 & 0.951 & 0.90 & 0.905 & 1.00 & 1.000 \\
 2 & 6 & 0.60 & 0.620 & 0.65 & 0.668 & 0.85 & 0.851 & 0.80 & 0.805 & 1.00 & 1.000 & 0.75 & 0.770 & 0.75 & 0.768 & 0.95 & 0.951 & 0.90 & 0.905 & 1.00 & 1.000 \\
 2 & 6 & 0.80 & 0.805 & 0.85 & 0.859 & 0.85 & 0.851 & 0.90 & 0.908 & 1.00 & 1.000 & 0.90 & 0.905 & 0.80 & 0.809 & 0.95 & 0.951 & 0.80 & 0.808 & 1.00 & 1.000 \\
 2 & 6 & 0.80 & 0.811 & 0.85 & 0.857 & 0.95 & 0.948 & 0.85 & 0.857 & 1.00 & 1.000 & 0.80 & 0.811 & 0.85 & 0.857 & 0.95 & 0.948 & 0.85 & 0.857 & 1.00 & 1.000 \\
 \hline
 3 & 8 & 0.90 & 0.903 & 0.90 & 0.903 & 0.95 & 0.948 & 0.90 & 0.900 & 1.00 & 1.000 & 0.80 & 0.797 & 0.80 & 0.797 & 0.85 & 0.841 & 0.85 & 0.844 & 1.00 & 1.000 \\
 3 & 8 & 0.80 & 0.802 & 0.80 & 0.800 & 0.85 & 0.847 & 0.80 & 0.802 & 1.00 & 1.000 & 0.90 & 0.896 & 0.95 & 0.943 & 0.95 & 0.941 & 0.90 & 0.896 & 1.00 & 1.000 \\
 3 & 8 & 0.60 & 0.619 & 0.65 & 0.667 & 0.85 & 0.847 & 0.80 & 0.802 & 1.00 & 1.000 & 0.75 & 0.767 & 0.75 & 0.764 & 0.95 & 0.941 & 0.90 & 0.896 & 1.00 & 1.000 \\
 3 & 8 & 0.80 & 0.802 & 0.85 & 0.855 & 0.85 & 0.847 & 0.90 & 0.903 & 1.00 & 1.000 & 0.90 & 0.896 & 0.80 & 0.800 & 0.95 & 0.941 & 0.80 & 0.797 & 1.00 & 1.000 \\
 3 & 8 & 0.80 & 0.808 & 0.85 & 0.853 & 0.95 & 0.943 & 0.85 & 0.853 & 1.00 & 1.000 & 0.80 & 0.803 & 0.85 & 0.847 & 0.95 & 0.935 & 0.85 & 0.847 & 1.00 & 1.000 \\
	\hline
	\end{tabular}
	\vspace{0.6em}
\end{table}

\begin{table}[hbt!]
	\vspace{-0.5em}	
	\caption{Comparison of true and estimated worker quality based on Algorithm~\ref{algo:ENN3}.}  \label{tab:quality_results_ENN3} 
	\vspace{+0.1em}	
	\setlength\tabcolsep{2.1pt} 
	\scriptsize
	\centering
	\begin{tabular}{lc|cc|cc|cc|cc|cc|cc|cc|cc|cc|cc}
	\hline
	Sim & d  &$a_1$&$\widehat{a}_1$&$a_2$&$\widehat{a}_2$&$a_3$&$\widehat{a}_3$&$a_4$&$\widehat{a}_4$&$a_5$&$\widehat{a}_5$&$b_1$&$\widehat{b}_1$&$b_2$&$\widehat{b}_2$&$b_3$&$\widehat{b}_3$&$b_4$&$\widehat{b}_4$&$b_5$&$\widehat{b}_5$ \\
	\hline
 1 & 4 & 0.90 & 0.893 & 0.90 & 0.893 & 0.95 & 0.936 & 0.90 & 0.889 & 1.00 & 0.980 & 0.80 & 0.811 & 0.80 & 0.811 & 0.85 & 0.857 & 0.85 & 0.858 & 0.90 & 0.903 \\
 1 & 4 & 0.80 & 0.796 & 0.80 & 0.792 & 0.85 & 0.840 & 0.80 & 0.796 & 0.80 & 0.789 & 0.90 & 0.905 & 0.95 & 0.952 & 0.95 & 0.951 & 0.90 & 0.905 & 1.00 & 0.992 \\
 1 & 4 & 0.60 & 0.614 & 0.65 & 0.662 & 0.85 & 0.837 & 0.80 & 0.793 & 0.80 & 0.790 & 0.75 & 0.771 & 0.75 & 0.770 & 0.95 & 0.953 & 0.90 & 0.907 & 0.95 & 0.955 \\
 1 & 4 & 0.80 & 0.793 & 0.85 & 0.848 & 0.85 & 0.837 & 0.90 & 0.895 & 0.80 & 0.790 & 0.90 & 0.907 & 0.80 & 0.811 & 0.95 & 0.953 & 0.80 & 0.809 & 0.95 & 0.955 \\
 1 & 4 & 0.80 & 0.800 & 0.85 & 0.843 & 0.95 & 0.931 & 0.85 & 0.843 & 0.90 & 0.887 & 0.80 & 0.813 & 0.85 & 0.859 & 0.95 & 0.951 & 0.85 & 0.859 & 0.90 & 0.905 \\
 \hline
 2 & 6 & 0.90 & 0.908 & 0.90 & 0.908 & 0.95 & 0.954 & 0.90 & 0.905 & 1.00 & 0.986 & 0.80 & 0.808 & 0.80 & 0.808 & 0.85 & 0.853 & 0.85 & 0.855 & 0.90 & 0.899 \\
 2 & 6 & 0.80 & 0.808 & 0.80 & 0.805 & 0.85 & 0.854 & 0.80 & 0.808 & 0.80 & 0.802 & 0.90 & 0.903 & 0.95 & 0.949 & 0.95 & 0.948 & 0.90 & 0.903 & 1.00 & 0.992 \\
 2 & 6 & 0.60 & 0.620 & 0.65 & 0.669 & 0.85 & 0.851 & 0.80 & 0.806 & 0.80 & 0.803 & 0.75 & 0.769 & 0.75 & 0.768 & 0.95 & 0.949 & 0.90 & 0.904 & 0.95 & 0.951 \\
 2 & 6 & 0.80 & 0.807 & 0.85 & 0.861 & 0.85 & 0.852 & 0.90 & 0.910 & 0.80 & 0.804 & 0.90 & 0.903 & 0.80 & 0.808 & 0.95 & 0.949 & 0.80 & 0.807 & 0.95 & 0.950 \\
 2 & 6 & 0.80 & 0.812 & 0.85 & 0.858 & 0.95 & 0.949 & 0.85 & 0.858 & 0.90 & 0.904 & 0.80 & 0.810 & 0.85 & 0.855 & 0.95 & 0.946 & 0.85 & 0.855 & 0.90 & 0.901 \\
 \hline
 3 & 8 & 0.90 & 0.895 & 0.90 & 0.895 & 0.95 & 0.939 & 0.90 & 0.892 & 1.00 & 0.983 & 0.80 & 0.803 & 0.80 & 0.803 & 0.85 & 0.849 & 0.85 & 0.851 & 0.90 & 0.894 \\
 3 & 8 & 0.80 & 0.800 & 0.80 & 0.798 & 0.85 & 0.845 & 0.80 & 0.800 & 0.80 & 0.795 & 0.90 & 0.898 & 0.95 & 0.945 & 0.95 & 0.943 & 0.90 & 0.898 & 1.00 & 0.992 \\
 3 & 8 & 0.60 & 0.618 & 0.65 & 0.665 & 0.85 & 0.843 & 0.80 & 0.798 & 0.80 & 0.796 & 0.75 & 0.769 & 0.75 & 0.767 & 0.95 & 0.946 & 0.90 & 0.901 & 0.95 & 0.948 \\
 3 & 8 & 0.80 & 0.798 & 0.85 & 0.850 & 0.85 & 0.842 & 0.90 & 0.898 & 0.80 & 0.795 & 0.90 & 0.901 & 0.80 & 0.804 & 0.95 & 0.946 & 0.80 & 0.801 & 0.95 & 0.949 \\
 3 & 8 & 0.80 & 0.803 & 0.85 & 0.847 & 0.95 & 0.935 & 0.85 & 0.847 & 0.90 & 0.891 & 0.80 & 0.807 & 0.85 & 0.852 & 0.95 & 0.942 & 0.85 & 0.852 & 0.90 & 0.897 \\
	\hline
	\end{tabular}
	\vspace{-0.6em}
\end{table}

\subsection{Proof of Theorem~\ref{thm:ENN_re}}\label{sec:pf_thm:ENN_re}
For the sake of simplicity, we omit $\bw_{n_j}$ in the subscript of such notations as $\widehat{\phi}_{n_j,s,\bw_j}^{E}$ and $S_{n_j,\bw_j}^{j}$. Write $\mathring{P}^0=\pi_1^0 P_1^0 - (1-\pi_1^0) P_0^0$. We have 
\begin{eqnarray*}
\textrm{Regret} (\widehat{\phi}_{n_j,s}^{E}) 
&=& {\mathbb E}[R(\widehat{\phi}_{n_j,s}^{E})]- R(\phi^{\ast})  \\
&=& \int_{{\cal R}} \pi_1^0 \big[  {\mathbb P}\big(\widehat{\phi}_{n_j,s}^{E}(x) =0 \big) -  \indi{\phi^{\ast}(x) =0} \big]  d P_1^0(x) \\
&~~~& + \int_{{\cal R}} (1-\pi_1^0) \big[  {\mathbb P}\big(\widehat{\phi}_{n_j,s}^{E}(x) =1 \big) - \indi{\phi^{\ast}(x) =1} \big]  d P_0^0(x) \\
&=& \int_{{\cal R}} \big[  {\mathbb P}\big(\widehat{\phi}_{n_j,s}^{E}(x) =0 \big) -  \indi{\eta^0(x)<1/2} \big] d \mathring{P}^0(x).
\end{eqnarray*}

Without loss of generality, we consider the $j$-th worker data of ${\cal D}^{C}$: ${\cal D}^{j}=\{(X_i^{j},Y_i^{j}), i=1,\ldots,n_j\}$. Given $X=x$, we define $(X_{(i)}^{j},Y_{(i)}^{j})$ such that $\|X_{(1)}^{j}-x\|\le \|X_{(2)}^{j}-x\|\le \ldots \le \|X_{(n)}^{j}-x\|$.

Denote the estimated regression function on the $j$-th enhanced worker data as \begin{equation*}
S_{n_j}^{E}(x)={\textstyle\sum}_{i=1}^{n_j}w_{j,i}\tilde{Y}_{(i)}^{j},
\end{equation*}
where
$\tilde{Y}_{(i)}^{j}=\frac{Y_{(i)}^{j}+b^j-1}{a^j+b^j-1}$ is the enhanced label. Denote the weighted average of estimated regression function from $s$ worker data as 
$$
S_{n_j,s}^{E}(x)={\textstyle\sum}_{j=1}^sW_j S_{n_j}^{E}(x)= {\textstyle\sum}_{j=1}^sW_j{\textstyle\sum}_{i=1}^{n_j}w_{j,i}\tilde{Y}_{(i)}^{j}.
$$
We can also write $S_{n_j,s}^{E}(x)$ as
\begin{eqnarray*}
S_{n_j,s}^{E}(x)={\textstyle\sum}_{j=1}^s W_j {\textstyle\sum}_{i=1}^{n_j} w_{j,i}\tilde{Y}_{(i)}^{j}={\textstyle\sum}_{j=1}^s {\textstyle\sum}_{i=1}^{n_j} W_jw_{j,i}\tilde{Y}_{(i)}^{j} ={\textstyle\sum}_{l=1}^N  w_{Nl}\tilde{Y}_l,
\end{eqnarray*}
where 
\begin{align*}
N=&{\textstyle\sum}_{j=1}^s n_j,\\
\{\tilde{Y}_1,\tilde{Y}_2,\ldots \tilde{Y}_N\} =&\{\tilde{Y}_{(1)}^{1},\tilde{Y}_{(2)}^{1},\ldots,\tilde{Y}_{(n_1)}^{(1)},\ldots,\tilde{Y}_{(1)}^{s},\tilde{Y}_{(2)}^{s},\ldots,\tilde{Y}_{(n_s)}^{s} \},\\
\{w_{N1},w_{N2},\ldots w_{NN}\}
=&\{W_1w_{1,1},W_1w_{1,2}\ldots,W_1w_{1,n_1},\ldots,W_{s}w_{s,1},W_{s}w_{s,2}\ldots,W_{s}w_{s,n_s} \}.
\end{align*}
The ENN classifier is defined as
$$
\widehat{\phi}_{n_j,s}^{E}(x)=\indi{S_{n_j,s}^{E}(x)\ge1/2}.
$$
Since ${\mathbb P}\big(\widehat{\phi}_{n_j,s}^{E}(x)=0\big)= {\mathbb P}\big(S_{n_j,s}^{E}(x)<1/2\big)$, the regret of ENN becomes
\begin{align*}
{\rm Regret}(\widehat{\phi}_{n_j,s}^{E}) &= \int_{{\cal R}} \big\{ {\mathbb P}(S_{n_j,s}^{E}(x) < 1/2) - \indi{\eta^0(x)<1/2}\big\} d\mathring{P}^0(x).
\end{align*}

In the expert data, denote the boundary ${\cal S}=\{x\in {\cal R}: \eta^0(x)=1/2\}$. For $\epsilon>0$, let ${\cal S}^{\epsilon\epsilon} = \{x\in {\mathbb R}^d: \eta^0(x)=1/2 ~\textrm{and}~ \textrm{dist}(x, {\cal S})<\epsilon \}$, where $\textrm{dist}(x, {\cal S})=\inf_{x_0\in {\cal S}} \|x-x_0\|$. We will focus on the set
$$
{\cal S}^{\epsilon} = \Big\{x_0 + t\frac{\dot{\eta}^0(x_0)}{\|\dot{\eta}^0(x_0)\|}: x_0 \in {\cal S}^{\epsilon\epsilon}, |t| < \epsilon \Big\}.
$$

Let $\mu_{n_j}^j(x)={\mathbb E}\{S_{n_j}^{E}(x)\}$, $[\sigma_{n_j}^j(x)]^2=\textrm{Var}\{S_{n_j}^{E}(x)\}$, and $\epsilon_{n_j}=n_j^{-\beta/(4d)}$. Denote $s_{n_j}^2=\sum_{i=1}^{n_j} w_{j,i}^2$ and $t_{n_j}=n_j^{-2/d}\sum_{i=1}^{n_j} \alpha_i w_{j,i}$. From Lemma~\ref{lemma:mu_sigma_j}, we have
uniformly for $\bw_{n_j}\in W_{n_j,\beta}$,
\begin{eqnarray*}
\sup_{x\in {\cal S}^{\epsilon_{n_j}}} |\mu_{n_j}^j(x) - \eta^0(x)-a^0(x)t_{n_j}| &=& o(t_{n_j}), \\
\sup_{x\in {\cal S}^{\epsilon_{n_j}}}  \big|[\sigma_{n_j}^j(x)]^2-\frac{1}{4}s_{n_j}^2\big| &=& o(s_{n_j}^2).
\end{eqnarray*}
Let $\mu_{n_j,s}(x)={\mathbb E}\{S_{n_j,s}^{E}(x)\}$, $\sigma_{n_j,s}^2(x)=\textrm{Var}\{S_{n_j,s}^{E}(x)\}$. We have 
\begin{align*}
\mu_{n_j,s}(x)&={\mathbb E}\{S_{n_j,s}^{E}(x)\} = {\mathbb E}\{  {\textstyle\sum}_{j=1}^sW_j S_{n_j}^{E}(x)\} ={\textstyle\sum}_{j=1}^s W_j\mu_{n_j}^j(x), \\
\sigma_{n_j,s}^2(x)&=\textrm{Var}\{S_{n_j,s}^{E}(x)\} = \textrm{Var}\{ {\textstyle\sum}_{j=1}^sW_j S_{n_j}^{E}(x)\} ={\textstyle\sum}_{j=1}^s (W_j)^2[\sigma_{n_j}^j(x)]^2.
\end{align*}

Denote $\epsilon_{n_j,s}=\underset{j\in\{1\dots s\}}{\min}\{\epsilon_{n_j}\}$, $s_{n_j,s}^2={\textstyle\sum}_{j=1}^sW_j^2s_{n_j}^2$ and $t_{n_j,s}={\textstyle\sum}_{j=1}^s W_jt_{n_j}$. From Lemma~\ref{lemma:mu_sigma_s}, we have, uniformly for $\bw_{n_j}\in W_{n_j,\beta}$,
\begin{eqnarray}
\sup_{x\in {\cal S}^{\epsilon_{{n_j},s}}} |\mu_{n_j,s}(x) - \eta^0(x)- a^0(x)t_{n_j,s}| &=& o(t_{n_j,s}), \label{step1:tnjs_ENN}\\
\sup_{x\in {\cal S}^{\epsilon_{{n_j},s}}} \big|\sigma_{n_j,s}^2(x)-\frac{1}{4}s_{n_j,s}^2\big| &=& o(s_{n_j,s}^2).\label{step1:snjs_ENN}
\end{eqnarray}

We organize our proof in three steps. In {\it Step 1}, we decompose the integral over ${\cal R} \cap {\cal S}^{\epsilon_{{n_j},s}}$ as an integral along ${\cal S}$ and an integral in the perpendicular direction; in {\it Step 2}, we focus on the complement set ${\cal R} \backslash {\cal S}^{\epsilon_{{n_j},s}} $; {\it Step 3} combines the results and applies a normal approximation in ${\cal S}^{\epsilon_{{n_j},s}}$ to yield the final conclusion.

{\it Step 1}: For $x_0\in {\cal S}$ and $t\in {\mathbb R}$, denote $x_0^t=x_0+t \dot{\eta}^0(x_0)/\|\dot{\eta}^0(x_0)\|$. Denote $\psi^0=\pi_1^0 f_1^0 - (1-\pi_1^0) f_0^0$, $\bar{f}=\pi_1^0 f_1^0 + (1-\pi^0_1) f_0^0$ as the Radon-Nikodym derivatives with respect to Lebesgue measure of the restriction of $\mathring{P}^0$ and $\bar{P}^0$ to ${\cal S}^{\epsilon_{{n_j},s}}$ for large $n_j$ respectively.

Similar to \citet{S12}, we consider a change of variable from $x$ to $x_0^t$. By the theory of integration on manifolds and Weyl's tube formula \citep{G04}, we have, uniformly for $\bw_{n_j}\in W_{n_j,\beta}$, 
\begin{align}
&\int_{{\cal R}\cap{{\cal S}^{\epsilon_{{n_j},s}}}} \big\{ {\mathbb P}(S_{n_j,s}^{E}(x) < 1/2) - \indi{\eta^0(x)<1/2} \big\} d \mathring{P}^0(x) \nonumber\\  
=&\int_{{\cal S}} \int_{-\epsilon_{{n_j},s}}^{\epsilon_{{n_j},s}}\psi(x_0^t) \big\{{\mathbb P}\big(S_{n_j,s}^{E}(x_0^t) < 1/2\big) 
- \indi{t<0}\big\} dt d\textrm{Vol}^{d-1}(x_0)\{1+o(1)\}. \nonumber
\end{align}

{\it Step 2}: Bound the contribution to regret from ${\cal R}\backslash {\cal S}^{\epsilon_{{n_j},s}}$. We show that 
\begin{align*}
\sup_{\bw_{n_j}\in W_{n_j,\beta}} \int_{{\cal R}\backslash {\cal S}^{\epsilon_{{n_j},s}}} \big\{ {\mathbb P}\big(S_{n_j,s}^{E}(x) < 1/2\big) 
- \indi{\eta^0(x)<1/2} \big\} d \mathring{P}^0(x) = o(s_{n_j,s}^2+t_{n_j,s}^2).
\end{align*}

Applying Hoeffding's inequality to $S_{n_j,s}^{E}(x)$,  uniformly for $\bw_{n_j}\in W_{n_j,\beta}$ and $x\in {\cal R}\backslash {\cal S}^{\epsilon_{{n_j},s}}$, we have
\begin{align*}
&|{\mathbb P}(S_{n_j,s}^{E}(x)< 1/2) - \indi{\eta^0(x)<1/2}| \\
\leq& \exp\Big(\frac{-2(\mu_{n_j,s}(x)-1/2)^2}{\sum_{l=1}^N (w_{Nl}-0)^2} \Big) 
\leq\exp\Big(  \frac{-2(c_{10}N^{-\beta/(4d)}/4)^2}{{\textstyle\sum}_{j=1}^s(W_j)^2 s_{n_j}^2)} \Big)\\
\leq& \exp\Big(\frac{-c_{10}^2 N^{-\beta/(2d)}}{8{\textstyle\sum}_{j=1}^s(W_j)^2n_j^{-\beta}} \Big)
= \exp\Big(\frac{-c_{10}^2 N^{-\beta/(2d)}}{8{\textstyle\sum}_{j=1}^s(n_j/N)^2n_j^{-\beta}} \Big)\\
=& \exp\Big(\frac{-c_{10}^2 N^{-\beta/(2d)}}{8N^{-\beta}{\textstyle\sum}_{j=1}^s(n_j/N)^{2-\beta}} \Big)
\leq \exp\Big(\frac{-c_{10}^2 N^{-\beta/(2d)}}{8N^{-\beta}} \Big)
= o(s_{n_j,s}^2+t_{n_j,s}^2).
\end{align*}
The second inequality holds by Lemma~\ref{lemma:Hoeff_bound1} and $c_{10}$ is a positive constant. The last inequality holds by generalized mean inequality and $\beta\in(0,1/2)$. This completes {\it Step 2}.

{\it Step 3}: In the end, we will show
\begin{eqnarray*}
&&\int_{{\cal S}} \int_{-\epsilon_{{n_j},s}}^{\epsilon_{{n_j},s}} \psi^0(x_0^t) \big\{{\mathbb P}\big(S_{n_j,s}^{E}(x_0^t) < 1/2\big) - \indi{t<0}\big\} dt d\textrm{Vol}^{d-1}(x_0) \nonumber\\
&=& B_1 s_{n_j,s}^2 + B_2 t_{n_j,s}^2 + o(s_{n_j,s}^2+t_{n_j,s}^2).
\end{eqnarray*}
Applying Taylor expansion, we have, for $x_0 \in {\cal S}$,
\begin{eqnarray*}
\psi^0(x_0^t) &=& \psi^0(x_0) + \dot{\psi}^0(x_0)^T(x_0^t-x_0) + o(x_0^t-x_0)  \\
&=& \dot{\psi}^0(x_0)^T\frac{\dot{\eta}^0(x_0)}{\|\dot{\eta}^0(x_0)\|} t + o(t) \nonumber\\
&=& \|\dot{\psi}^0(x_0)\| t + o(t), \nonumber
\end{eqnarray*}
where the above second equality holds by definition of $x_0^t$, and the third equality holds by Lemma~\ref{lemma:dot_E}.

Hence,
\begin{eqnarray}
&&\int_{{\cal S}} \int_{-\epsilon_{{n_j},s}}^{\epsilon_{{n_j},s}} \psi^0(x_0^t) \big\{{\mathbb P}\big(S_{n_j,s}^{E}(x_0^t) < 1/2\big)  - \indi{t<0}\big\} dt d\textrm{Vol}^{d-1}(x_0)\label{eq:Taylor1_re_E} \\
&=& \int_{{\cal S}} \int_{-\epsilon_{{n_j},s}}^{\epsilon_{{n_j},s}} t\|\dot{\psi}^0(x_0)\|  \big\{{\mathbb P}\big(S_{n_j,s}^{E}(x_0^t) < 1/2\big)  \nonumber\\
&&\qquad\qquad\qquad\qquad\qquad
 - \indi{t<0}\big\} dt d\textrm{Vol}^{d-1}(x_0) \{1+o(1) \}. \nonumber
\end{eqnarray}

Next, we decompose
\begin{eqnarray}
&&\int_{{\cal S}} \int_{-\epsilon_{{n_j},s}}^{\epsilon_{{n_j},s}} t\|\dot{\psi}^0(x_0)\| \big\{{\mathbb P}\big(S_{n_j,s}^{E}(x_0^t) < 1/2\big)- \indi{t<0}\big\} dt d\textrm{Vol}^{d-1}(x_0) \label{eq:decompose_11_re_E} \\
&=& \int_{{\cal S}}\int_{-\epsilon_{{n_j},s}}^{\epsilon_{{n_j},s}}t\|\dot{\psi}^0(x_0)\| \big\{\Phi\big(\frac{1/2 - \mu_{n_j,s}(x_0^t)}{\sigma_{n_j,s}(x_0^t)} \big) \nonumber \\
&&\qquad\qquad\qquad\qquad\qquad
- \indi{t<0}\big\} dt d\textrm{Vol}^{d-1}(x_0) + R_{11}. \nonumber
\end{eqnarray}

Let $Z_l= (w_{Nl}\tilde{Y}_l - w_{Nl} \mathbb{E} [\tilde{Y}_l])/\sigma_{n_j,s}(x)$ and $V=\sum_{l=1}^N Z_l$. Note that $\mathbb{E}(Z_l)=0$, $\textrm{Var}(Z_l)<\infty$, and $\textrm{Var}(V)=1$. 
The nonuniform Berry-Esseen Theorem \citep{grigor2012upper} implies that there exists a constant $c_{11}>0$, such that
$$
\Big|\mathbb{P}(V\le By) - \Phi(y)\Big| \le \frac{c_{11}A}{B^3(1 + |y|^3)},
$$
where $A=\sum_{l=1}^N E|Z_l|^3$ and $\big(\sum_{l=1}^N E|Z_l|^2)^{1/2}$. In the case of ENN, we have
\begin{align*}
A&=\sum_{l=1}^N \mathbb{E}|\frac{w_{Nl}Y_l - w_{Nl} \mathbb{E} [Y_l]}{\sigma_{n_j,s}^3(x)}|^3 \leq \sum_{l=1}^N \frac{16|w_{Nl}|^3}{s_{n_j,s}^3}=\frac{16\sum_{l=1}^N w_{Nl}^3}{s_{n_j,s}^3}, \\
B&=({\textstyle\sum}_{l=1}^N \textrm{Var}(Z_l))^{1/2}=\sqrt{ \textrm{Var}(V)}=1.
\end{align*}
Denote $c_{12}=16c_{11}$, we have
\begin{eqnarray*}
&&\sup_{x_0\in {\cal S}}\sup_{t\in [-\epsilon_{{n_j},s},\epsilon_{{n_j},s}]}\Big|\mathbb{P}\Big(\frac{S_{n_j,s}^{E}(x_0^t)-\mu_{n_j,s}(x_0^t)}{\sigma_{n_j,s}(x_0^t)}\le y \Big)-\Phi(y)\Big| \\
&\le& \frac{\sum_{l=1}^Nw_{Nl}^3}{s_{n_j,s}^3}\frac{c_{12}}{ 1 + |y|^3}. \nonumber
\end{eqnarray*}

Similar to \cite{S12}, by \eqref{step1:tnjs_ENN} and \eqref{step1:snjs_ENN}, we have there exists constants $c_{13},c_{14}>0$ such that, uniformly for $\bw_n\in W_{n,\beta}$,
\begin{equation*}
\inf_{x_0 \in {\cal S}}\inf_{c_{13} t_{n_j,s}\leq|t|\leq \epsilon_{{n_j},s}} \Big| \frac{1/2-\mu_{n_j,s}(x_0^t)}{\sigma_{n_j,s}(x_0^t)} \Big| \geq \frac{c_{14}|t|}{s_{n_j,s}}. 
\end{equation*}
Therefore, we have
\begin{align*}
& \int_{-\epsilon_{{n_j},s}}^{\epsilon_{{n_j},s}}  |t|\|\dot{\psi}^0(x_0)\| \Big| {\mathbb P}\big(S_{n_j,s}^{E}(x_0^t) < 1/2\big) - \Phi\Big(\frac{1/2 - \mu_{n_j,s}(x_0^t)}{\sigma_{n_j,s}(x_0^t)} \Big)\Big| dt \\
\leq &  \int_{|t| \leq c_{13} t_{n_j,s}} |t| \|\dot{\psi}^0(x_0)\| \frac{ c_{13}\sum_{l=1}^N w_{Nl}^3}{s_{n_j,s}^3} dt  \\
& +  \int_{c_{13} t_{n_j,s} \leq |t| \leq \epsilon_{{n_j},s}} \frac{ c_{13}\sum_{l=1}^N w_{Nl}^3}{s_{n_j,s}^3}\frac{|t| \|\dot{\psi}^0(x_0)\|}{1+ c_{14}^3 |t|^3/s_{n_j,s}^3}  dt  \\
\leq &  \frac{ c_{13}\sum_{l=1}^N w_{Nl}^3}{s_{n_j,s}^3} \int_{|t| \leq c_{13}t_{n_j,s}} |t|\|\dot{\psi}^0(x_0)\|  dt  \\
& + \frac{c_{13}\sum_{l=1}^N w_{Nl}^3}{s_{n_j,s}^3} \int_{c_{13} t_{n_j,s} \leq |t| \leq \epsilon_{{n_j},s}}  \frac{\|\dot{\psi}^0(x_0)\||t|}{ c_{14}^2 |t|^2/s_{n_j,s}^2}  dt=o(s_{n_j,s}^2+t_{n_j,s}^2). 
\end{align*}
The inequality above leads to $|R_{11}|=o(s_{n_j,s}^2+t_{n_j,s}^2)$.

Next, we decompose
\begin{align}
&\int_{{\cal S}}  \int_{-\epsilon_{{n_j},s}}^{\epsilon_{{n_j},s}} t\|\dot{\psi}^0(x_0)\| \big\{\Phi\big(\frac{1/2 - \mu_{n_j,s}(x_0^t)}{\sigma_{n_j,s}(x_0^t)} \big) - \indi{t<0}\big\} dt d\textrm{Vol}^{d-1}(x_0) \label{eq:decompose_12_re_E}\\
=& \int_{{\cal S}} \int_{-\epsilon_{{n_j},s}}^{\epsilon_{{n_j},s}} t\|\dot{\psi}^0(x_0)\| \big\{\Phi\big(\frac{-2t\|\dot{\eta}^0(x_0)\|- 2a^0(x_0)t_{n_j,s}}{s_{n_j,s}} \big) \nonumber \\
&\qquad\qquad\qquad\qquad\qquad\qquad
- \indi{t<0}\big\} dt d\textrm{Vol}^{d-1}(x_0) + R_{12}. \nonumber
\end{align}
Denote $r=t/s_{n_j,s}$ and $r_{x_0}=\frac{-a^0(x_0)t_{n_j,s}}{\|\dot{\eta}^0(x_0)\|s_{n_j,s}}$. According to Lemma~\ref{lemma:mu_sigma_s} , for a sufficiently small $\epsilon\in (0,\inf_{x_0\in {\cal S}} \|\dot{\eta}^0(x_0)\|)$ and a large $n_j$, for all $\bw_{n_j}\in W_{{n_j},\beta}$, $x_0\in {\cal S}$ and $r\in[-\epsilon_{{n_j},s}/s_{n_j,s},\epsilon_{{n_j},s}/s_{n_j,s}]$, similar to \citet{S12}, we have
$$
\Big| \frac{1/2-\mu_{n_j,s}(x_0^{rs_{n_j,s}})}{\sigma_{n_j,s}(x_0^{rs_{n_j,s}})} -[-2\|\dot{\eta}^0(x_0)\|(r-r_{x_0})] \Big| \le \epsilon^2(|r|+t_{n_j,s}/s_{n_j,s}).
$$
In addition, when $|r|\le \epsilon t_{n_j,s}/s_{n_j,s}$,
\begin{eqnarray*}
\Big| \Phi\Big(\frac{1/2-\mu_{n_j,s}(x_0^{rs_{n_j,s}})}{\sigma_{n_j,s}(x_0^{rs_{n_j,s}})}\Big) -\Phi\big(-2\|\dot{\eta}^0(x_0)\|(r-r_{x_0})\big) \Big|\le 1,
\end{eqnarray*}
and when $\epsilon t_{n_j,s}/s_{n_j,s} < |r| < \epsilon_{{n_j},s}/s_{n_j,s}$,
\begin{eqnarray*}
&&\Big| \Phi\Big(\frac{1/2-\mu_{n_j,s}(x_0^{rs_{n_j,s}})}{\sigma_{n_j,s}(x_0^{rs_{n_j,s}})}\Big) -\Phi\big(-2\|\dot{\eta}^0(x_0)\|(r-r_{x_0})\big) \Big| \\
&\le& \epsilon^2(|r|+t_{n_j,s}/s_{n_j,s})\phi(\|\dot{\eta}^0(x_0)\||r-r_{x_0}|),
\end{eqnarray*}
where $\phi$ is the density function of standard normal distribution. Therefore, we have
\begin{align} 
&\int_{-\epsilon_{{n_j},s}}^{\epsilon_{{n_j},s}} |t|\|\dot{\psi}^0(x_0)\| \Big| \Phi\Big(\frac{1/2 - \mu_{n_j,s}(x_0^t)}{\sigma_{n_j,s}(x_0^t)}\Big) 
- \Phi\Big(\frac{-2t\|\dot{\eta}^0(x_0)\|- 2a^0(x_0)t_{n_j,s}}{s_{n_j,s}} \Big)\Big| dt \nonumber\\
=& \|\dot{\psi}^0(x_0)\| s_{n_j,s}^2  \int_{-\epsilon_{{n_j},s}/s_{n_j,s}}^{\epsilon_{{n_j},s}/s_{n_j,s}} |r| \big|\Phi\Big(\frac{1/2-\mu_{n_j}(x_0^{rs_{n_j,s}})}{\sigma_{n_j}(x_0^{rs_{n_j,s}})}\Big) \nonumber \\
&\qquad\qquad\qquad\qquad\qquad
-\Phi\big(-2\|\dot{\eta}^0(x_0)\|(r-r_{x_0})\big)  \big| dr  \nonumber \\
\le & \| \dot{\psi}^0(x_0) \| s_{n_j,s}^2 \Big[ \int_{|r|\le \epsilon t_{n_j,s}/s_{n_j,s}} |r| dr \nonumber\\
&+ \epsilon^2 \int_{-\infty}^{\infty} |r|(|r|+t_{n_j,s}/s_{n_j,s})\phi(\|\dot{\eta}^0(x_0)\||r-r_{x_0}|) dr \Big] =o(s_{n_j,s}^2+t_{n_j,s}^2). \nonumber 
\end{align} 
The inequality above leads to $R_{12}=o(s_{n_j,s}^2+t_{n_j,s}^2)$.

By \eqref{eq:Taylor1_re_E}, \eqref{eq:decompose_11_re_E} and  \eqref{eq:decompose_12_re_E}, we have
\begin{eqnarray}
&&\int_{{\cal S}} \int_{-\epsilon_{{n_j},s}}^{\epsilon_{{n_j},s}} \psi^0(x_0^t) \big\{{\mathbb P}\big(S_{n_j,s}^{E}(x_0^t) < 1/2\big) - 
\indi{t<0}\big\} dt d\textrm{Vol}^{d-1}(x_0) \label{eq:exp1_re_ENN} \\
\quad &=& \int_{{\cal S}} \int_{-\epsilon_{{n_j},s}}^{\epsilon_{{n_j},s}} t\|\dot{\psi}^0(x_0)\|  \big\{\Phi\big(\frac{-2t[\|\dot{\eta}^0(x_0)\|- 2a^0(x_0)t_{n_j,s}}{s_{n_j,s}} \big) \nonumber\\
&&\qquad\qquad\qquad\qquad
- \indi{t<0}\big\} dt d\textrm{Vol}^{d-1}(x_0) + o(s_{n_j,s}^2+t_{n_j,s}^2). \nonumber
\end{eqnarray}

Finally, after replacing $t=us_{n_j,s}/2$ in \eqref{eq:exp1_re_ENN}, we have, up to $o(s_{n_j,s}^2+t_{n_j,s}^2)$ difference,
\begin{align}
{\rm Regret}(\widehat{\phi}_{n_j,s}^{E})
&=\frac{s_{n_j,s}^2}{4}\int_{{\cal S}} \int_{-\infty}^{\infty} \|\dot{\psi}^0(x_0)\| u \big\{\Phi\big(-[\|\dot{\eta}^0(x_0)\|u-\frac{2a^0(x_0)t_{n_j,s}}{s_{n_j,s}} \big) \nonumber\\
&\qquad\qquad\qquad\qquad\qquad
- \indi{u<0}\big\} du d\textrm{Vol}^{d-1}(x_0) \nonumber\\
&=\frac{s_{n_j,s}^2}{2}\int_{{\cal S}} \int_{-\infty}^{\infty} \|\dot{\eta}^0(x_0)\| \bar{f}(x_0) u \big\{\Phi\big(-\|\dot{\eta}^0(x_0)\|  u-\frac{2a^0(x_0)t_{n_j,s}}{s_{n_j,s}} \big)   \label{psi_eta_2_re_ENN}\\
&\qquad\qquad\qquad\qquad\qquad
 -\indi{u<0} \big\} du d\textrm{Vol}^{d-1}(x_0) \nonumber\\
&=B_1s_{n_j,s}^2+ B_2t_{n_j,s}^2\label{II_re_ENN}\\
&=B_1{\textstyle\sum}_{j=1}^s \Big(\frac{n_j}{N}\Big)^2 {\textstyle\sum}_{i=1}^{n_j} w_{j,i}^2+ B_2\Big({\textstyle\sum}_{j=1}^s \frac{n_j}{N}{\textstyle\sum}_{i=1}^{n_j} \frac{\alpha_i w_{j,i}}{n_j^{2/d}}\Big)^2.\nonumber
\end{align}
\eqref{psi_eta_2_re_ENN} holds by Lemma~\ref{lemma:dot_E}, and \eqref{II_re_ENN} can be calculated by Lemma~\ref{lemma:G_E}. This completes the proof of Theorem~\ref{thm:ENN_re}.\hfill $\blacksquare$

\subsection{Proof of Theorem~\ref{thm:ENN_rr} }\label{sec:pf_thm:ENN_rr}
From Theorem~\ref{thm:ENN_re} and Proposition~\ref{thm:WNN_re_E}, we have, for large $n_j$,
\begin{align*}
&\frac{{\rm Regret}(\widehat{\phi}_{n,s,\bw_{n}}^{E})}{{\rm Regret}(\widehat{\phi}_{N,\bw_{N}}^{0})} \\
=& \frac{\big[ B_1 \sum_{j=1}^s\Big(\frac{n_j}{N}\Big)^2 \sum_{i=1}^{n_j} w_{j,i}^2+ B_2 \Big(\sum_{j=1}^s\frac{n_j}{N}\sum_{i=1}^{n_j} \frac{\alpha_i w_{j,i}}{n_j^{2/d}}\Big)^2 \big]\{1+o(1)\} }{\big[ B_1\sum_{i=1}^N w_{Ni}^2 + B_2 \big(\sum_{i=1}^N \frac{\alpha_i w_{Ni}}{N^{2/d}} \big)^2 \big]\{1+o(1)\}} \\
\rightarrow& 1, \;as\; n_j\rightarrow \infty. \nonumber
\end{align*}
The last equality holds by \eqref{eq:ENN_rr_weight1} and \eqref{eq:ENN_rr_weight2}. This completes the proof of Theorem~\ref{thm:ENN_rr}. \hfill $\blacksquare$

\subsection{Proof of Corollary~\ref{thm:opt_ENN}}
\label{sec:pf_thm:opt_ENN} 
Denote $a_n\succeq b_n$ if $b_n=O(a_n)$, $a_n\succ b_n$ if $b_n=o(a_n)$, $a_n\asymp b_n$ if $a_n\succeq b_n$ and $b_n \succeq a_n$. To find the optimal value of \eqref{eq:ENN_re}, we write its Lagrangian as 
$$
L(\bw_{n_j})=\Big(\sum_{j=1}^s\frac{n_j}{N}\sum_{i=1}^{n_j} \frac{\alpha_i w_{j,i}}{n_j^{2/d}}\Big)^2 + \lambda \sum_{j=1}^s\Big(\frac{n_j}{N}\Big)^2\sum_{i=1}^n w_{j,i}^2 + \sum_{j=1}^{s}\nu_j(\sum_{i=1}^{n_j}w_{j,i}-1),
$$
where $\lambda=B_1/B_2$. Since all the weights are nonnegative, we denote $l_j^{*}=\max\{i:w_{j,i}^{*}>0\}$. Setting the derivative of $L(\bw_{n_j})$ to be $0$, we have 
\begin{equation}
\frac{\partial L(\bw_{n_j})}{\partial w_{j,i}} = 2\frac{n_j}{N} \frac{\alpha_i}{n_j^{2/d}}\Big(\sum_{j=1}^s\frac{n_j}{N}\sum_{i=1}^{l_j^{*}} \frac{\alpha_i w_{j,i}}{n_j^{2/d}}\Big) + 2\lambda \Big(\frac{n_j}{N}\Big)^2 w_{j,i} + \nu_j =0. \label{eq:derivative_ENN_re1}
\end{equation}
Dividing $n_j/N$ on both sides of \eqref{eq:derivative_ENN_re1}, we have 
\begin{equation}
 2\frac{\alpha_i}{n_j^{2/d}}\Big({\textstyle\sum}_{j=1}^s\frac{n_j}{N}{\textstyle\sum}_{i=1}^{l_j^{*}} \frac{\alpha_i w_{j,i}}{n_j^{2/d}}\Big) + 2\lambda \frac{n_j}{N}w_{j,i} +\tilde{\nu}_j =0, \label{eq:derivative_ENN_re2}
\end{equation}
where $\tilde{\nu}_j=\nu_j/\frac{n_j}{N}$.

(i) Summing \eqref{eq:derivative_ENN_re2} from $i=1$ to $l_j^{*}$, (ii) multiplying \eqref{eq:derivative_ENN_re2} by $\alpha_i$ and then summing from $i=1$ to $l_j^{*}$, we have
\begin{align}
2 n_j^{-2/d}(l_j^{*})^{1+2/d}\Big({\textstyle\sum}_{j=1}^s\frac{n_j}{N}{\textstyle\sum}_{i=1}^{l_j^{*}} \frac{\alpha_i w_{j,i}}{n_j^{2/d}}\Big) + 2\lambda \frac{n_j}{N}+\tilde{\nu}_j l_j^{*} =& 0,\label{eq:derivative_ENN_re_i}\\
2 n_j^{-2/d}{\textstyle\sum}_{i=1}^{l_j^{*}} \alpha_i^2\Big({\textstyle\sum}_{j=1}^s\frac{n_j}{N}{\textstyle\sum}_{i=1}^{l_j^{*}} \frac{\alpha_i w_{j,i}}{n_j^{2/d}}\Big) + 2\lambda \frac{n_j}{N} {\textstyle\sum}_{i=1}^{l_j^{*}}\alpha_i w_{j,i} + \tilde{\nu}_j (l_j^{*})^{1+2/d} =& 0. \label{eq:derivative_ENN_re_ii}
\end{align}
(iii) Multiplying \eqref{eq:derivative_ENN_re_i} by ${l_j^{*}}^{2/d}$ and then subtracting \eqref{eq:derivative_ENN_re_ii}, we have
\begin{align}
n_j^{-2/d}\big((l_j^{*})^{1+4/d}-{\textstyle\sum}_{i=1}^{l_j^{*}} \alpha_i^2\big)\Big({\textstyle\sum}_{j=1}^s\frac{n_j}{N}{\textstyle\sum}_{i=1}^{l_j^{*}} \frac{\alpha_i w_{j,i}}{n_j^{2/d}}\Big) + \lambda \frac{n_j}{N} {l_j^{*}}^{2/d}-\lambda \frac{n_j}{N} {\textstyle\sum}_{i=1}^{l_j^{*}}\alpha_i w_{j,i}=& 0. \label{eq:derivative_ENN_re_iii}
\end{align}
(iv) Multiplying \eqref{eq:derivative_ENN_re_iii} by $n_j^{-2/d}$ and then summing from $j=1$ to $s$, we have
\begin{align*}
\Big({\textstyle\sum}_{j=1}^sn_j^{-4/d}\big((l_j^{*})^{1+4/d}-{\textstyle\sum}_{i=1}^{l_j^{*}} \alpha_i^2\big)-\lambda\Big)\Big({\textstyle\sum}_{j=1}^s\frac{n_j}{N}{\textstyle\sum}_{i=1}^{l_j^{*}} \frac{\alpha_i w_{j,i}}{n_j^{2/d}}\Big) + \lambda{\textstyle\sum}_{j=1}^s\frac{n_j}{N}n_j^{-2/d}{l_j^{*}}^{2/d}=& 0.  
\end{align*}
Therefore, we have 
\begin{align}
\sum_{j=1}^s\frac{n_j}{N}\sum_{i=1}^{l_j^{*}} \frac{\alpha_i w_{j,i}}{n_j^{2/d}}=& \frac{\lambda\sum_{j=1}^s\frac{n_j}{N}n_j^{-2/d}{l_j^{*}}^{2/d}}{\lambda-\sum_{j=1}^sn_j^{-4/d}\big((l_j^{*})^{1+4/d}-\sum_{i=1}^{l_j^{*}} \alpha_i^2\big)}. \label{eq:derivative_ENN_re_v}
\end{align}
Plugging \eqref{eq:derivative_ENN_re_v} back into \eqref{eq:derivative_ENN_re_i}, we have
\begin{align}
\tilde{\nu}_j =&-\frac{2\lambda n_j^{-2/d}(l_j^{*})^{2/d}\sum_{j=1}^s\frac{n_j}{N}n_j^{-2/d}{l_j^{*}}^{2/d}}{\lambda+\sum_{j=1}^sn_j^{-4/d}\big(\sum_{i=1}^{l_j^{*}} \alpha_i^2-(l_j^{*})^{1+4/d}\big)}-2\lambda \frac{n_j}{N}(l_j^{*})^{-1}. \label{eq:derivative_ENN_re_vi}
\end{align}
Plugging \eqref{eq:derivative_ENN_re_v} and \eqref{eq:derivative_ENN_re_vi} back into \eqref{eq:derivative_ENN_re2}, we have
\begin{align}
 w_{j,i}^*=&\frac{1}{l_j^{*}}+\frac{((l_j^{*})^{2/d}-\alpha_i)(N/n_j) n_j^{-2/d}\sum_{j=1}^s\frac{n_j}{N}n_j^{-2/d}{l_j^{*}}^{2/d}}{\lambda+\sum_{j=1}^sn_j^{-4/d}\big(\sum_{i=1}^{l_j^{*}} \alpha_i^2-(l_j^{*})^{1+4/d}\big)}. \label{eq:optweight_ENN}
\end{align}

Here $w_{j,i}^{*}$ is decreasing in $i$, since $\alpha_i$ is increasing in $i$ and $\lambda+\sum_{j=1}^sn_j^{-4/d}\big(\sum_{i=1}^{l_j^{*}} \alpha_i^2-(l_j^{*})^{1+4/d}\big)>0$ from Lemma~\ref{lemma:alpha}. Next we solve for $l_j^{*}$. According to the definition of $l_j^{*}$, we only need to find the last $l_j$ such that $w_{j,l}^{*}>0$. Using the results from Lemma~\ref{lemma:alpha}, solving this equation reduces to finding the $l_j^{*}$ such that
\begin{align}
(1+\frac{2}{d})(l_j^{*}-1)^{2/d} \le  \frac{\lambda+\frac{4}{d(d+4)}\sum_{j=1}^sn_j^{-4/d}(l_j^{*})^{1+4/d}\{1+O(\frac{1}{l_j^{*}})\}}{l_j^{*}(N/n_j) n_j^{-2/d}\sum_{j=1}^s\frac{n_j}{N}n_j^{-2/d}{l_j^{*}}^{2/d}}+(l_j^{*})^{2/d}\qquad\nonumber\\ 
\qquad\qquad\qquad\qquad\qquad\qquad\le (1+\frac{2}{d})(l_j^{*})^{2/d}. \label{eq:sandwich_inequality_1}
\end{align}
Dividing both sides of \eqref{eq:sandwich_inequality_1} by $(l_j^{*})^{2/d}$, we have for large $n_j$
\begin{align}
(1+\frac{2}{d})\Big(\frac{l_j^{*}-1}{l_j^{*}}\Big)^{2/d} \le \frac{\lambda+\frac{4}{d(d+4)}\sum_{j=1}^sn_j(l_j^{*}/n_j)^{1+4/d}\{1+O(\frac{1}{l_j^{*}})\}}{N(l_j^{*}/n_j)^{1+2/d} \sum_{j=1}^s\frac{n_j}{N}n_j^{-2/d}{l_j^{*}}^{2/d}}+1\le 1+\frac{2}{d}. \label{eq:sandwich_inequality_2}
\end{align}
By the squeeze theorem,  the value of $l_j^{*}/n_j$ doesn't depend on $j$. Therefore, \eqref{eq:sandwich_inequality_2} can be simplied to 
\begin{align*}
(1+\frac{2}{d})\Big(\frac{l_j^{*}-1}{l_j^{*}}\Big)^{2/d} \le \frac{\lambda}{N(l_j^{*}/n_j)^{1+4/d}}+\frac{4}{d(d+4)}\{1+O(\frac{1}{l_j^{*}})\}+1\le 1+\frac{2}{d}. 
\end{align*}
Therefore, for large $n_j$, we have
\begin{align}
&l_j^{*}=\Big\lceil \Big\{\frac{d(d+4)}{2(d+2)}\Big\}^{\frac{d}{d+4}}\lambda^{\frac{d}{d+4}}\frac{n_j}{N^{\frac{d}{d+4}}}\Big\rceil=\Big\lceil \Big\{\frac{d(d+4)}{2(d+2)}\Big\}^{\frac{d}{d+4}}\Big(\frac{B_1}{B_2}\Big)^{\frac{d}{d+4}}\frac{n_j}{N^{\frac{d}{d+4}}}\Big\rceil,\nonumber\\
&{\textstyle\sum}_{i=1}^{l_j^{*}} \alpha_i w_{j,i} \asymp (l_j^{*})^{2/d}, \mbox{   and   }{\textstyle\sum}_{i=1}^{l_j^{*}} w_{j,i}^2 \asymp \frac{1}{l_j^{*}}. \nonumber
\end{align}
Due to Assumption (w.1) in Section~\ref{sec:defwnb_E}, we have $l_j^{*} \rightarrow \infty$ as $n_j \rightarrow \infty$. When $n_j\succ N^{d/(d+4)}$, plugging $l_j^{*}$ and \eqref{eq:sum_alpha} into \eqref{eq:optweight_ENN} yields the optimal weight and ${{\rm Regret}(\widehat{\phi}_{n_j,s,\bw_{n_j}^*}^{E})}/{\rm Regret}(\widehat{\phi}_{N,\bw_N^*}^{0}) \rightarrow 1.$

Denote $H(\bw_{n_j})$ as the Hessian matrix of $L(\bw_{n_j})$. We have 
\begin{align*}
\frac{\partial^2 L(\bw_{n_j})}{\partial w_{j,i}^2} =& 2\Big(\frac{n_j}{N}\Big)^2\Big(\frac{\alpha_i}{n_j^{2/d}}\Big)^2 + 2\lambda\Big(\frac{n_j}{N}\Big)^2\mbox{, and}\\ \frac{\partial^2 L(\bw_{n_j})}{\partial w_{j,i}\partial w_{j',i'}} =& 2 \Big(\frac{n_j}{N}\Big)\Big(\frac{n_{j'}}{N}\Big)\Big(\frac{\alpha_i}{n_j^{2/d}}\Big)\Big(\frac{\alpha_{i'}}{n_{j'}^{2/d}}\Big)\mbox{, if } (j,i)\neq (j',i').
\end{align*}
For any nonzero vector $X_{l^{*}}=(x_{1,1},...,x_{1,l_1^{*}},\dots,x_{s,1},...,x_{s,l_s^{*}})^T$, we have
\begin{align*}
&X_{l^{*}}^T H(\bw_n) X_{l^{*}}\\
=&\sum_{j=1}^s\sum_{j'=1}^s\Big[2\Big(\frac{n_j}{N}\Big)\Big(\frac{n_{j'}}{N}\Big) \sum_{i=1}^{l_j^{*}}\sum_{i'=1}^{l_{j'}^{*}} \Big(\frac{\alpha_i}{n_j^{2/d}}\Big)\Big(\frac{\alpha_{i'}}{n_{j'}^{2/d}}\Big) x_{j,i}x_{j',i'} \Big]+ 2\lambda\sum_{j=1}^s\Big(\frac{n_j}{N}\Big)^2\sum_{i=1}^{l_j^{*}}x_{j,i}^2 \\
=& 2\Big[\sum_{j=1}^s\Big(\frac{n_j}{N}\Big) \Big(\sum_{i=1}^{l_j^{*}}\frac{\alpha_i}{n_j^{2/d}}\Big)x_{j,i}\Big]^2 + 2\lambda\sum_{j=1}^s\Big(\frac{n_j}{N}\Big)^2\sum_{i=1}^{l_j^{*}}x_{j,i}^2>0.
\end{align*}
Therefore, $H(\bw_{n_j})$ is positive definite, and this verifies that the above optimal value achieves the global minimum.

Next, we analyze the case of $n_j=O( N^{d/(d+4)})$. Due to Assumption (w.1) in Section~\ref{sec:defwnb_E}, we have $l_j^{*} \rightarrow \infty$ as $n_j \rightarrow \infty$. Therefore, we have as $n_j\rightarrow \infty$, 
\begin{align*}
\Big(\sum_{j=1}^s\frac{n_j}{N}\sum_{i=1}^{n_j} \frac{\alpha_i w_{j,i}}{n_j^{2/d}}\Big)^2 \asymp& \Big(\sum_{j=1}^s \frac{n_j}{N} \frac{(l_j^{*})^{2/d}}{n_j^{2/d}}\Big)^2\succ\Big(\sum_{j=1}^s \frac{n_j}{N} \frac{1}{N^{\frac{d}{(d+4)}\frac{2}{d}}}\Big)^2 \succeq  N^{-4/(d+4)}.
\end{align*}
\citet{S12} showed that 
\begin{align}
{\rm Regret }(\widehat{\phi}_{N,\bw_N^*}) \asymp N^{-4/(d+4)}. \label{ownn_re_E}
\end{align}
Therefore, applying \eqref{ownn_re_E}, we have, as $n_j\rightarrow \infty$, 
\begin{align*}
\frac{{\rm Regret}(\widehat{\phi}_{n_j,s,w_{n_j}}^{E})}{{\rm Regret}(\widehat{\phi}_{N,\bw_N^*})} \asymp& \frac{ B_1 \sum_{j=1}^s\big(\frac{n_j}{N}\big)^2 \sum_{i=1}^{n_j} w_{j,i}^2+ B_2 \big(\sum_{j=1}^s\frac{n_j}{N}\sum_{i=1}^{n_j} \frac{\alpha_i w_{j,i}}{n_j^{2/d}}\big)^2}{N^{-4/(d+4)}} \\
\succeq &\frac{B_2 \big(\sum_{j=1}^s\frac{n_j}{N}\sum_{i=1}^{n_j} \frac{\alpha_i w_{j,i}}{n_j^{2/d}}\big)^2}{N^{-4/(d+4)}} \rightarrow \infty. \nonumber  
\end{align*}
This completes the proof of Corollary~\ref{thm:opt_ENN}. \hfill $\blacksquare$

\subsection{Proof of Theorem~\ref{thm:ENN2_re}}
\label{sec:pf_thm:ENN2_re} 
In this section, we apply similar notations as those in Section~\ref{sec:pf_thm:ENN_re}. For the sake of simplicity, we omit $\bw_{n_j}$ in the subscript of such notations as $\widehat{\phi}_{{n_j},s,\bw_{n_j}}^{E2}$ and $S_{n_j,s,\bw_{n_j}}^{E2}$. We have
\begin{eqnarray*}
\textrm{Regret} (\widehat{\phi}_{n_j,s}^{E2}) 
&=& \int_{{\cal R}} \big[ {\mathbb P}\big(\widehat{\phi}_{n_j,s}^{E2}(x) =0 \big) -  \indi{\eta^0(x)<1/2} \big] d \mathring{P}^0(x).
\end{eqnarray*}
Denote the estimated regression function on the $j$-th enhanced worker data with estimated worker quality as 
\begin{equation*}
S_{n_j}^{E2}(x)={\textstyle\sum}_{i=1}^{n_j}w_{j,i}\check{Y}_{(i)}^{j},
\end{equation*}
where
$\check{Y}_{(i)}^{j}=\frac{Y_{(i)}^{j}+\widehat{b}^j-1}{\widehat{a}^j+\widehat{b}^j-1}$ is the enhanced label with estimated worker quality from Algorithm~\ref{algo:ENN2}. Similarly, denote the weighted average of estimated regression function from $s$ enhanced worker data as 
$$
S_{n_j,s}^{E2}(x)={\textstyle\sum}_{j=1}^sW_j S_{n_j}^{E2}(x)= {\textstyle\sum}_{j=1}^sW_j{\textstyle\sum}_{i=1}^{n_j}w_{j,i}\check{Y}_{(i)}^{j}.
$$
Therefore, the ENN2 classifier is defined as
$$
\widehat{\phi}_{n_j,s}^{E2}(x)=\indi{S_{n_j,s}^{E2}(x)\ge1/2}.
$$
Since ${\mathbb P}\big(\widehat{\phi}_{n_j,s}^{E2}(x)=0\big)= {\mathbb P}\big(S_{n_j,s}^{E2}(x)<1/2\big)$, the regret of ENN2 becomes
\begin{align*}
{\rm Regret}(\widehat{\phi}_{n_j,s}^{E2}) &= \int_{{\cal R}} \big\{ {\mathbb P}(S_{n_j,s}^{E2}(x) < 1/2) - \indi{\eta^0(x)<1/2}\big\} d\mathring{P}^0(x).
\end{align*}

Let $\check{\mu}_{n_j}^j(x)={\mathbb E}\{S_{n_j}^{E2}(x)\}$, $[\check{\sigma}_{n_j}^j(x)]^2=\textrm{Var}\{S_{n_j}^{E2}(x)\}$. Denote $\check{s}_{n_j}^2=s_{n_j}^2+(s_{\tilde{n}}^2+t_{\tilde{n}}^2)$ and $\check{t}_{n_j}=t_{n_j}+(s_{\tilde{n}}^2+t_{\tilde{n}}^2)$, where $\tilde{n}=\underset{j\in\{1,\dots s-1\}}{\max}n_j$. From Lemma~\ref{lemma:check_mu_j} and Lemma~\ref{lemma:check_sigma_j}, we have
uniformly for $\bw_{n_j}\in W_{n_j,\beta}$,
\begin{eqnarray*}
\sup_{x\in {\cal S}^{\epsilon_{n_j}}} |\check{\mu}_{n_j}^j(x) - \eta^0(x)-a^0(x)t_{n_j}| &=& o(\check{t}_{n_j}), \\
\sup_{x\in {\cal S}^{\epsilon_{n_j}}}  \big|[\check{\sigma}_{n_j}^j(x)]^2-\frac{1}{4}s_{n_j}^2\big| &=& o(\check{s}_{n_j}^2).
\end{eqnarray*}
Let $\check{\mu}_{n_j,s}(x)={\mathbb E}\{S_{n_j,s}^{E2}(x)\}$, $\check{\sigma}_{n_j,s}^2(x)=\textrm{Var}\{S_{n_j,s}^{E2}(x)\}$. We have 
\begin{align*}
\check{\mu}_{n_j,s}(x)&={\mathbb E}\{S_{n_j,s}^{E2}(x)\} = {\mathbb E}\{ {\textstyle\sum}_{j=1}^sW_j S_{n_j}^{E2}(x)\} ={\textstyle\sum}_{j=1}^s W_j\check{\mu}_{n_j}^j(x), \\
\check{\sigma}_{n_j,s}^2(x)&=\textrm{Var}\{S_{n_j,s}^{E2}(x)\} = \textrm{Var}\{ {\textstyle\sum}_{j=1}^sW_j S_{n_j}^{E2}(x)\} ={\textstyle\sum}_{j=1}^s (W_j)^2[\check{\sigma}_{n_j}^j(x)]^2.
\end{align*}
Denote $\check{s}_{n_j,s}^2={\textstyle\sum}_{j=1}^sW_j^2\check{s}_{n_j}^2$ and $\check{t}_{n_j,s}={\textstyle\sum}_{j=1}^s W_j\check{t}_{n_j}$. From Lemma~\ref{lemma:check_mu_sigma_s}, we have, uniformly for $\bw_{n_j}\in W_{n_j,\beta}$,
\begin{eqnarray}
\sup_{x\in {\cal S}^{\epsilon_{{n_j},s}}} |\check{\mu}_{n_j,s}(x) - \eta^0(x)- a^0(x)t_{n_j,s}| &=& o(\check{t}_{n_j,s}), \label{step1:check_tnjs_ENN}\\
\sup_{x\in {\cal S}^{\epsilon_{{n_j},s}}} \big|\check{\sigma}_{n_j,s}^2(x)-\frac{1}{4}s_{n_j,s}^2\big| &=& o(\check{s}_{n_j,s}^2).\label{step1:check_snjs_ENN}
\end{eqnarray}

Comparing \eqref{step1:check_tnjs_ENN} and  \eqref{step1:tnjs_ENN}, we find that $\check{\mu}_{n_j,s}(x)$ and $\mu_{n_j,s}(x)$ have a similar property. In addition, comparing \eqref{step1:check_snjs_ENN} and  \eqref{step1:snjs_ENN}, $\check{\sigma}_{n_j,s}^2(x)$ and $\sigma_{n_j,s}^2(x)$ also have a similar property. Therefore, after substituting $\mu_{n_j,s}(x)$ and $\sigma_{n_j,s}^2(x)$ by $\check{\mu}_{n_j,s}(x)$ and $\check{\sigma}_{n_j,s}^2(x)$ in {\it Step 1}, {\it Step 2} and {\it Step 3} of Section~\ref{sec:pf_thm:ENN_re}, we have, up to $o(\check{s}_{n_j,s}^2+\check{t}_{n_j,s}^2)$ difference,
\begin{align}
{\rm Regret}(\widehat{\phi}_{n_j,s}^{E2})
&=B_1s_{n_j,s}^2+ B_2t_{n_j,s}^2.\label{II_re_ENN2}
\end{align}
Therefore, applying Lemma~\ref{lemma:remainder_check} and \eqref{II_re_ENN2}, we have
\begin{align*}
{\rm Regret}(\widehat{\phi}_{n_j,s}^{E2})
&=[B_1s_{n_j,s}^2+ B_2t_{n_j,s}^2]\{1+o(1)\}.
\end{align*}
This completes the proof of Theorem~\ref{thm:ENN2_re}. \hfill $\blacksquare$

\subsection{Lemmas}\label{sec:main_lemma_E}
In this section, we provide some lemmas. 
\begin{itemize}
    \item Lemma~\ref{lemma:tilde_eta_a_j}--Lemma~\ref{lemma:G_E} are used for proving Theorem~\ref{thm:ENN_re}.
    \item Lemma~\ref{lemma:alpha} is used for proving Corollary~\ref{thm:opt_ENN}.
    \item Lemma~\ref{lemma:check_mu_j}--Lemma~\ref{lemma:remainder_check} are used for proving Theorem~\ref{thm:ENN2_re}.
\end{itemize}

\begin{lemma}
\label{lemma:tilde_eta_a_j} 
We have \begin{align*}
\tilde{\eta}^j(x)= \eta^0(x)
\mbox{   and   } \tilde{a}^j(x)= a^0(x).
\end{align*}
\end{lemma}
\noindent {Proof of Lemma~\ref{lemma:tilde_eta_a_j}:}
First, we have
\begin{align*}
\tilde{\eta}^j(x)&={\mathbb E}^j(\tilde{Y}^j|X^j=x)\\
&={\mathbb E}^j(\frac{Y^{j}+b^j-1}{a^j+b^j-1}|X^j=x)\\
&=\frac{{\mathbb E}^j(Y^{j}|X^j=x)+b^j-1}{a^j+b^j-1}\\
&=\frac{\eta^j(x)+b^j-1}{a^j+b^j-1}\\
&=\frac{a^j\eta^0(x)+(1-b^j)(1-\eta^0(x))+b^j-1}{a^j+b^j-1}\\
&= \eta^0(x).
\end{align*}
Next, we have
\begin{align*}
\tilde{a}^j(x)&=\sum_{m=1}^d  \frac{c_{m,d}\{\tilde{\eta}_{m}^j(x)\bar{f}_m(x) + 1/2 \tilde{\eta}_{mm}^j(x)\bar{f}(x)\}}{a_d^{1+2/d} \bar{f}(x)^{1+2/d}}\\
&=\sum_{m=1}^d  \frac{c_{m,d}\{\eta_{m}^0(x)\bar{f}_m(x) + 1/2 \eta_{mm}^0(x)\bar{f}(x)\}}{a_d^{1+2/d} \bar{f}(x)^{1+2/d}}\\
&= a^0(x). \hfill \blacksquare
\end{align*}

\begin{lemma}
\label{lemma:mu_sigma_j} 
Uniformly for $\bw_{n_j}\in W_{n_j,\beta}$, we have
\begin{eqnarray*}
\sup_{x\in {\cal S}^{\epsilon_{n_j}}} |\mu_{n_j}^j(x) - \eta^0(x)-a^0(x)t_{n_j}| &=& o(t_{n_j}), \\
\sup_{x\in {\cal S}^{\epsilon_{n_j}}}  \big|[\sigma_{n_j}^j(x)]^2-\frac{1}{4}s_{n_j}^2\big| &=& o(s_{n_j}^2).
\end{eqnarray*}
\end{lemma}
\noindent {Proof of Lemma~\ref{lemma:mu_sigma_j}:} We have
\begin{align}
\mu_{n_j}^j(x)=&{\textstyle\sum}_{i=1}^{n_j}w_{j,i}{\mathbb E}_X[\tilde{\eta}^j (X_{(i)}^j)]\nonumber\\
=&{\textstyle\sum}_{i=1}^{n_j}w_{j,i}{\mathbb E}_X[\eta^0 (X_{(i)}^j)]\nonumber\\
=&\mu_{n_j}^0(x). \label{eq:mu_n_j}
\end{align}
The second equality holds by Lemma~\ref{lemma:tilde_eta_a_j}. Similarly, we have
\begin{align}
[\sigma_{n_j}^j(x)]^2=&{\textstyle\sum}_{i=1}^{n_j}w_{j,i}^2\big[{\mathbb E}_X[\tilde{\eta}^j (X_{(i)}^j)]-({\mathbb E}_X[\tilde{\eta}^j (X_{(i)}^j)])^2\big]\nonumber\\
=&{\textstyle\sum}_{i=1}^{n_j}w_{j,i}^2\big[{\mathbb E}_X[\tilde{\eta}^0 (X_{(i)}^j)]-({\mathbb E}_X[\tilde{\eta}^0 (X_{(i)}^j)])^2\big]\nonumber\\
=&[\sigma_{n_j}^0(x)]^2. \label{eq:sigma_n_j}
\end{align}
In addition, \citet{S12} showed that, uniformly for $\bw_n\in W_{n,\beta}$,
\begin{eqnarray}
\sup_{x\in {\cal S}^{\epsilon_{n_j}}} |\mu_{n_j}^0(x) - \eta^0(x) - a^0(x)t_{n_j}| &=& o(t_{n_j}),\label{step1:tn0}\\
\sup_{x\in {\cal S}^{\epsilon_{n_j}}}  \big|[\sigma_{n_j}^0(x)]^2-\frac{1}{4}s_{n_j}^2\big| &=& o(s_{n_j}^2). \label{step1:sn0}
\end{eqnarray}
Therefore, applying \eqref{eq:mu_n_j}, \eqref{eq:sigma_n_j}, \eqref{step1:tn0} and \eqref{step1:sn0}, we have, uniformly for $\bw_{n_j}\in W_{n_j,\beta}$,
\begin{eqnarray*}
\sup_{x\in {\cal S}^{\epsilon_{n_j}}} |\mu_{n_j}^j(x) - \eta^0(x)-a^0(x)t_{n_j}| &=& o(t_{n_j}), \\
\sup_{x\in {\cal S}^{\epsilon_{n_j}}}  \big|[\sigma_{n_j}^j(x)]^2-\frac{1}{4}s_{n_j}^2\big| &=& o(s_{n_j}^2). \hfill \blacksquare
\end{eqnarray*}

\begin{lemma}
\label{lemma:mu_sigma_s} 
Uniformly for $\bw_{n_j}\in W_{n_j,\beta}$, we have
\begin{eqnarray*}
\sup_{x\in {\cal S}^{\epsilon_{{n_j},s}}} |\mu_{n_j,s}(x) - \eta^0(x)- a^0(x)t_{n_j,s}| &=& o(t_{n_j,s}), \\
\sup_{x\in {\cal S}^{\epsilon_{{n_j},s}}} \big|\sigma_{n_j,s}^2(x)-\frac{1}{4}s_{n_j,s}^2\big| &=& o(s_{n_j,s}^2).
\end{eqnarray*}
\end{lemma}
\noindent {Proof of Lemma~\ref{lemma:mu_sigma_s}:} We have
\begin{eqnarray*}
&&\sup_{x\in {\cal S}^{\epsilon_{{n_j},s}}} |\mu_{n_j,s}(x) - \eta^0(x)- a^0(x)t_{n_j,s}|\nonumber\\
&=&\sup_{x\in {\cal S}^{\epsilon_{{n_j},s}}} |{\textstyle\sum}_{j=1}^s W_j\mu_{n_j}^j(x) - {\textstyle\sum}_{j=1}^s W_j\eta^0(x)- a^0(x){\textstyle\sum}_{j=1}^s W_jt_{n_j}|\nonumber\\
&=&\sup_{x\in {\cal S}^{\epsilon_{{n_j},s}}} |{\textstyle\sum}_{j=1}^s W_j[\mu_{n_j}^j(x) - \eta^0(x) - a^0(x)t_{n_j}]| \nonumber\\ 
&\le& {\textstyle\sum}_{j=1}^s W_j\sup_{x\in {\cal S}^{\epsilon_{n_j}}}|[\mu_{n_j}^j(x) - \eta^0(x) - a^0(x)t_{n_j}]| \nonumber\\
&=& o({\textstyle\sum}_{j=1}^s W_jt_{n_j}) = o(t_{n_j,s}).
\end{eqnarray*}
The last inequality holds by triangle inequality and $\epsilon_{n_j,s}=\underset{j\in\{1\dots s\}}{\min}\{\epsilon_{n_j}\}$. The last second equality holds by Lemma~\ref{lemma:mu_sigma_j}. Next, we have
\begin{eqnarray*}
\sup_{x\in {\cal S}^{\epsilon_{{n_j},s}}} \big|\sigma_{n_j,s}^2(x)-\frac{1}{4}s_{n_j,s}^2\big|
&=&\sup_{x\in {\cal S}^{\epsilon_{{n_j},s}}} \Big|{\textstyle\sum}_{j=1}^s(W_j)^2[\sigma_{n_j}^2(x)-\frac{1}{4}s_{n_j}^2]\Big| \nonumber\\
&\le&{\textstyle\sum}_{j=1}^s(W_j)^2\sup_{x\in {\cal S}^{\epsilon_{n_j}}} |\sigma_{n_j}^2(x)-\frac{1}{4}s_{n_j}^2| \nonumber\\
&=& o({\textstyle\sum}_{j=1}^s(W_j)^2 s_{n_j}^2) = o(s_{n_j,s}^2). 
\end{eqnarray*}
The last inequality holds by triangle inequality and $\epsilon_{n_j,s}=\underset{j\in\{1\dots s\}}{\min}\{\epsilon_{n_j}\}$. The last second equality holds by Lemma~\ref{lemma:mu_sigma_j}. \hfill $\blacksquare$

\begin{lemma}
\label{lemma:Hoeff_bound1} 
There exists a constant $c_{10}>0$ such that, for a sufficiently large $n_j$, and uniformly for $\bw_{N}\in W_{N,\beta}$, we have
\begin{eqnarray*}
\inf_{x\in {\cal R}\backslash {\cal S}^{\epsilon_{n_j,s}} } \big| \mu_{n_j,s}(x) -1/2 \big| \geq c_{10} N^{\beta/(4d)}/4.
\end{eqnarray*}
\end{lemma}
\noindent {Proof of Lemma~\ref{lemma:Hoeff_bound1} :}
\citet{S12} showed that, there exists a constant $c_{10}>0$ such that, for a sufficiently large $N$, and uniformly for $\bw_{N}\in W_{N,\beta}$,
\begin{eqnarray*}
\inf_{x\in {\cal R}\backslash {\cal S}^{\epsilon_{N}} } \big| \mu_{n_j,s}(x) -1/2 \big| \geq c_{10} \epsilon_N/4,
\end{eqnarray*}
where $\epsilon_N=N^{\beta/(4d)}$.
Therefore, we have
\begin{eqnarray*}
\inf_{x\in {\cal R}\backslash {\cal S}^{\epsilon_{N}} } \big| \mu_{n_j,s}(x) -1/2 \big| \geq c_{10} N^{\beta/(4d)}/4.
\end{eqnarray*}
As $N\ge n_j\geq\epsilon_{n_j,s}=\underset{j\in\{1\dots s\}}{\min}\{\epsilon_{n_j}\}$, we have
\begin{eqnarray*}
\inf_{x\in {\cal R}\backslash {\cal S}^{\epsilon_{n_j,s}} } \big| \mu_{n_j,s}(x) -1/2 \big| \geq c_{10} N^{\beta/(4d)}/4. \hfill \blacksquare
\end{eqnarray*}

\begin{lemma}
\label{lemma:dot_E} 
For $x_0 \in {\cal S} $, we have
\begin{eqnarray*}
2\bar{f}(x_0) \Vert\dot{\eta}^0(x_0)\Vert = \Vert\dot{\psi}^0(x_0)\Vert \;\;{\rm and}\;\; \dot{\psi}^0(x_0)^T\dot{\eta}^0(x_0) = \Vert \dot{\eta}^0(x_0) \Vert  \Vert\dot{\psi}^0(x_0)\Vert.
\end{eqnarray*}
\end{lemma}
\noindent {Proof of Lemma~\ref{lemma:dot_E}:}  
By $\eta^0 = {\mathbb P}^0(Y^0=1|X^0=x) = \frac{\pi_1^0 f_1^0}{ \pi_1^0 f_1^0 + (1-\pi^0_1) f_0^0}$, we have 
$$
\dot{\eta}^0 = \frac{\pi_1^0(1-\pi_1^0) (\dot{f_1^0}f_0^0-f_1^0\dot{f_0^0})}{ (\pi_1^0 f_1^0 + (1-\pi_1^0) f_0^0)^2}.
$$
For $x_0 \in {\cal S}$, $\pi_1^0 f_1^0(x_0) = (1-\pi_1^0) f_0^0(x_0) = \frac{1}{2} \bar{f}(x_0)$, we have 
\begin{eqnarray*}
\dot{\eta}^0(x_0) &=& \frac{\pi_1^0(1-\pi_1^0) (\dot{f_1^0}(x_0)f_0^0(x_0)-f_1^0(x_0)\dot{f_0^0}(x_0))}{ [\pi_1^0 f_1^0(x_0) + (1-\pi_1^0) f_0^0(x_0)]^2} \\
 &=& \frac{1/2(\pi_1^0\dot{f_1^0}(x_0)-(1-\pi_1^0)\dot{f_0^0}(x_0))}{\bar{f}(x_0)} = \frac{ \dot{\psi}^0(x_0)}{2\bar{f}(x_0)}. 
 \end{eqnarray*}
Therefore, 
\begin{align*}
2\bar{f}(x_0) \Vert\dot{\eta}^0(x_0)\Vert =& \Vert\dot{\psi}^0(x_0)\Vert\;\;{\rm and}\;\; \\ \dot{\psi}^0(x_0)^T\dot{\eta}^0(x_0) =& 2\bar{f}(x_0)\dot{\eta}^0(x_0)^T\dot{\eta}^0(x_0) = \Vert \dot{\eta}^0(x_0) \Vert  \Vert\dot{\psi}^0(x_0)\Vert. \hfill \blacksquare
\end{align*}

\begin{lemma}
\label{lemma:G_E} 
\citep{SQC16} For any distribution function $G$, constant $a$, and constant $b>0$, we have
\begin{align*}
&\int_{-\infty}^{\infty} \big\{G(-bu- a ) - \indi{u<0}\big\} du = -\frac{1}{b}\big\{ a + \int_{-\infty}^{\infty} t dG(t) \big\}, \nonumber\\
&\int_{-\infty}^{\infty} u \big\{G(-bu- a ) - \indi{u<0}\big\} du \nonumber\\ 
&\qquad\qquad\qquad
= \frac{1}{b^2}\big\{ \frac{1}{2}a^2 + \frac{1}{2}\int_{-\infty}^{\infty} t^2dG(t) + a\int_{-\infty}^{\infty} t dG(t) \big\}. \nonumber \hfill \blacksquare
\end{align*}
\end{lemma}

\begin{lemma}
\label{lemma:alpha}\citep{SQC16}
Given $\alpha_i=i^{1+2/d}-(i-1)^{1+2/d}$, we have
\begin{eqnarray}
&&(1+\frac{2}{d})(i-1)^{\frac{2}{d}}\le \alpha_i \le (1+\frac{2}{d})i^{\frac{2}{d}},\nonumber \\
&&\sum_{j=1}^k \alpha_j^2 = \frac{(d+2)^2}{d(d+4)}k^{1+4/d} \big\{1+O(\frac{1}{k})\big\}. \label{eq:sum_alpha} \blacksquare
\end{eqnarray} 
\end{lemma}

\begin{lemma}
\label{lemma:check_mu_j} 
Uniformly for $\bw_{n_j}\in W_{n_j,\beta}$, we have
\begin{eqnarray*}
\sup_{x\in {\cal S}^{\epsilon_{n_j}}} |\check{\mu}_{n_j}^j(x) - \eta^0(x)-a^0(x)t_{n_j}| &=& o(\check{t}_{n_j}).
\end{eqnarray*}
\end{lemma}
\noindent {Proof of Lemma~\ref{lemma:check_mu_j}:} First, we decompose
\begin{align}
&\sup_{x\in {\cal S}^{\epsilon_{n_j}}} |\check{\mu}_{n_j}^j(x) - \eta^0(x)-a^0(x)t_{n_j}|\nonumber\\
\le&\sup_{x\in {\cal S}^{\epsilon_{n_j}}} |\check{\mu}_{n_j}^j(x) - \mu_{n_j}^j(x)|+\sup_{x\in {\cal S}^{\epsilon_{n_j}}} |\mu_{n_j}^j(x) - \eta^0(x)-a^0(x)t_{n_j}|\nonumber\\
=&\sup_{x\in {\cal S}^{\epsilon_{n_j}}}|{\mathbb E}^j\big[{\textstyle\sum}_{i=1}^{n_j}w_{j,i}\check{Y}_{(i)}^{j}\big]-{\mathbb E}^j\big[{\textstyle\sum}_{i=1}^{n_j}w_{j,i}\tilde{Y}_{(i)}^{j}\big]| + o(t_{n_j})\nonumber\\
=&\sup_{x\in {\cal S}^{\epsilon_{n_j}}}|{\textstyle\sum}_{i=1}^{n_j}w_{j,i}{\mathbb E}^j(\check{Y}_{(i)}^{j}-\tilde{Y}_{(i)}^{j})| + o(t_{n_j})=R_{81} + o(t_{n_j}).\label{eq:check_mu_j_decompose}
\end{align}
The above first equality holds by Lemma~\ref{lemma:mu_sigma_j}.

Next, we have
\begin{align*}
&|{\mathbb E}^j(\check{Y}_{(i)}^{j}-\tilde{Y}_{(i)}^{j})|\\
=&\Big|{\mathbb E}^j\Big( \frac{Y_{(i)}^{j}+\widehat{b}^j-1}{\widehat{a}^j+\widehat{b}^j-1}-\frac{Y_{(i)}^{j}+b^j-1}{a^j+b^j-1}\Big)\Big|\\
\le&\Big|{\mathbb E}^j\Big( \frac{Y_{(i)}^{j}+\widehat{b}^j-1}{\widehat{a}^j+\widehat{b}^j-1}-\frac{Y_{(i)}^{j}+\widehat{b}^j-1}{a^j+b^j-1}\Big)\Big|
+
\Big|{\mathbb E}^j\Big( \frac{Y_{(i)}^{j}+\widehat{b}^j-1}{a^j+b^j-1}-\frac{Y_{(i)}^{j}+b^j-1}{a^j+b^j-1}\Big)\Big|\\
=&R_{811}+R_{812}.
\end{align*}
Next, we have
\begin{align}
|R_{811}|=&\Big|{\mathbb E}^j\Big( \frac{Y_{(i)}^{j}+\widehat{b}^j-1}{\widehat{a}^j+\widehat{b}^j-1}-\frac{Y_{(i)}^{j}+\widehat{b}^j-1}{a^j+b^j-1}\Big)\Big|\nonumber\\
\le& \Big|{\mathbb E}^j\Big( \frac{1}{\widehat{a}^j+\widehat{b}^j-1}-\frac{1}{a^j+b^j-1}\Big)\Big|\nonumber\\
=&\Big|{\mathbb E}^j\Big( \frac{1}{(\widehat{a}^j+\widehat{b}^j-1)(a^j+b^j-1)}(\widehat{a}^j-a^j+\widehat{b}^j-b^j)\Big|\nonumber\\
=&O\big(|{\mathbb E}^j(\widehat{a}^j-a^j)|+|{\mathbb E}^j(\widehat{b}^j-b^j)|\big)=O(t_{n_s}^2+s_{n_s}^2)=o(t_{\tilde{n}}^2+s_{\tilde{n}}^2),\label{eq:R811}
\end{align}
where the last second equality holds by Lemma~\ref{lemma:hat_a_hat_b}.

Next, we have
\begin{align}
|R_{812}|=&|{\mathbb E}^j\Big( \frac{Y_{(i)}^{j}+\widehat{b}^j-1}{a^j+b^j-1}-\frac{Y_{(i)}^{j}+b^j-1}{a^j+b^j-1}\Big)|\nonumber\\
=&\frac{1}{a^j+b^j-1}\big|{\mathbb E}^j( \widehat{b}^j-b^j)\big|=O(t_{n_s}^2+s_{n_s}^2)=o(t_{\tilde{n}}^2+s_{\tilde{n}}^2).\label{eq:R812}
\end{align}
The last second equality holds by Lemma~\ref{lemma:hat_a_hat_b}.
Combining \eqref{eq:R811} and \eqref{eq:R812}, we have 
\begin{align}
|R_{81}|
=&\sup_{x\in {\cal S}^{\epsilon_{n_j}}}|{\textstyle\sum}_{i=1}^{n_j}w_{j,i}{\mathbb E}^j(\check{Y}_{(i)}^{j}-\tilde{Y}_{(i)}^{j})|=o(t_{\tilde{n}}^2+s_{\tilde{n}}^2).\label{eq:R81}
\end{align}

Therefore, combining \eqref{eq:check_mu_j_decompose} and \eqref{eq:R81}, we have
\begin{align*}
\sup_{x\in {\cal S}^{\epsilon_{n_j}}} |\check{\mu}_{n_j}^j(x) - \eta^0(x)-a^0(x)t_{n_j}| =o(t_{n_j}+t_{\tilde{n}}^2+s_{\tilde{n}}^2) =o(\check{t}_{n_j}).
\end{align*}
This completes the proof of Lemma~\ref{lemma:check_mu_j}. \hfill $\blacksquare$

\begin{lemma}
\label{lemma:hat_a_hat_b} 
Given $\widehat{a}$ and $\widehat{b}$ derived from Algorithm~\ref{algo:ENN2}, we have, uniformly for $\bw_{n_j}\in W_{n_j,\beta}$, 
\begin{align*}
|{\mathbb E}^j(\widehat{a}^j-a^j)|=&O(t_{n_s}^2+s_{n_s}^2), \mbox{   and   }\\ 
|{\mathbb E}^j(\widehat{b}^j-b^j)|=&O(t_{n_s}^2+s_{n_s}^2). 
\end{align*}
\end{lemma}
\noindent {Proof of Lemma~\ref{lemma:hat_a_hat_b}:} We decompose
\begin{align*}
&|{\mathbb E}^j(\widehat{a}^j-a^j)|\\
=&|{\mathbb E}^j\Big(\frac{{\textstyle\sum}_{i=1}^{n_j}\indi{\widehat{\phi}_{n_s,\tilde{k}}(X_i^j)=1,Y_i^j=1}}{{\textstyle\sum}_{i=1}^{n_j}\indi{\widehat{\phi}_{n_s,\tilde{k}}(X_i^j)=1}}-\frac{{\mathbb E}^j(\indi{Y^j=1,Y^0=1}}{{\mathbb E}^0(Y^0)}\Big)\Big|\\
=&\frac{1}{n_j}|{\mathbb E}^j\Big(\frac{{\textstyle\sum}_{i=1}^{n_j}\indi{\widehat{\phi}_{n_s,\tilde{k}}(X_i^j)=1,Y_i^j=1}{\mathbb E}^0(Y^0)-{\textstyle\sum}_{i=1}^{n_j}\indi{\widehat{\phi}_{n_s,\tilde{k}}(X_i^j)=1}{\mathbb E}^j(\indi{Y^j=1,Y^0=1}}
{(1/n_j){\textstyle\sum}_{i=1}^{n_j}\indi{\widehat{\phi}_{n_s,\tilde{k}}(X_i^j)=1}{\mathbb E}^0(Y^0)}\Big)\Big|\\
\le&c_{9}|{\mathbb E}^j((1/n_j){\textstyle\sum}_{i=1}^{n_j}\indi{\widehat{\phi}_{n_s,\tilde{k}}(X_i^j)=1,Y_i^j=1}){\mathbb E}^0(Y^0)\\
&-{\mathbb E}^j((1/n_j){\textstyle\sum}_{i=1}^{n_j}\indi{\widehat{\phi}_{n_s,\tilde{k}}(X_i^j)=1}){\mathbb E}^j(\indi{Y^j=1,Y^0=1}\big)\Big|\\
=&c_{9}|{\mathbb E}^j(\indi{\widehat{\phi}_{n_s,\tilde{k}}(X^j)=1,Y^j=1}){\mathbb E}^0(Y^0)-{\mathbb E}^j(\widehat{\phi}_{n_s,\tilde{k}}(X_i^j)){\mathbb E}^j(\indi{Y^j=1,Y^0=1}\big)\Big|\\
\le&c_{9}|{\mathbb E}^j(\indi{\widehat{\phi}_{n_s,\tilde{k}}(X^j)=1,Y^j=1}){\mathbb E}^0(Y^0)-{\mathbb E}^j(\indi{\widehat{\phi}_{n_s,\tilde{k}}(X^j)=1,Y^j=1}){\mathbb E}^j(\widehat{\phi}_{n_s,\tilde{k}}(X_i^j))|\\
&+c_{9}|{\mathbb E}^j(\indi{\widehat{\phi}_{n_s,\tilde{k}}(X^j)=1,Y^j=1}){\mathbb E}^j(\widehat{\phi}_{n_s,\tilde{k}}(X_i^j)))-{\mathbb E}^j(\widehat{\phi}_{n_s,\tilde{k}}(X_i^j)){\mathbb E}^j(\indi{Y^j=1,Y^0=1})\Big|\\
=&R_{91}+R_{92},
\end{align*}
where $c_{9}$ is a positive constant.

Next, we have 
\begin{align}
&|R_{91}|\nonumber\\
=&
\Big|{\mathbb E}^j(\indi{\widehat{\phi}_{n_s,\tilde{k}}(X^j)=1,Y^j=1}){\mathbb E}^0(Y^0)-{\mathbb E}^j(\indi{\widehat{\phi}_{n_s,\tilde{k}}(X^j)=1,Y^j=1}){\mathbb E}^j(\widehat{\phi}_{n_s,\tilde{k}}(X^j))\Big|\nonumber\\
=&{\mathbb E}^j(\indi{\widehat{\phi}_{n_s,\tilde{k}}(X^j)=1,Y^j=1})|{\mathbb E}^j(\widehat{\phi}_{n_s,\tilde{k}}(X^j))-{\mathbb E}^0(Y^0)|\nonumber\\
\le&|{\mathbb E}^j(\widehat{\phi}_{n_s,\tilde{k}}(X^j))-{\mathbb E}^0(Y^0)|\nonumber\\
=&|{\mathbb E}_X[{\mathbb P}(S_{n_s}^{0}(X)<1/2)-\indi{ \eta^0(X)<1/2}]|=O(t_{n_s}^2+s_{n_s}^2). \label{eq:R91}
\end{align}
The above last equality holds by applying Proposition~\ref{thm:WNN_re_E} for the $k$NN classifier on the $s$-th worker data.

Next, we have 
\begin{align}
&|R_{92}|\nonumber\\
=&\Big|{\mathbb E}^j(\indi{\widehat{\phi}_{n_s,\tilde{k}}(X^j)=1,Y^j=1}){\mathbb E}^j(\widehat{\phi}_{n_s,\tilde{k}}(X^j))-{\mathbb E}^j(\indi{Y^j=1,Y^0=1}{\mathbb E}^j(\widehat{\phi}_{n_s,\tilde{k}}(X^j))\Big|\nonumber\\
\le&\Big|{\mathbb E}^j(\indi{\widehat{\phi}_{n_s,\tilde{k}}(X^j)=1,Y^j=1})-{\mathbb E}^j(\indi{Y^j=1,Y^0=1})\Big|\nonumber\\
=&|{\mathbb E}^j(\indi{\widehat{\phi}_{n_s,\tilde{k}}(X^j)=1,Y^j=1,Y^0=0}+\indi{\widehat{\phi}_{n_s,\tilde{k}}(X^j)=0,Y^j=1,Y^0=1})|\nonumber\\
\le&|{\mathbb E}^j(\indi{\widehat{\phi}_{n_s,\tilde{k}}(X^j)=1,Y^0=0}+\indi{\widehat{\phi}_{n_s,\tilde{k}}(X^j)=0,Y^0=1})|\nonumber\\
=&|{\mathbb E}^j(\indi{\widehat{\phi}_{n_s,\tilde{k}}(X^j)=1}-\indi{Y^0=1})|\nonumber\\
=&|{\mathbb E}_X[{\mathbb P}(S_{n_s}^{0}(X)<1/2)-\indi{\eta^0(X)<1/2}]|=O(t_{n_s}^2+s_{n_s}^2). \label{eq:R92}
\end{align}
The above last equality holds by applying Proposition~\ref{thm:WNN_re_E} for the $k$NN classifier on the $s$-th worker data.

Therefore, combining \eqref{eq:R91} and \eqref{eq:R92}, we have
\begin{align*}
|{\mathbb E}^j(\widehat{a}^j-a^j)|=O(t_{n_s}^2+s_{n_s}^2).
\end{align*}
Similarly, we have  
\begin{align*}
|{\mathbb E}^j(\widehat{b}^j-b^j)|=O(t_{n_s}^2+s_{n_s}^2). 
\end{align*}
This completes the proof of Lemma~\ref{lemma:hat_a_hat_b}. \hfill $\blacksquare$

\begin{lemma}
\label{lemma:check_sigma_j} 
Uniformly for $\bw_{n_j}\in W_{n_j,\beta}$, we have
\begin{eqnarray*}
\sup_{x\in {\cal S}^{\epsilon_{n_j}}}  \big|[\check{\sigma}_{n_j}^j(x)]^2-\frac{1}{4}s_{n_j}^2\big| &=& o(\check{s}_{n_j}^2).
\end{eqnarray*}
\end{lemma}
\noindent {Proof of Lemma~\ref{lemma:check_sigma_j}:} First, we decompose
\begin{align*}
&\sup_{x\in {\cal S}^{\epsilon_{n_j}}} |[\check{\sigma}_{n_j}^j(x)]^2-\frac{1}{4}s_{n_j}^2|\\
\le&\sup_{x\in {\cal S}^{\epsilon_{n_j}}} |[\check{\sigma}_{n_j}^j(x)]^2 - [\sigma_{n_j}^j(x)]^2|+\sup_{x\in {\cal S}^{\epsilon_{n_j}}} |[\sigma_{n_j}^j(x)]^2-\frac{1}{4}s_{n_j}^2|\\
=& R_{101} + o(s_{n_j}^2).    
\end{align*}
The above last equality holds by Lemma~\ref{lemma:mu_sigma_j}.

Next, we have
\begin{align*}
&|R_{101}| \\
=&\sup_{x\in {\cal S}^{\epsilon_{n_j}}} |[\check{\sigma}_{n_j}^j(x)]^2 - [\sigma_{n_j}^j(x)]^2|\\
=&\sup_{x\in {\cal S}^{\epsilon_{n_j}}}|\textrm{Var}\big[{\textstyle\sum}_{i=1}^{n_j}w_{j,i}\check{Y}_{(i)}^{j}\big]-\textrm{Var}\big[{\textstyle\sum}_{i=1}^{n_j}w_{j,i}\tilde{Y}_{(i)}^{j}\big]|\\
=&\sup_{x\in {\cal S}^{\epsilon_{n_j}}}\big|{\mathbb E}^j\big[\big({\textstyle\sum}_{i=1}^{n_j}w_{j,i}\check{Y}_{(i)}^{j}\big)^2\big]-\big[{\mathbb E}^j\big({\textstyle\sum}_{i=1}^{n_j}w_{j,i}\check{Y}_{(i)}^{j}\big)\big]^2 \\
&- \big\{ {\mathbb E}^j\big[\big({\textstyle\sum}_{i=1}^{n_j}w_{j,i}\tilde{Y}_{(i)}^{j}\big)^2\big]-\big[{\mathbb E}^j\big({\textstyle\sum}_{i=1}^{n_j}w_{j,i}\tilde{Y}_{(i)}^{j}\big)\big]^2 \big\}\big|\\
\le&\sup_{x\in {\cal S}^{\epsilon_{n_j}}}\big|{\mathbb E}^j\big[\big({\textstyle\sum}_{i=1}^{n_j}w_{j,i}\check{Y}_{(i)}^{j}\big)^2\big]- {\mathbb E}^j\big[\big({\textstyle\sum}_{i=1}^{n_j}w_{j,i}\tilde{Y}_{(i)}^{j}\big)^2\big]\big|\\
&+\sup_{x\in {\cal S}^{\epsilon_{n_j}}}\big|\big[{\mathbb E}^j\big({\textstyle\sum}_{i=1}^{n_j}w_{j,i}\check{Y}_{(i)}^{j}\big)\big]^2-\big[{\mathbb E}^j\big({\textstyle\sum}_{i=1}^{n_j}w_{j,i}\tilde{Y}_{(i)}^{j}\big)\big]^2 \big|\\
=&\sup_{x\in {\cal S}^{\epsilon_{n_j}}}\big|{\mathbb E}^j\big[\big({\textstyle\sum}_{i=1}^{n_j}w_{j,i}\check{Y}_{(i)}^{j}-{\textstyle\sum}_{i=1}^{n_j}w_{j,i}\tilde{Y}_{(i)}^{j}\big)\big({\textstyle\sum}_{i=1}^{n_j}w_{j,i}\check{Y}_{(i)}^{j}+{\textstyle\sum}_{i=1}^{n_j}w_{j,i}\tilde{Y}_{(i)}^{j}\big)\big]\big|\\
&+\sup_{x\in {\cal S}^{\epsilon_{n_j}}}\big|\big[{\mathbb E}^j\big({\textstyle\sum}_{i=1}^{n_j}w_{j,i}\check{Y}_{(i)}^{j}\big)-{\mathbb E}^j\big({\textstyle\sum}_{i=1}^{n_j}w_{j,i}\tilde{Y}_{(i)}^{j}\big)\big]\big[{\mathbb E}^j\big({\textstyle\sum}_{i=1}^{n_j}w_{j,i}\check{Y}_{(i)}^{j}\big)+{\mathbb E}^j\big({\textstyle\sum}_{i=1}^{n_j}w_{j,i}\tilde{Y}_{(i)}^{j}\big)\big] \big|\\
\le&4\sup_{x\in {\cal S}^{\epsilon_{n_j}}}\big|{\mathbb E}^j\big[\big({\textstyle\sum}_{i=1}^{n_j}w_{j,i}\check{Y}_{(i)}^{j}-{\textstyle\sum}_{i=1}^{n_j}w_{j,i}\tilde{Y}_{(i)}^{j}\big)\big]\big|\\
=&O(t_{n_s}^2+s_{n_s}^2)=o(t_{\tilde{n}}^2+s_{\tilde{n}}^2).
\end{align*}
The above last second inequality holds by \eqref{eq:R81}.
Therefore, we have
\begin{align*}
\sup_{x\in {\cal S}^{\epsilon_{n_j}}}  \big|[\check{\sigma}_{n_j}^j(x)]^2-\frac{1}{4}s_{n_j}^2\big| =o(s_{n_j}^2+t_{\tilde{n}}^2+s_{\tilde{n}}^2)=o(\check{s}_{n_j}^2).
\end{align*}
This completes the proof of Lemma~\ref{lemma:check_sigma_j}. \hfill $\blacksquare$

\begin{lemma}
\label{lemma:check_mu_sigma_s} 
Uniformly for $\bw_{n_j}\in W_{n_j,\beta}$, we have
\begin{eqnarray*}
\sup_{x\in {\cal S}^{\epsilon_{{n_j},s}}} |\check{\mu}_{n_j,s}(x) - \eta^0(x)- a^0(x)t_{n_j,s}| &=& o(\check{t}_{n_j,s}),\\
\sup_{x\in {\cal S}^{\epsilon_{{n_j},s}}} \big|\check{\sigma}_{n_j,s}^2(x)-\frac{1}{4}s_{n_j,s}^2\big| &=& o(\check{s}_{n_j,s}^2).
\end{eqnarray*}
\end{lemma}
\noindent {Proof of Lemma~\ref{lemma:check_mu_sigma_s}:} We have
\begin{eqnarray*}
&&\sup_{x\in {\cal S}^{\epsilon_{{n_j},s}}} |\check{\mu}_{n_j,s}(x) - \eta^0(x)- a^0(x)t_{n_j,s}|\nonumber\\
&=&\sup_{x\in {\cal S}^{\epsilon_{{n_j},s}}} |{\textstyle\sum}_{j=1}^s W_j\check{\mu}_{n_j}^j(x) - {\textstyle\sum}_{j=1}^s W_j\eta^0(x)- a^0(x){\textstyle\sum}_{j=1}^s W_jt_{n_j}|\nonumber\\
&\le& {\textstyle\sum}_{j=1}^s W_j\sup_{x\in {\cal S}^{\epsilon_{n_j}}}|[\check{\mu}_{n_j}^j(x) - \eta^0(x) - a^0(x)t_{n_j}]| \nonumber\\
&=& o({\textstyle\sum}_{j=1}^s W_j\check{t}_{n_j}) = o(\check{t}_{n_j,s}).
\end{eqnarray*}
The last inequality holds by triangle inequality and $\epsilon_{n_j,s}=\underset{j\in\{1\dots s\}}{\min}\{\epsilon_{n_j}\}$. The last second equality holds by Lemma~\ref{lemma:check_mu_j}.

Next, we have
\begin{eqnarray*}
\sup_{x\in {\cal S}^{\epsilon_{{n_j},s}}} \big|\check{\sigma}_{n_j,s}^2(x)-\frac{1}{4}s_{n_j,s}^2\big|
&=&\sup_{x\in {\cal S}^{\epsilon_{{n_j},s}}} \Big|{\textstyle\sum}_{j=1}^s(W_j)^2[\check{\sigma}_{n_j}^2(x)-\frac{1}{4}s_{n_j}^2]\Big| \nonumber\\
&\le&{\textstyle\sum}_{j=1}^s(W_j)^2\sup_{x\in {\cal S}^{\epsilon_{n_j}}} |\check{\sigma}_{n_j}^2(x)-\frac{1}{4}s_{n_j}^2| \nonumber\\
&=& o({\textstyle\sum}_{j=1}^s(W_j)^2 \check{s}_{n_j}^2) = o(\check{s}_{n_j,s}^2). 
\end{eqnarray*}
The last inequality holds by triangle inequality and $\epsilon_{n_j,s}=\underset{j\in\{1\dots s\}}{\min}\{\epsilon_{n_j}\}$. The last second equality holds by Lemma~\ref{lemma:check_sigma_j}. \hfill $\blacksquare$

\begin{lemma}
\label{lemma:remainder_check} 
Uniformly for $\bw_{n_j}\in W_{n_j,\beta}$, we have
\begin{eqnarray*}
\check{s}_{n_j,s}^2+\check{t}_{n_j,s}^2=O(s_{n_j,s}^2+t_{n_j,s}^2).
\end{eqnarray*}
\end{lemma}
\noindent {Proof of Lemma~\ref{lemma:remainder_check}:} We have
\begin{align*}
&\check{s}_{n_j,s}^2+\check{t}_{n_j,s}^2\\
=&{\textstyle\sum}_{j=1}^s(W_j)^2 \check{s}_{n_j}^2+({\textstyle\sum}_{j=1}^s W_j\check{t}_{n_j})^2\\
=&{\textstyle\sum}_{j=1}^s(W_j)^2 [s_{n_j}^2+(s_{\tilde{n}}^2+t_{\tilde{n}}^2)]+({\textstyle\sum}_{j=1}^s W_j[t_{n_j}+(s_{\tilde{n}}^2+t_{\tilde{n}}^2)])^2\\
=&{\textstyle\sum}_{j=1}^s(W_j)^2 s_{n_j}^2+({\textstyle\sum}_{j=1}^s W_jt_{n_j})^2+{\textstyle\sum}_{j=1}^s(W_j)^2(s_{\tilde{n}}^2+t_{\tilde{n}}^2)\\
&+2{\textstyle\sum}_{j=1}^s W_j(s_{\tilde{n}}^2+t_{\tilde{n}}^2)t_{n_j}+[{\textstyle\sum}_{j=1}^s W_j(s_{\tilde{n}}^2+t_{\tilde{n}}^2)]^2\\
=&s_{n_j,s}^2+t_{n_j,s}^2+O\big({\textstyle\sum}_{j=1}^s(W_j)^2(s_{n_j}^2+t_{n_j}^2)\big)\\
&+O\big({\textstyle\sum}_{j=1}^s W_j(s_{n_j}+t_{n_j})t_{n_j}\big)+O\big([{\textstyle\sum}_{j=1}^s W_j(s_{n_j}+t_{n_j})]^2\big)\\
=&O(s_{n_j,s}^2+t_{n_j,s}^2).
\end{align*}
The last second equality holds by $n_j/n_s=O(1)$, $\tilde{n}=\underset{j\in\{1,\dots s-1\}}{\max}n_j$ and $t_{n_s}^2+s_{n_s}^2=O(s_{n_j}^2+t_{n_j}^2)=o(s_{n_j}+t_{n_j})$. \hfill $\blacksquare$

\end{document}